\documentclass{aa}
\usepackage{graphicx}
\usepackage{latexsym}
\usepackage{color}
\usepackage{ulem}

\usepackage[utf8]{inputenc}
\usepackage[english]{babel}

\usepackage[dvipsnames]{xcolor}
\usepackage{amssymb}
\usepackage{longtable}    
\usepackage{lscape}
\usepackage{url}
\usepackage{natbib}
\bibpunct{(}{)}{;}{a}{}{,} % to follow the A&A style
\newcommand{\lamost}{{\sf LAMOST}}

\newcommand{\kepler}{{\it Kepler}}
\newcommand{\ktwo}{{\it K2}}
\newcommand{\degree}{$^{\circ}$}
\newcommand{\teff}{$T_{\rm eff}$}
\newcommand{\tefflasp}{$T_{\rm eff}^{\sf LASP}$}
\newcommand{\logg}{$\log g$}
\newcommand{\vsini}{$v\sin i$}

\newcommand{\feh}{[Fe/H]}

\newcommand{\project}{LK project}
\newcommand{\projectlrs}{LK-LRS project}
\newcommand{\projectmrs}{LK-MRS project}

\newcommand{\cool}{{\it cool}}
\newcommand{\hot}{{\it hot}}
\newcommand{\unclass}{{\it unclass}}
\newcommand{\rotfit}{{\sf ROTFIT}}

\newcommand{\lasp}{{\sf LASP}}
\newcommand{\kms}{km\,s$^{-1}$}
\newcommand{\halpha}{H$\alpha$}
\newcommand{\Whalpha}{$W^{\rm res}_{\rm H\alpha}$}
\newcommand{\WLi}{$W_{\rm Li}$}

\begin{document}

\title{Characterization of {\it Kepler} targets based on medium-resolution LAMOST spectra analyzed with ROTFIT\thanks{Based on observations collected with the Large Sky Area Multi-Object Fiber Spectroscopic Telescope (\lamost) located at the Xinglong observatory, China.}~\fnmsep\thanks{Tables~\ref{Tab:data_red}, \ref{Tab:data_blue}, \ref{Tab:APs}, \ref{Tab:active}, and \ref{Tab:Lithium} are only available at the CDS via anonymous ftp to {\tt cdsarc.u-strasbg.fr (130.79.128.5)} or via {\tt http://cdsarc.u-strasbg.fr/viz-bin/qcat?J/A+A/?/?}. } }

\author{A. Frasca\inst{1}\and 
        J. Molenda-\.Zakowicz\inst{2}\and
        J. Alonso-Santiago\inst{1}\and
        G. Catanzaro\inst{1}\and
        P. De Cat\inst{3}\and
        J.~N. Fu\inst{4}\and
        W. Zong\inst{4}\and
        J. X. Wang\inst{4}\and
        T. Cang\inst{4}\and
        J. T. Wang\inst{4,5}
 }

\offprints{A. Frasca\\ \email{antonio.frasca@inaf.it}}

\institute{INAF - Osservatorio Astrofisico di Catania, via S. Sofia, 78, 95123 Catania, Italy
\and
University of Wroc{\l}aw, Faculty of Physics and Astronomy, Astronomical Institute, ul.~Kopernika 11, 51-622 Wroc{\l}aw, Poland
\and
Royal observatory of Belgium, Ringlaan 3, B-1180 Brussel, Belgium
\and
Department of Astronomy, Beijing Normal University, Beijing~100875, P.~R.~China
\and
Key Lab for Optical Astronomy, National Astronomical Observatories, Chinese Academy of Sciences, Beijing~100101, P.~R.~China
}

\date{Received 4 February 2022 / Accepted 28 April 2022}

% \abstract{}{}{}{}{} 
% 5 {} token are mandatory
 
\abstract 
  % context heading (optional)
{} %leave it empty if necessary  
  % aims heading (mandatory)
{In this work we present the results of our analysis of 16\,300 medium-resolution LAMOST spectra of late-type stars in the \kepler\ field
with the aim of determining the stellar parameters, activity level, lithium atmospheric content, and binarity.}
  % methods heading (mandatory)
{We have used a version of the code \rotfit\  specifically developed for the \lamost\ medium-resolution spectra to determine
stellar parameters via the adoption of a grid of spectra of real stars.  We provide a catalog with the atmospheric parameters (\teff, \logg, and \feh), radial velocity (RV), and projected rotation velocity (\vsini).
For cool stars (\teff$\le 6500$\,K), we also calculated the \halpha\ and \ion{Li}{i}$\lambda$6708  equivalent width, which are important indicators of chromospheric activity and evolutionary stage, respectively.}
  % results heading (mandatory)
{From the sample of 16\,300 spectra, we have derived the RV and atmospheric parameters for 14\,300 spectra of 7443 stars. 
Literature data (mainly from high- or medium-resolution spectra) were used for a quality control of the results and to assess the accuracy of the derived parameters. The \teff\ and \logg\ values are in good agreement with the literature, although their distribution displays some clustering effects, which may be the result of the nonuniform distribution of the templates in the parameter space. 
The most relevant differences are found for \feh, which appears to be overestimated for metal-poor stars; this overestimation is also likely due to the template grid. We propose a relation to correct the \feh\ values derived with \rotfit.
We were able to identify interesting objects, such as double-lined binaries, stars with variable RVs, lithium-rich giants, and emission-line objects. 
Based on the \halpha\ flux, we found 327 active stars. We were able to detect the \ion{Li}{i}$\lambda$6708 line and measure its equivalent width for 1657 stars, both giants and stars on the main sequence.
Regarding the latter, we performed a discrete age classification based on the atmospheric lithium abundance and the upper envelopes of a few open clusters. 
Among the giants, we found 195 Li-rich stars, 161 of which are reported here for the first time. No relationship is found between stellar rotation and lithium abundance, which allows us to rule out merger scenarios as the predominant explanation of the enrichment of Li in our sample. The fraction of Li-rich giants, $\approx$\,4\%, is higher than expected.
}
 % conclusions heading (optional), leave it empty if necessary 
{}
% MAXIMUM 6 KEYWORDS !! 
   \titlerunning{Analysis of \lamost\ medium-resolution spectra of \kepler\ targets}
      \authorrunning{A. Frasca et al.}

\maketitle

%===================================================================
\section{Introduction}
\label{Sec:Intro}

The Large Sky Area Multi-Object Fiber Spectroscopic Telescope \citep[\lamost;][]{Cui12} is a National Major Scientific Project undertaken by the Chinese Academy of Science. It is a unique instrument, located at the Xinglong station and situated south of the main peak of the Yanshan mountains in Hebei province (China).\ \lamost\  combines a large aperture (4-meter telescope) with a wide field of view (circular region with a diameter of 5 degrees on the sky) that is covered by 4000 optical fibers. These fibers are connected to 16 multi-object optical spectrometers with 250 fibers each \citep{wang1996,xing1998}, making this instrument the ideal tool for obtaining spectroscopic observations for a large number of targets in an efficient way. The data acquired with the \lamost\ instrument allow multi-fold analyses of the observed objects to be conducted, including a homogeneous determination of the atmospheric parameters (APs): the effective temperature \teff, surface gravity \logg, metallicity \feh, radial velocity RV, and projected rotational velocity \vsini.

The \lamost\ Extra-GAlactic Survey (LEGAS) and the \lamost\ Experiment for Galactic Understanding and Exploration (LEGUE) were the two initial scientific driving forces behind the \lamost\ project \citep{Zhao2012}. However, it was soon realized that observations of the field of view of the nominal \kepler\ mission with \lamost\ would be a scientific gold mine, being a win-win opportunity for both communities: it would provide the \kepler\ community with the data needed for a homogeneous spectroscopic determination of stellar parameters for objects observed by \kepler\ while the \lamost\ community could benefit from high-precision results derived from data obtained elsewhere for \kepler\ objects to calibrate the \lamost\ results. Therefore, the proposal of the \lamost-\kepler\ project (hereafter \project) was well received in 2010. A detailed description is given by \citet[][hereafter Paper~I]{decat2015}.\ The first observations for the project were carried out during the test phase of \lamost\ in early 2011. The scientific observations of the pilot survey of \lamost\ began on October 24, 2011, while the first 5-year regular survey started about one year later, on September 28, 2012. During the first regular survey of \lamost, only single-shot low-resolution spectra (LRS) with a spectral resolution $R \sim 1800$ covering optical wavelengths ranging from 370 to 900\,nm were gathered. 

The purpose of the observations collected during the \projectlrs\ is multifarious. First, the APs yielded by the \project\ complement and can serve as a test bench for the content of the \kepler\ Input Catalog \citep[KIC;][]{brown2011}. As such, they provide firm bases for asteroseismic and evolutionary modeling of stars in the \kepler\ field. Second, the gathered data enable us to flag interesting objects as they allow us to identify 
fast-rotating stars and objects for which the variability in radial velocity (RV) exceeds $\sim 20$~\kms; such objects are good candidates for spectroscopic binaries or pulsating stars. 
Similarly, stars that show strong emission in their spectral lines or display other relevant spectral features can be identified and used for further research that reaches beyond asteroseismic analysis. 
The analysis of the LK LRS has been performed by three teams with different methodologies. Their results have been presented by \citet{Ren2016} and \citet{Zong2018} for the Asian group, \citet{Frasca2016} for the European group, and \citet{Gray2016} for the American group.

In 2015, the \project\ was extended to include targets within the fields observed by the \ktwo\ space mission \citep{Wang2020ApJS..251...27W}. 
Since the start of the second phase of the regular survey of \lamost\ in September 2018, medium-resolution spectrographs both in single-shot and time-series mode have also been used. 
The corresponding medium-resolution spectra (MRS) have a spectral resolution $R \sim 7500$.

Within the \projectmrs, time series of MRS \lamost\ spectra are being gathered for 4 footprints in the \kepler\ field and 16 footprints distributed within the northern \ktwo\ campaigns. 
\citet{Zong2020ApJS..251...15Z} present the first results of the \projectmrs\ based on the analysis of the data up to June 2019, with the \lamost\ stellar parameter pipeline (\lasp) adapted to the resolution of the MRS, $R \sim 7500$  \citep{luo2015,Wang2019ApJS..244...27W}.

In this paper we focus on the MRS of objects in the \kepler\ field that were gathered during the test phase of the medium-resolution spectrographs in the year between the end of the first and the start of the second regular survey of \lamost\ (September 2017 -- May 2018). 
They were made available to the general public in the sixth data release (DR6) of \lamost\ on September 30, 2020.
We adapted the code {\sf ROTFIT}, developed by \citet{Frasca2003, Frasca2006} and discussed in detail by \citet{Molenda2013}, to MRS \lamost\ spectra. Then we applied it to the spectra collected with plates that intersect with the \kepler\ field of view (both from the \projectmrs\ and the regular survey), selecting spectra of sufficient quality to derive the \teff, \logg, \feh, RV, and \vsini\ of the observed stars.
Moreover, we determined the residual equivalent width of the Balmer \halpha\ line \Whalpha\ to search for active stars, and the equivalent width of the \ion{Li}{i} $\lambda$6708\,\AA\ absorption line \WLi\ to detect young stars. 

This paper is organized as follows. In Sect.~\ref{Sec:Data} we briefly describe the selection of our sample and the observations. Section~\ref{Sec:Analysis} presents the methods of analysis and discusses the accuracy of the data. That section also includes a brief description of the {\sf ROTFIT} pipeline, the procedure for the measure of the activity indicators, and a comparison of the RVs and APs derived in this work with values from the literature. The results from the chromospheric activity indicators and the lithium abundance are presented in Sect.~\ref{Sec:Chromo}.
We summarize the main findings of this work in Sect.~\ref{Sec:Concl}.

%====================================================================
\section{Observations and sample selection}
\label{Sec:Data}

\begin{figure*}
\begin{center}
\includegraphics[width=12.5cm]{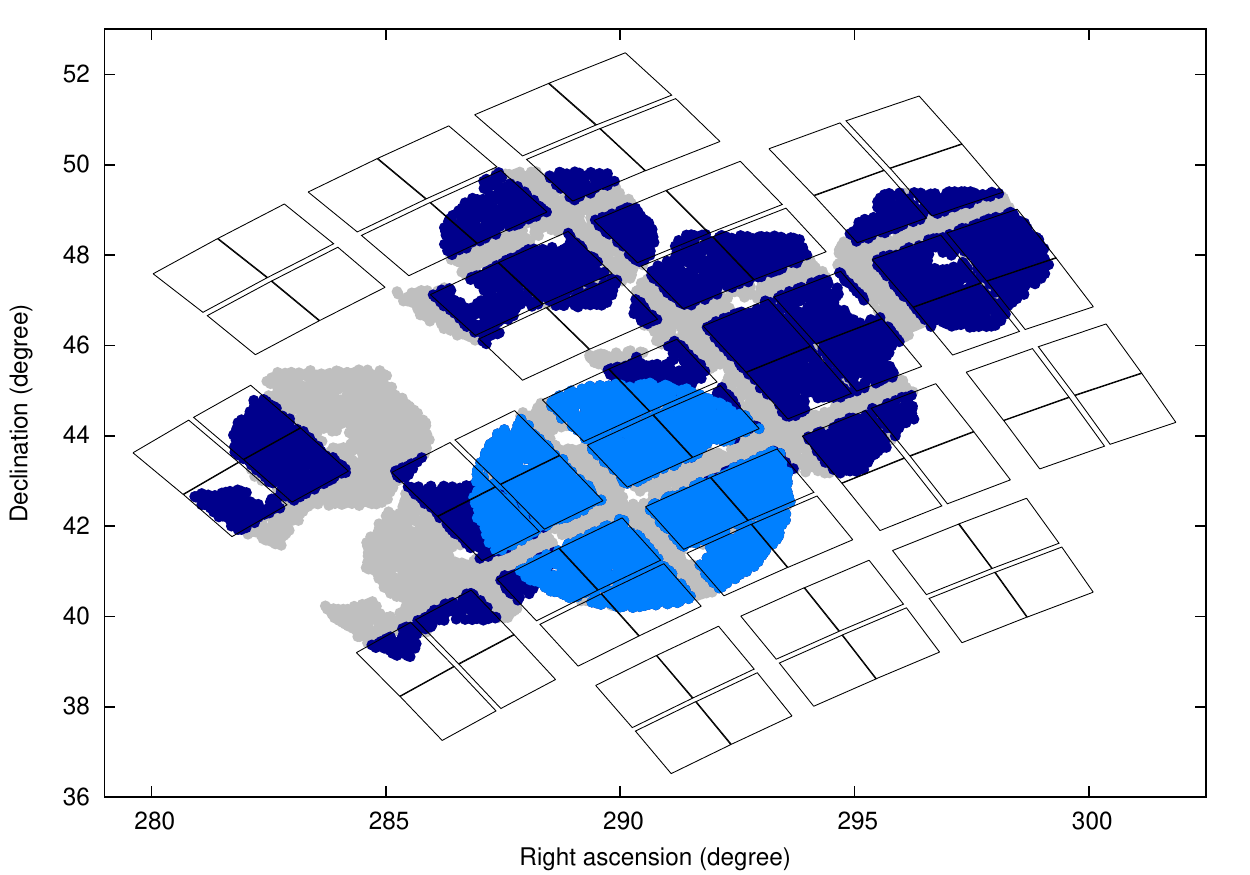}
\caption{Selection of the DR6 \projectmrs\ data in the \kepler\ field. The stars observed on plates that have overlap with the \kepler\ CCDs are shown as dots (blue for stars on a \kepler\ CCD, dark blue for observations in 2017, and light blue for observations in 2018). 
}
\label{Fig:plates}
\end{center}
\end{figure*}

The time in between the first (September 2012 -- June 2017) and second (September 2018 -- June 2023) phase of the regular survey of \lamost\ was used to test new medium-resolution spectrographs.
They have a spectral resolution of $R \sim 7500$, covering wavelengths of the visible spectrum ranging from 495 to 535\,nm and from 630 to 680\,nm with a blue and red arm, respectively \citep{Liu2019RAA....19...75L}.
In this paper, we use MRS retrieved from DR6 that were collected in the above mentioned test phases from plates that have objects in common with the \kepler\ field.
This led to an initial sample of 22,629 MRS \lamost\ spectra (Table\,\ref{Tab:plates}).
Their position on the sky is visualized in Fig.\,\ref{Fig:plates}.
Only seven medium-resolution spectrographs were available for the observations in 2017. They were fed by fibers positioned as displayed by 
%(\#sp02, \#sp03, \#sp04, \#sp05, \#sp08, \#sp09, \#sp15; 
the dark blue areas in Fig.\,\ref{Fig:plates}. \lamost\ was fully equipped with 16 medium-resolution spectrographs by the time the observation in 2018 were done (light blue area in Fig.\,\ref{Fig:plates}).
The results of the \lasp\ pipeline  \citep{luo2015,Wang2019ApJS..244...27W} and of the convolutional neural network method \citep{Wang2020ApJ...891...23W} are given in the \lamost\ MRS Parameter Catalog of DR6, which can be downloaded from {\tt http://dr6.lamost.org}.

\begin{table*}
\begin{center}
\caption{Overview of the LK--MRS plates observed in the period September 2017 -- June 2018  that have objects in common with the \kepler\ field. For each plate, we give the identifier, the right ascension and declination of the central star, the seeing during the observations, the exposure times, and the number of targets.}
\begin{tabular}{lccccccccc} 
\hline
\noalign{\smallskip}
Plan ID      & RA (2000) & Dec. (2000) & Date       & Seeing & Exposure      &\multicolumn{4}{c}{Number of targets} \\
             & hh:mm:ss.ss &  dd:mm:ss.ss & yyyy-mm-dd &   (arc sec)  &   (s)         & Total & \hot & 
\cool & \unclass \\ 
\noalign{\smallskip}
\hline
\noalign{\smallskip}
HIP9645901   & 19:36:37.98 & +44:41:41.77 & 2017-09-28 &   2.9  &  600$\times$4 &  1337 &  97 & 1023 &  217   \\ 
{\it HIP9286201}   & 18:55:20.10 & +43:56:45.93 & 2017-10-03 &   3.4  &  600$\times$3 &  1320 &  19 &  745 &  556   \\ 
{\it HIP9737201}   & 19:47:26.83 & +47:54:27.24 & 2017-10-03 &   3.3  &  600$\times$3 &  1329 &  78 &  782 &  469   \\ 
{\it HIP9380801}   & 19:06:17.04 & +41:24:49.61 & 2017-10-04 &   2.5  &  600$\times$3 &  1287 &  55 &  965 &  267   \\
{\it HIP9587901}   & 19:29:58.91 & +46:56:47.36 & 2017-10-04 &   2.9  &  600$\times$3 &  1288 &  73 &  944 &  271   \\
{\it HIP9448701}   & 19:13:53.55 & +48:20:57.43 & 2017-10-05 &   3.3  &  600$\times$3 &  1219 &  36 &  793 &  390   \\
HIP9511901   & 19:21:02.82 & +42:41:13.07 & 2018-05-24 &   2.8  &  900$\times$5 &  2968 & 126 & 1806 &  1036   \\
HIP95119KP01 & 19:21:02.82 & +42:41:13.07 & 2018-05-28 &   3.0  &  900$\times$7 &  2961 & 122 & 1814 &  1025   \\
HIP95119KP01 & 19:21:02.82 & +42:41:13.07 & 2018-05-29 &   2.3  &  600$\times$9 &  2980 & 119 & 1821 &  1040   \\
HIP95119KP01 & 19:21:02.82 & +42:41:13.07 & 2018-05-30 &   2.4  &  900$\times$5 &  2971 & 124 & 1784 &  1063  \\
HIP95119KP01 & 19:21:02.82 & +42:41:13.07 & 2018-05-31 &   2.3  & 1200$\times$4 &  2969 & 127 & 1787 &  1055   \\ 
\noalign{\smallskip}
\hline
\noalign{\smallskip}
TOTAL        &             &              &            &        &               & 22629 & 976 & 14264 &  7389   \\
\noalign{\smallskip}
\hline
\end{tabular}
\label{Tab:plates}
\end{center}
\textbf{Notes.} The identifiers of the plates observed during the regular \lamost\ survey are written in italic.
\end{table*}

In this paper, we used the \rotfit\ code (see Sect. \ref{Sec:Analysis}) for the determination of the stellar parameters, radial and projected rotational velocity (\vsini), chromospheric activity indicators and lithium abundance. We note that APs were already provided for most targets in other works based on different codes \citep[e.g.,][]{Zong2020ApJS..251...15Z}. For this task we used the stacked MRS, which are the sum of all the individual spectra of the same object acquired in the same night.
We decided to analyze these data with our code with the aim to have homogeneously determined parameters, using both the blue- and red-arm spectra, for all the FGKM-type stars for which we derive other parameters of interest, such as the \vsini, H$\alpha$ emission, and lithium abundance.
Since \rotfit\ is optimized for cool stars, we selected
 the spectra to be analyzed based on the results of the \lasp\ pipeline.
We subdivided the initial sample into a ``hot'' sample (\tefflasp\,$>$\,7000\,K), a ``cool'' sample (\tefflasp\,$\leq$\,7000\,K), and an ``unclass'' sample (no \tefflasp\ available), which contain 
976, 14264, and 7389 spectra, respectively (see ``Number of targets'' in Table\,\ref{Tab:plates}).
The spectra in the hot sample are not considered in the present work and will be included in a forthcoming paper (Catanzaro et al., in preparation), which will present a customized analysis based on a spectral-synthesis approach. 
A quick-look analysis of the spectra in the cool and unclass samples allowed us to discard bad spectra (e.g., missing parts or too many strong spikes) and those with clear signatures of hot (A-type or earlier) stars. This selection resulted in about 14,250 and 2050 spectra for the cool and unclass sample, respectively.
We used a signal-to-noise ratio (S/N) \,$\ge$\,10 in at least one arm as a threshold to select the reliable parameters from the cool sample. The success rate for this sample of spectra was very high, as we could determine parameters for 13,630 and 13,793 red- and blue-arm spectra, respectively.
However, the unclass spectra with S/N\,$\ge$\,10 are only a small fraction of the selected sample (about 200 and 1200 spectra for the blue and the red arm, respectively). Therefore, after a visual scrutiny, we decided to consider also the results from spectra with a S/N as low as 5 in one of the arms for the unclass sample, because no stellar parameters are available yet for these data. We ended up with 692 and 618 unclass spectra with APs in the red and blue arm, respectively.
The whole sample selected for a detailed analysis, which includes both cool and unclass, consists of 16,300 MRS \lamost\ spectra. For 14,300 of these spectra we were able to determine the APs.

The cross-identification with \kepler\ sources was based on the fiber coordinates, adopting a radius of 3.7 arcsec \citep[c.f.,][]{Zong2020ApJS..251...15Z,Wang2020ApJS..251...27W}.

%==================================================================
\section{Data analysis}
\label{Sec:Analysis}

\subsection{Radial velocity, projected rotation velocity, and atmospheric parameters}
\label{Sec:APs}

\begin{figure}[th]
\hspace{-0.5cm}
%\vspace{0cm}
%\begin{center}
\includegraphics[width=9.5cm]{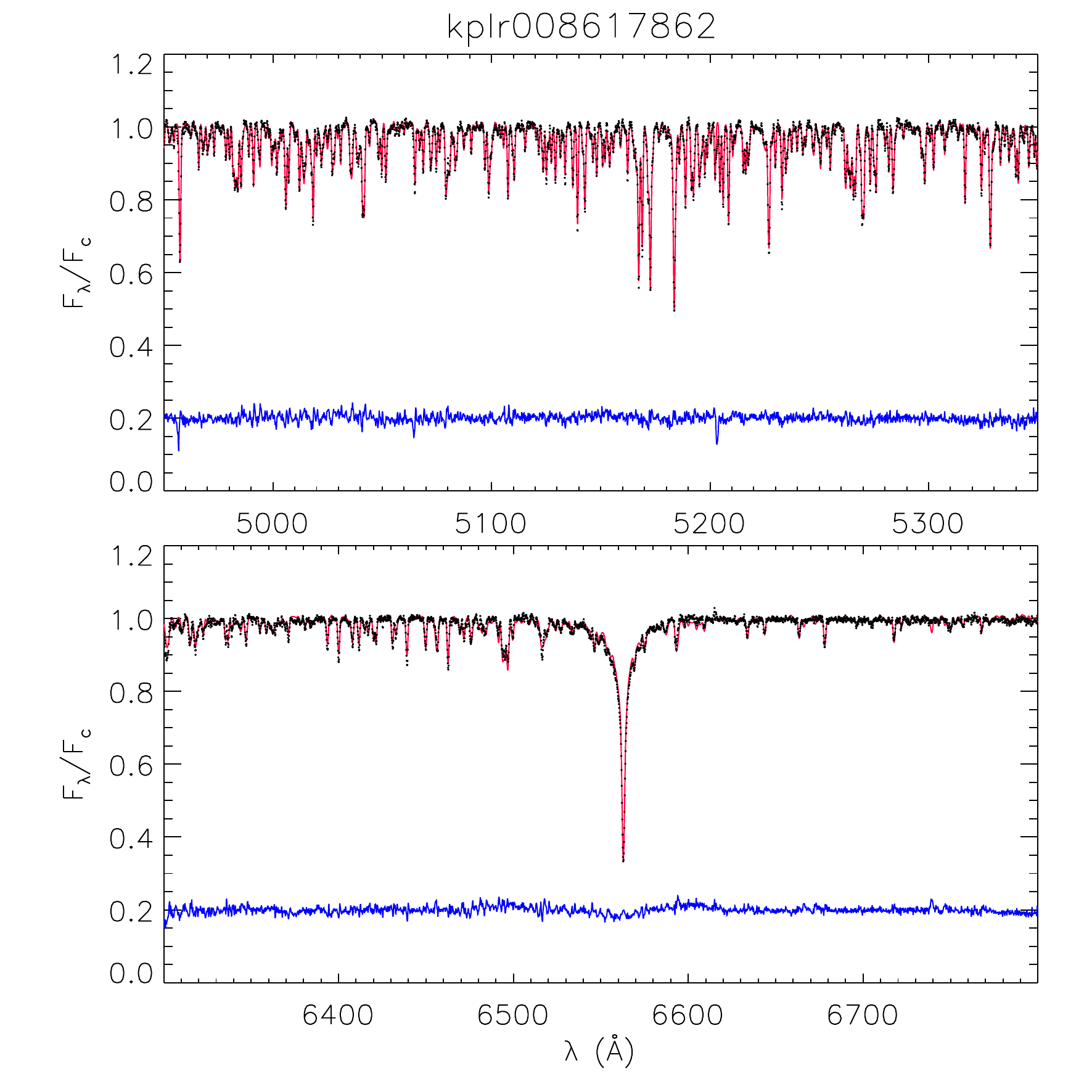}
\caption{Example of the continuum-normalized \lamost\ spectrum of a slowly rotating mid-F star (dots) in the blue arm ({\it upper panel}) and in 
the red arm ({\it lower panel}). The best template found by \rotfit\ is overplotted with a thin red line. The difference between the two spectra is shown in the bottom of each panel with a blue line shifted upward by 0.2.}
\label{Fig:spectrum}
%\end{center}
\end{figure}

We measured the RV, the projected rotation velocity, \vsini, and derived the APs -- \teff, \logg, and \feh\ -- by applying the code \rotfit\  \citep[e.g.,][]{Frasca2006,Frasca2015}, which was purposely modified to fit with the \lamost\ MRS. 
We adopted, as templates, a grid of high-resolution spectra of slowly rotating stars (\vsini\,$\leq3$\,\kms) with a low activity level that were retrieved from the ELODIE archive (R$\simeq$42,000; \citealt{Moultaka2004}). 
This is the same grid as that used for the analysis of young stars within the \textit{Gaia}-ESO survey by the OACT (Osservatorio Astrofisico di Catania) node \citep{Frasca2015}.
It contains spectra of 388 different stars, which sufficiently cover the space of the APs, although the density of templates is not uniform especially in the regime of metal-poor stars. 

We prefer real-star spectra  over synthetic ones because the former reproduce the unknown photospheric spectrum better, which can be subtracted from the target one to leave the chromospheric core contribution and to clean the \ion{Li}{i} line from blended neighbor lines. 
Some photospheric lines may be missing in the synthetic spectra, or the  depths and widths of some of them may be poorly reproduced due to uncertain intensity values, Land\'e factors and broadening effects. 

\begin{figure} 
\hspace{-.5cm}
\vspace{-.3cm}
  \includegraphics[width=9.3cm]{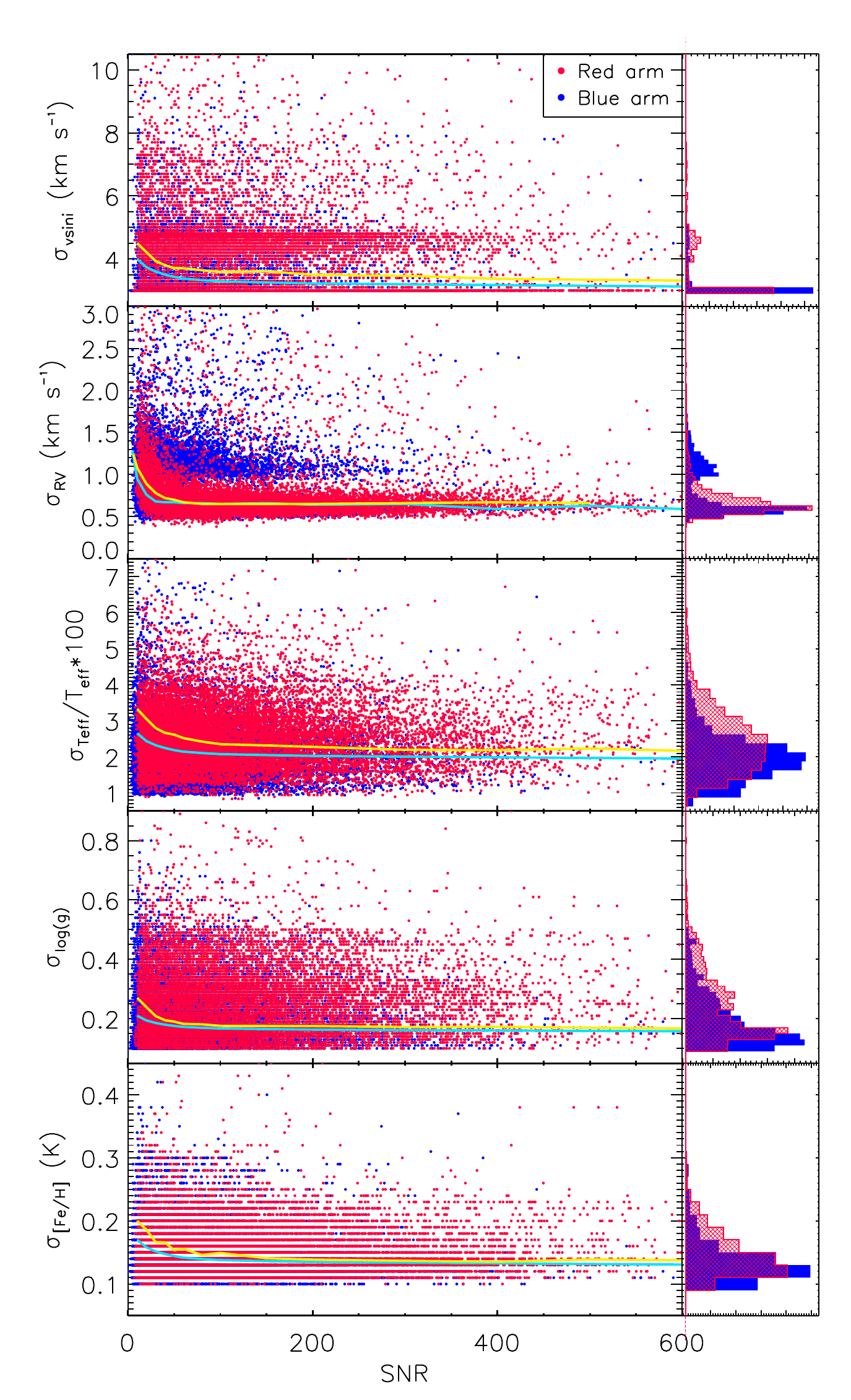}  
  \caption{Errors of RV, \teff , \logg, and \feh\ (from top to bottom) as a function  of the S/N for the blue-arm and  red-arm spectra. The full cyan and yellow lines in each box are the median of the blue-arm and red-arm errors, respectively. The blue and red histograms in the right panels display the distributions of these errors for the blue-arm and red-arm spectra, respectively.}
  \label{Fig:scatter} 
\end{figure}

The first step of the analysis performed by \rotfit\ is the normalization of the \lamost\ spectra to the local continuum, which was accomplished by the fit of a low-order polynomial.
After the normalization, the RV was measured by cross-correlating the target spectrum with spectra of four stars of different spectral type (SpT) selected from the library of ELODIE templates. In particular, we considered HD117176, HD166620, HD95735, and HD25329, whose SpTs are G5V, K2V, M2V, and K1 metal-poor, respectively.
%a few FGKM-type stars selected from the ELODIE template library.
For measuring RV, we chose the cross-correlation function (CCF) with the highest peak, which is the one obtained with the best-matching template.
This task was executed both for the blue and red arm 
of each spectrum, after excluding the lithium line and the core of the \halpha\ line ($\pm$1\,\AA\ around the line center), which would broaden the CCF peak and could be contaminated by chromospheric activity.
To measure the centroid and full width at half maximum of the CCF peak, we fitted it with a Gaussian. 
The RV error, $\sigma_{\rm RV}$, was estimated as the error of the center of the Gaussian fitted to the CCF peak using the procedure \textsc{curvefit} \citep{Bevington}, taking the CCF noise, $\sigma_{\rm CCF}$, into account. The CCF noise was evaluated as the standard deviation of the CCF values in two windows of about 600\,\kms\ each at the two sides of the peak. We adopted the value $w=1/\sigma_{\rm CCF}^2$ as the instrumental weight for all the points of the CCF peak that were considered for the Gaussian fit.
For the spectra with S/N$\ge 50$ we find median RV errors of $\approx$\,0.65\,\kms\ for the blue-arm and $\approx$\,0.70\,\kms\ for the red-arm, respectively (see Fig.\,\ref{Fig:scatter}). Although for a small fraction of sources ($\approx$3\%) the RV error is $\sigma_{\rm RV}>3$\,\kms, the median values reach a maximum of about 1.3\,\kms\ for the spectra with a low S/N (S/N$<$10). The RV errors in the blue arm display a two-fold behavior, with a larger fraction of them being similar to those of the red arm and barely visible under the red points in Fig.\,\ref{Fig:scatter} and a second branch with larger values. 
This double-peaked distribution for the RV errors in the blue arm is clearly visible in the right panel of Fig.\,\ref{Fig:scatter}, where the histogram displays two distinct peaks at about 0.7 and 1.1\,\kms. The peak with RV errors around 1.1\,\kms\ corresponds to stars with \teff$\leq 5500$\,K, for which the spectral features included in this range broaden, giving rise to larger RV errors.
We note that the RV errors evaluated with \rotfit\ are in agreement with \citet{Liu2019RAA....19...75L} and \citet{Zhang2021}, who report typical values of about 1\,\kms\ at S/N=20. As a further check of the RV errors, we used the task {\sc fxcor} of the IRAF\footnote{IRAF is distributed by the National Optical Astronomy Observatory, which is operated by the Association of the Universities for Research in Astronomy, inc. (AURA) under cooperative agreement with the National Science Foundation.} package and a set of synthetic templates for a small subsample of about 100 \lamost\  MRS of FGK-type stars. We find an excellent agreement between the RV values derived with both codes, both for the blue- and red-arm spectra. The errors calculated by {\sc fxcor} on the basis of the CCF peak's height and the antisymmetric noise \citep{Tonry1979} are, on average, similar to those derived with \rotfit. 

As pointed out by \citet{Wang2019ApJS..244...27W} and \citet{Zong2020ApJS..251...15Z}, the RV measured on \lamost\  MRS can be affected by systematic offsets in different runs that are related to the wavelength calibration. The largest offset of about 6.5 \kms\ is found for spectra acquired before May 2018 that were calibrated with Sc lamps, compared to the following ones for which Th-Ar lamps have been used. To account for these offsets and correct the RVs, we used a method similar to that adopted by \citet{Zong2020ApJS..251...15Z}, which is based on the RV measured by {\it Gaia} for a subset of stars with non-variable RVs enclosed in the \lamost-MRS plates. For each plate and observing date, these RV differences display a regular distribution that can be fitted with a Gaussian, whose center is a measure of the instrumental offset for the given plate and date (see Figs.\,\ref{Fig:RV_corr} and \ref{Fig:RV_corr_red} for some examples). These corrections are reported in Table\,\ref{Tab:RV_corr}. 

The resulting RV and $\sigma_{\rm RV}$ values are given in Cols. 16 and 17 of Tables\,\ref{Tab:data_red} and \,\ref{Tab:data_blue} for the red-arm and blue-arm spectra, respectively. The RVs corrected for instrumental offsets are reported in Col. 18 of the same tables.

For some spectra we noted two or three CCF peaks at a level larger than 5$\sigma_{\rm CCF}$ that we considered as significant (see Fig.~\ref{Fig:CCF_SB2} for an example). In these cases, we classified the object as a double-lined spectroscopic binary (SB2) or a triple system (SB3), respectively, and flagged its spectra accordingly in Tables\,\ref{Tab:data_red} and \ref{Tab:data_blue}. 
For these systems we discarded the RVs and APs from the above tables,
%Table\,\ref{Tab:data}, 
even if they were obtained near the conjunctions, when the lines of the two components are superimposed. The way the RVs of the components of SB2 systems are derived is described in Sect.\,\ref{Sec:SB2}.

For the determination of APs, the reference spectra were first aligned onto the target spectrum thanks to the RV measured as described above and were brought to the resolution of the \lamost\  MRS, $R_{\rm MRS}\simeq 7500$, by convolving them with a Gaussian kernel of width $W=\sqrt{1/R_{\rm MRS}^2 - 1/R_{\rm ELODIE}^2}$.
Then, each template was broadened by the convolution with a rotational profile of increasing \vsini\ (in steps of 1\,km\,s$^{-1}$) until a minimum $\chi^2$ was attained.

We point out that the resolution and sampling of the MRS \lamost\ spectra do not allow us to measure \vsini\ values smaller than 8\,\kms. This threshold was found by means of  Monte Carlo simulations with artificially broadened spectra that are described in Appendix\,\ref{Appendix:MonteCarlo}.
Therefore, the \vsini\ values smaller than 8\,\kms\ in Tables\,\ref{Tab:data_red} and \,\ref{Tab:data_blue} must be regarded as ``non-detection.'' Whenever the mean \vsini\ value is smaller than 8\,\kms\ in at least one arm, the final \vsini\ has
been replaced with ``$<8$\,\kms'' and flagged as an upper limit in the table containing the average APs (Table\,\ref{Tab:APs}).  

 As mentioned above, for each target we applied \rotfit\ on both the blue- and red-arm spectra that were analyzed independently. 
The templates were sorted in a decreasing order of $\chi^2$, giving the highest score to the best-fitting template. The MK SpT of the template with the highest score was assigned to the target star. 
Two examples of the application of \rotfit\ for slowly rotating stars of mid-F and K0\,III SpTs are shown in Figs.\,\ref{Fig:spectrum} and~\ref{Fig:spectrum3}, respectively. An example of the application of \rotfit\ for a rapidly rotating (\vsini$\simeq$\,110\,\kms) F5\,V star is shown in Fig.\,\ref{Fig:spectrum2}.

For each arm, we derived the values of \teff, \logg, \feh, and \vsini\ as the weighted averages of the parameters 
of the ten best-matching templates, using $1/\chi^2$ as the weight. As uncertainties of these parameters we took the standard errors of the weighted means to which we added in quadrature the average uncertainties of the APs of the templates ($\sigma_{T_{\rm eff}}=\pm$\,50\,K, $\sigma_{\log g}=\pm$\,0.1 dex, $\sigma_{\rm [Fe/H]}=\pm$\,0.1 dex). For the \vsini, we added in quadrature a value of $\sigma_{v\sin i}=\pm$\,3\,\kms, to take the uncertainty related to the spectra sampling into account.

As mentioned in Sect.~\ref{Sec:Data}, we were able to determine APs and RVs for 97\,\% of the blue-arm spectra in the cool sample,
while only 30\,\% of the spectra in the unclass sample could be successfully analyzed.
Regarding the red arm, we could measure stellar parameters for 
96\,\% of the spectra in the cool sample, and 34\,\% of the unclass-sample spectra.
The S/N ranges from 10 to 900 for the analyzed spectra in the cool sample and from about 5 to 450 in the unclass one.
However, the median S/N is much smaller in the latter sample, and this explains the lower success rate in determining the parameters. 
We remark that the samples of stars with parameters in the red and blue arm are not the same. In particular, for the cool and faint stars, the S/N in the red-arm can be sufficient to derive reliable parameters, while it is too low in the blue arm.

The errors of \vsini, \teff, \logg, and \feh\ are displayed as a function of S/N in Fig.~\ref{Fig:scatter}.
 As shown in this figure, the \vsini\ errors range from about 3 to 10\,\kms, with median values rising to about 4 and 5\,\kms\ for the blue and red arm, respectively, at the lowest S/Ns. The scatter plot for the percent error of \teff\ shows values ranging from about 1\% to 9\%, with an increasing trend for decreasing S/N. The median values of about 2\% and 2.5\% for the blue and red arm, respectively, increase slightly with the decrease of S/N. The errors of \logg\ are smaller than 0.5\,dex for nearly all the spectra, with median values of about 0.20 and 0.25 dex for the blue and red arm, respectively. The \logg\ errors increase slightly with the decrease of S/N, as shown by the scatter plot and the median, with the red-arm ones being larger, as also indicated by higher tail of the red histogram.
[Fe/H] values have uncertainties smaller than 0.3\,dex for most of the spectra, with median values of 0.15 dex for both the blue and the red arm. As for \teff\ and \logg, the \feh\ errors increase slightly with the decrease of S/N; this effect is larger for the red-arm values, as also shown by the histograms.  

\begin{figure} 
  \centering         
  \includegraphics[width=8.5cm]{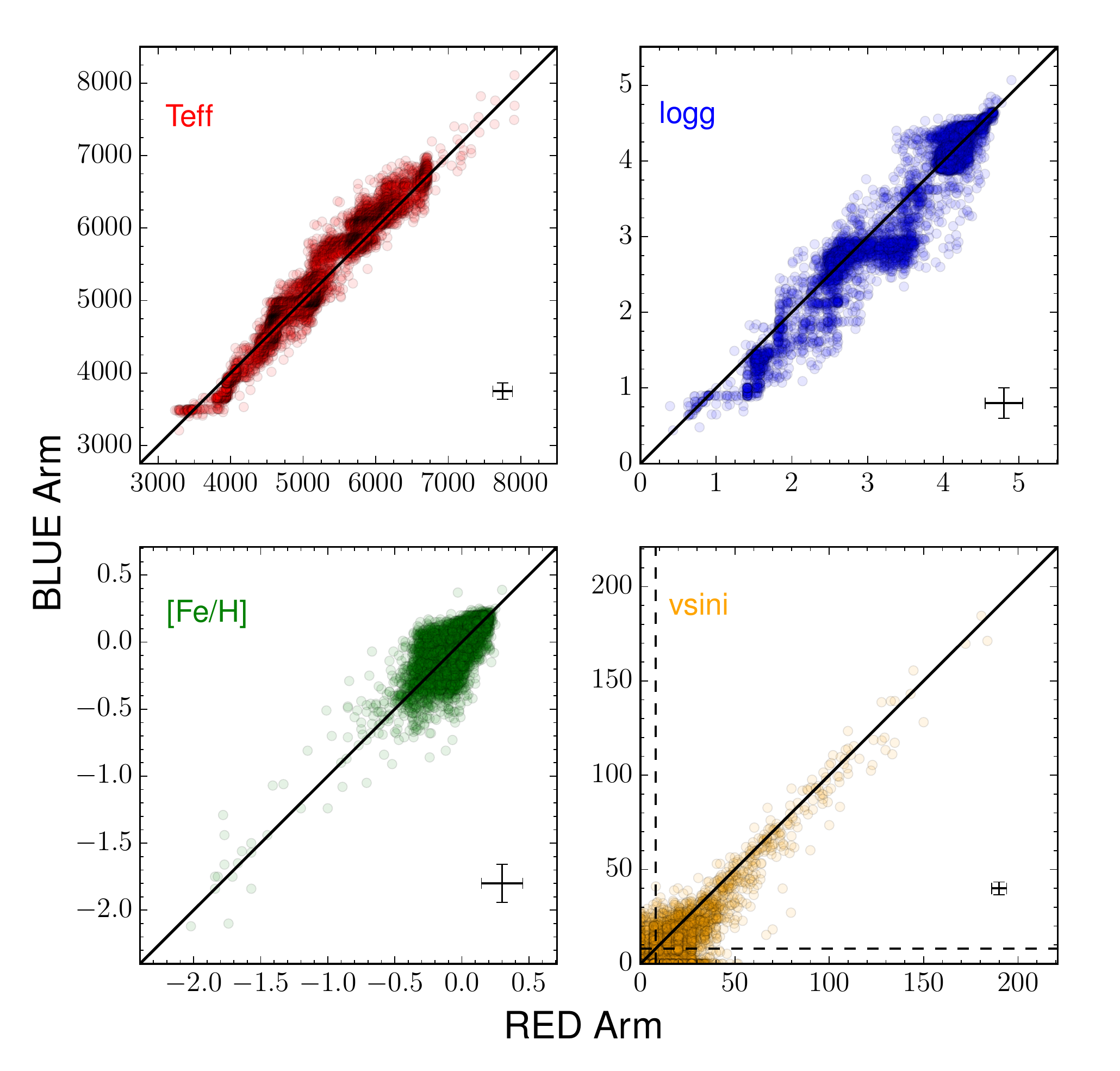}  
  \caption{Comparison of the 
   parameters obtained from both arms. Typical errors are displayed in the bottom-right corner of each panel. In the panel corresponding to the \vsini, the dashed lines show the upper limit at 8\,\kms.
  } 
  \label{fig_comp_param_int} 
\end{figure}

\begin{figure} 
  \includegraphics[width=9cm]{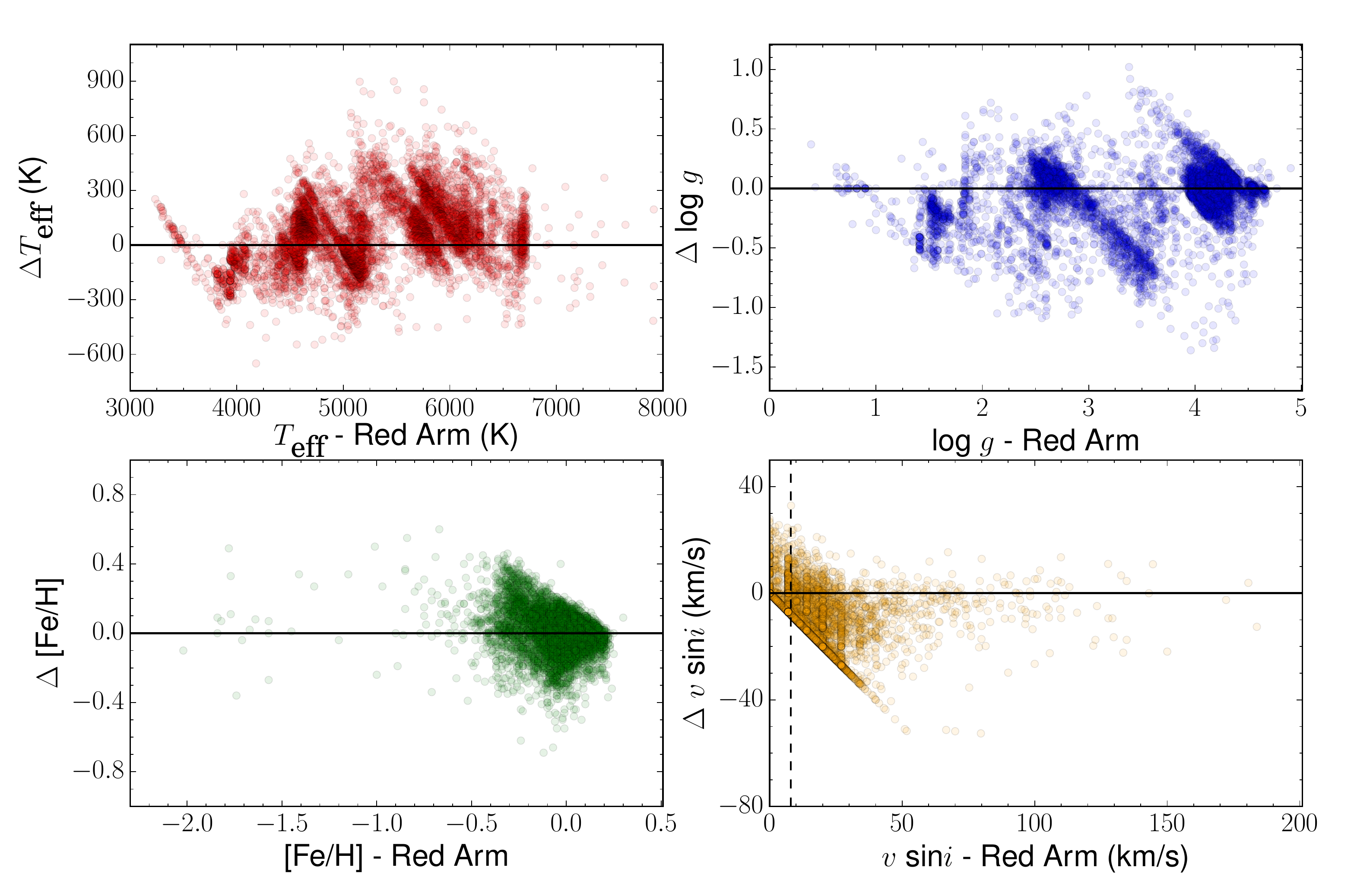}  
  \caption{Differences obtained from both arms (blue and red) for each parameter. In the panel corresponding to the \vsini, the dashed line shows the upper limit at 8\,\kms.} 
  \label{fig_err_param_int} 
\end{figure}

We remark that stellar parameters obtained from both arms are consistent for the vast majority of the stars contained in our sample (see Fig.~\ref{fig_comp_param_int}). 
If we compare the RVs measured by us in the blue and red arm, which are simultaneous, we find an offset of about $-0.1$\,\kms\ between blue and red RVs and an rms dispersion of 2.3 \kms\ (see Fig.\,\ref{Fig:RV_blue_red}). 
This can provide us with a further estimate of the average RV accuracy. If we assume that the blue- and red-arm RVs have, on average, the same uncertainty, we should divide this number by $\sqrt{2}$, getting an estimate of about 1.6\,\kms\ for the RV accuracy in both arms.
On average, we find differences (blue arm minus red arm) of: $\Delta\,T_{\rm eff}$\,=\,65$\pm$175\,K, $\Delta\log g$\,=\,$-$0.08$\pm$0.26,  $\Delta$[Fe/H]\,=\,0.00$\pm$0.12 and $\Delta v\sin i$\,=\,$-$8.7$\pm$9.5\,km\,s$^{-1}$. For the \vsini, the differences are $\Delta v\sin i$\,=\,$-$3.8$\pm$3.4\,\kms, if we exclude the upper limits.
For a better display of the results, the distribution of these differences for each parameter, taking the red arm as a reference, is shown in Fig.~\ref{fig_err_param_int}.
As seen above for the $RV$, the difference between the APs in the blue and red arm enables us to estimate average uncertainties, which turn out to be (dividing by $\sqrt{2}$) about 120\,K, 0.18\,dex, 0.09\,dex, and 2.5\,\kms for \teff, \logg, \feh, and \vsini, respectively. 
The ``zigzag'' pattern shown by the \teff\ and \logg\ plots in Fig.~\ref{fig_comp_param_int}, which is more evident in Fig.~\ref{fig_err_param_int}, is the effect of the clustering of the average parameters around those of the best (minimum $\chi^2$) template. This is not surprising, because, unlike other analysis codes, \rotfit\ does not apply any kind of interpolation or regularization between the parameters of the closest templates.
We note the floor of low \vsini\ values for the blue-arm spectra (\vsini$_{\rm BLUE}$), which extends up to \vsini$_{\rm RED}=50-60$\,\kms\ and  translates into the tilted strip in Fig.~\ref{fig_err_param_int}. This indicates cases in which the red arm provided a poor constrain to \vsini\ due to the few absorption lines or to the presence of molecular bands in their red-arm spectra.

The final APs were obtained as the weighted mean of those of the two arms, whenever measures in both arms were available.
As described in Sect.~\ref{Sec:Data}, in total (from the cool and unclass samples) we selected about 16\,300 spectra
for our analysis. These spectra, once their coordinates 
were cross-matched, correspond to 8268 different stars. For 821 of such sources, it was not possible to find reliable parameters because of the binary nature of some sources (see Sect.~\ref{Sec:SB2}) or due to the poor quality of the spectra in other cases. In total, we provide parameters (see Table~\ref{Tab:APs}) for  7443 stars, the vast majority of which (7146 stars) has been obtained from both arms. Most of the stars were observed only once (5487) while the rest (1956) have been observed up to six times.

The final parameters (i.e., the average values for each star obtained from all its individual measurements) were calculated by employing a S/N-weighted mean, while the uncertainties 
were quantified by using the average of individual errors. For the stars observed once, with both arms we averaged both values, whereas for those stars with multiple observations 
we first calculated a single value per arm and then performed the average of both. 
We remark that the average value of the RV for stars with multiple observations can hide a genuine RV variation ascribable to pulsations or to the presence of an unseen close companion that would make the source a single-lined spectroscopic binary (SB1). Therefore, we do not list this value in Table~\ref{Tab:APs}, but, for stars with multiple observations, we calculated the reduced $\chi^2$ and the probability $P(\chi^2)$ that the RV variations have a random occurrence \citep[e.g.,][]{Press1992}. Whenever $P(\chi^2)<0.05$ we considered the RV variation as significant and flagged the corresponding source with ``RVvar'' in Table\,\ref{Tab:APs}. However, the individual values of RV measured in each spectrum can be found in Tables\,\ref{Tab:data_blue} and \ref{Tab:data_red} for the blue and red arm, respectively.
In general, the APs derived from different arms or observations are in good agreement with each other.
Just a few outliers have been detected. In these cases, a visual check of the spectra helped us to improve the results by leaving out the worst ones (i.e., the noisiest, those with artifacts, etc.).

Also, $Gaia$ parallaxes have been used to discern the proper values when large discrepancies 
in \logg-values were found.
This was particularly helpful for low-signal spectra of M-type stars. 
Typical errors from different arms or observations are one or two spectral subtypes. Lastly, the final S/N was taken as the average of individual 
values. This number is not intended to be an actual ratio but only an indication of the overall quality of the spectra for each target.
These final values (\teff, \logg, \feh, $v$\,sin\,$i$, SpT and S/N) are displayed in Table~\ref{Tab:APs} for the stars in our sample along with their equatorial coordinates, 
KIC designation, number of observations
(N) and an indication of the arm(s) with which they have been observed (Arm): blue (b), red (r) or both (br). 
The last column (Rem) reports any useful remark, such as 
%SB2 for double-lined spectroscopic binaries, 
RVvar for stars with variable RV, Li for stars with a detection of the \ion{Li}{i}\,$\lambda$6708 absorption line, and Haf or Hae for stars with the H$\alpha$ line filled in by core emission or with a pure H$\alpha$ emission profile above the local continuum.

\renewcommand{\tabcolsep}{0.15cm}

\begin{table*}[ht]
  \caption{Final APs of the investigated sources.}
\begin{center}
\begin{tabular}{lccclcrccccc}
\hline\hline
\noalign{\smallskip}
RA   &  Dec.  &     KIC      &  N  & ~~~~\teff  & \logg       &        \feh~~~~      &  \vsini   &  SpT        & S/N  & Arm & Rem \\   % RV
(J2000)  &  (J2000) &    &  & ~~~~(K) &  &  &  (\kms) &  &   & \\  
\noalign{\smallskip}
\hline
\noalign{\smallskip}
280.925900  &  42.657299  &  KIC07090703   &  1  &  5949 $\pm$ 138  &  4.26 $\pm$ 0.12  &    0.09 $\pm$ 0.11  &       $\leq$8.0   &    G1.5\,V  &   5  &   b & \dots \\ % $-$32.77 $\pm$ 5.23  &        
280.952640  &  42.641399  &  KIC07090759   &  1  &  5852 $\pm$ 120  &  4.25 $\pm$ 0.17  &    0.06 $\pm$ 0.15  &       $\leq$8.0   &       G2.5  &  30  &  br & \dots  \\ % 3.97 $\pm$ 3.85  &    
281.054810  &  42.621838  &  KIC07090977   &  1  &  6322 $\pm$ 141  &  3.98 $\pm$ 0.15  & $-$0.07 $\pm$ 0.13  &  18.4 $\pm$ 3.0  &     F6I\,V  &  35  &  br & \dots \\ % 3.90 $\pm$ 4.18  &  
281.074070  &  42.474880  &  KIC06921876   &  1  &  5253 $\pm$ 231  &  3.36 $\pm$ 0.48  & $-$0.40 $\pm$ 0.16  &  11.5 $\pm$ 3.6  &    G8\,III  &  13  &  br & \dots \\ % $-$114.60 $\pm$ 4.30  &   
281.143590  &  42.475670  &  KIC06922004   &  1  &  6915 $\pm$ 221  &  4.11 $\pm$ 0.19  & $-$0.26 $\pm$ 0.19  &  20.3 $\pm$ 3.7  &     A8\,Vs  &  16  &   b & \dots \\ % $-$86.98 $\pm$ 5.04  &   
281.174930  &  42.476238  &  KIC06922059   &  1  &  5829 $\pm$ ~\,70  &  4.31 $\pm$ 0.12  &    0.06 $\pm$ 0.12  &   9.7 $\pm$ 3.7  &    G1.5\,V  &  41  &  br & Li  \\ % $-$61.00 $\pm$ 3.73  &   
281.212860  &  42.691792  &  KIC07091248   &  1  &  5255 $\pm$ 133  &  3.76 $\pm$ 0.40  &    0.01 $\pm$ 0.14  &       $\leq$8.0   & G2.5\,IIIb  &  17  &  br & \dots \\ % $-$8.83 $\pm$ 4.11  &
281.235410  &  42.620834  &  KIC07091292   &  1  &  5812 $\pm$ 141  &  4.09 $\pm$ 0.18  & $-$0.22 $\pm$ 0.14  &  28.0 $\pm$ 3.9  &      G3\,V  &  85  &  br & Li \\ % $-$26.17 $\pm$ 4.12  &   
281.258420  &  42.605980  &  KIC07091357   &  1  &  5701 $\pm$ 114  &  4.43 $\pm$ 0.12  & $-$0.05 $\pm$ 0.16  &       $\leq$8.0   &    G1.5\,V  &  11  &   b & \dots \\ % $-$20.50 $\pm$ 6.76  &    
281.260650  &  42.478771  &  KIC06922204   &  1  &  5843 $\pm$ 140  &  4.21 $\pm$ 0.20  & $-$0.13 $\pm$ 0.15  &  10.8 $\pm$ 3.9  &      G3\,V  &  45  &  br & Li \\ % $-$29.03 $\pm$ 3.85  &      
\dots & \dots & \dots & \dots & \dots & \dots & \dots & \dots & \dots & \dots & \dots & \dots \\
\noalign{\smallskip}
\hline
\end{tabular}
\end{center}
{\bf Notes.} The full table is only available in electronic form.
\label{Tab:APs}
\end{table*}

\subsection{Comparison with the literature}
\label{Subsec:external}

\begin{figure*}[htb]
\includegraphics[width=9cm]{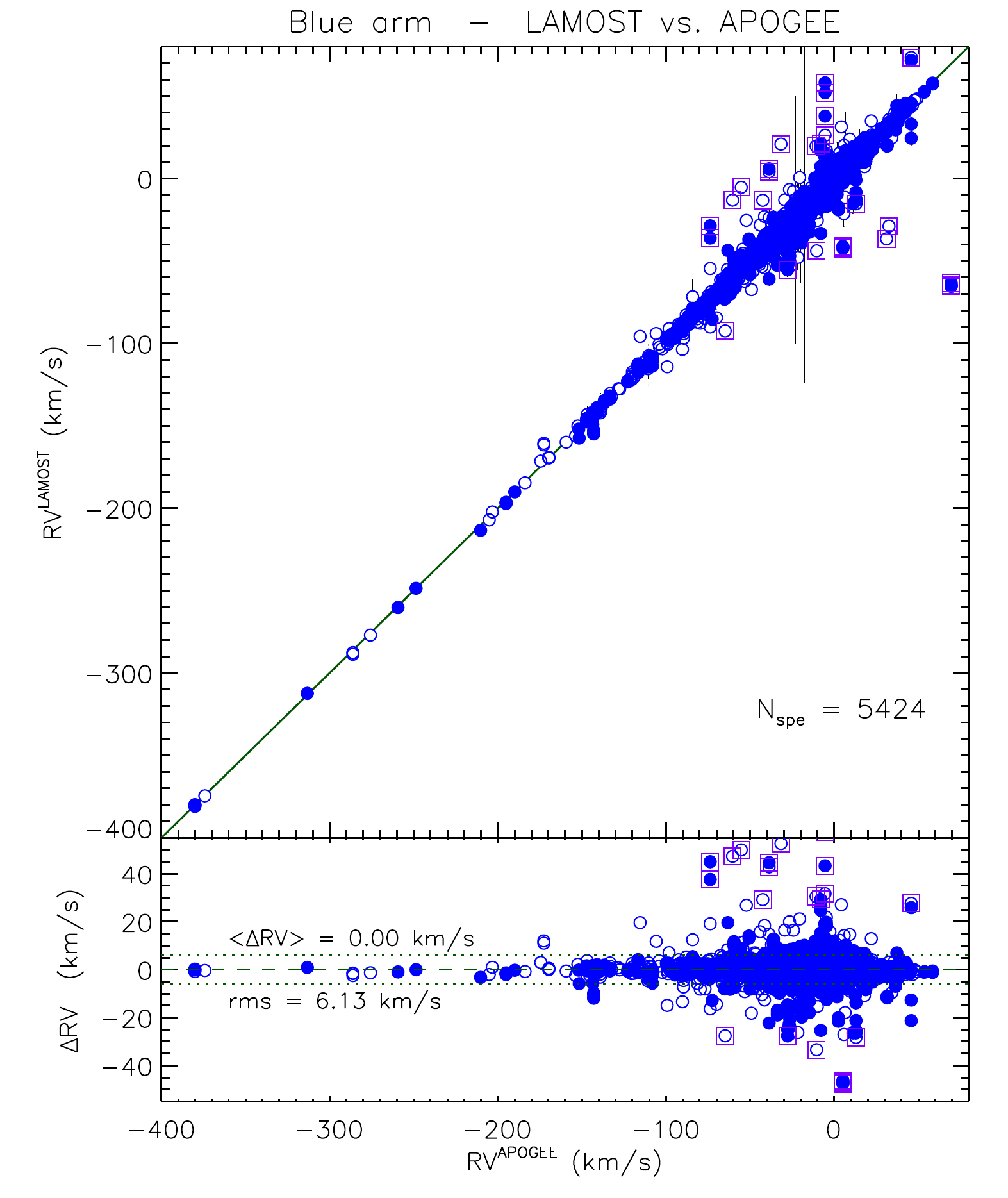}
\includegraphics[width=9cm]{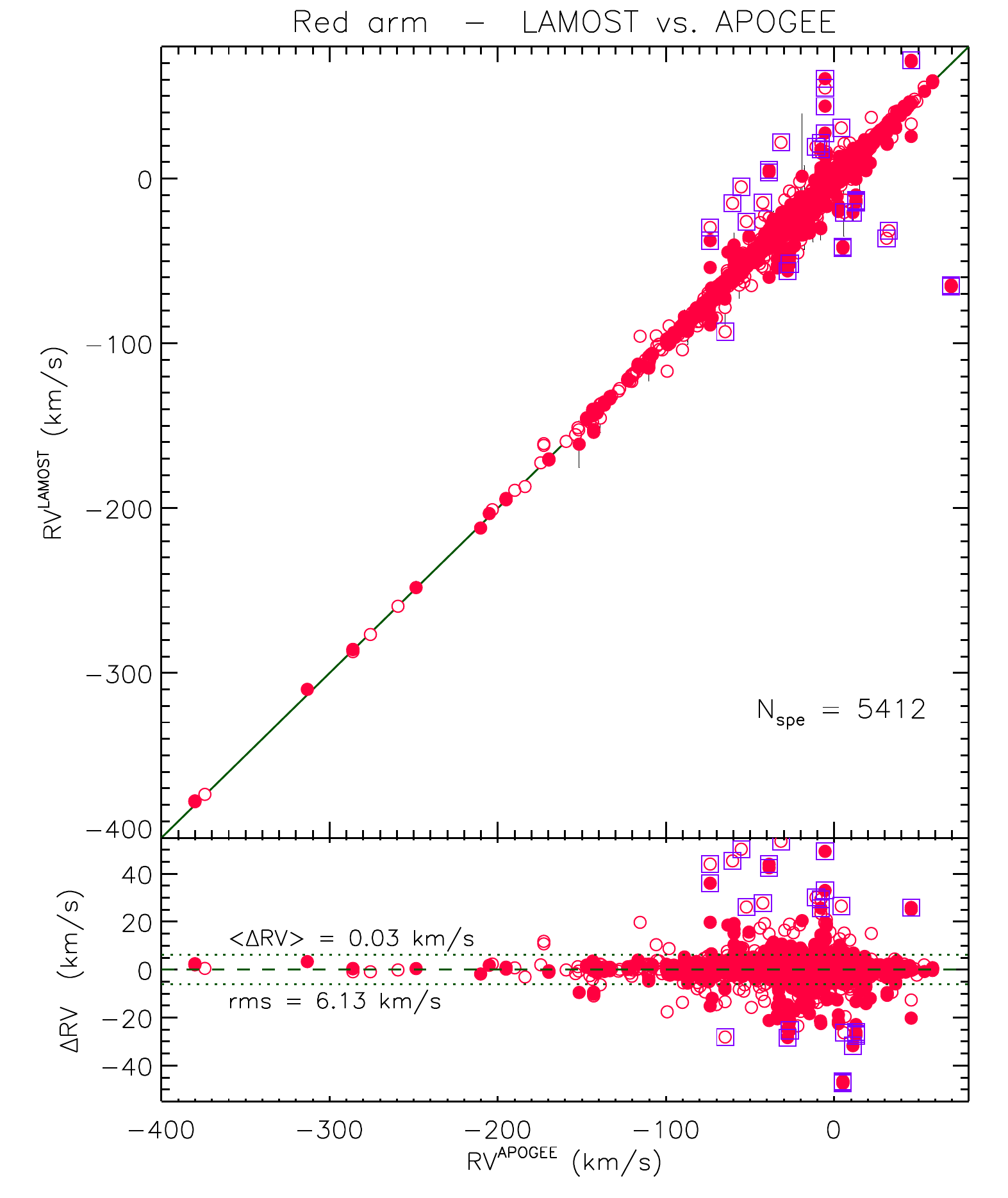}
\includegraphics[width=9.2cm,height=5cm]{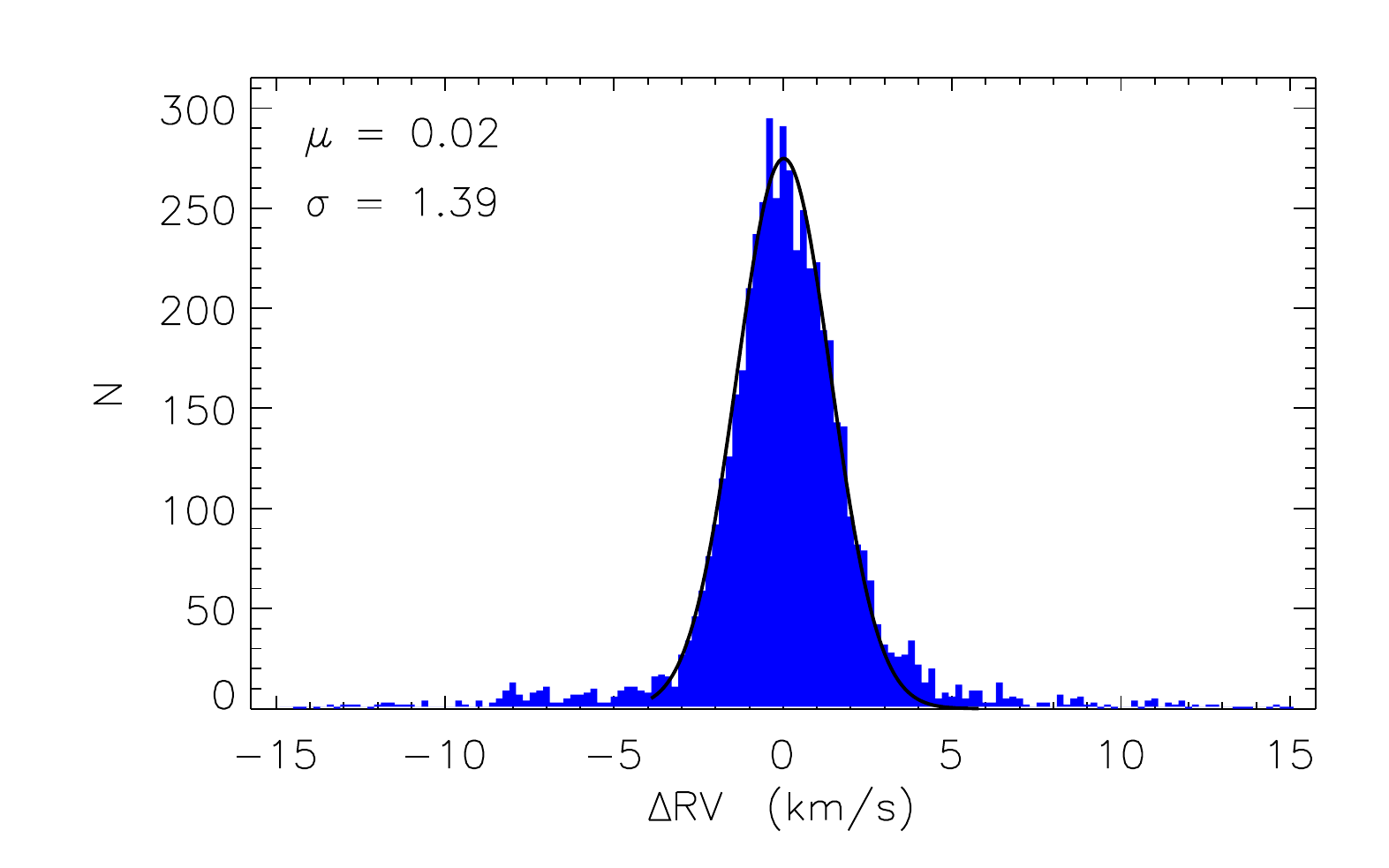}
\includegraphics[width=9.2cm,height=5cm]{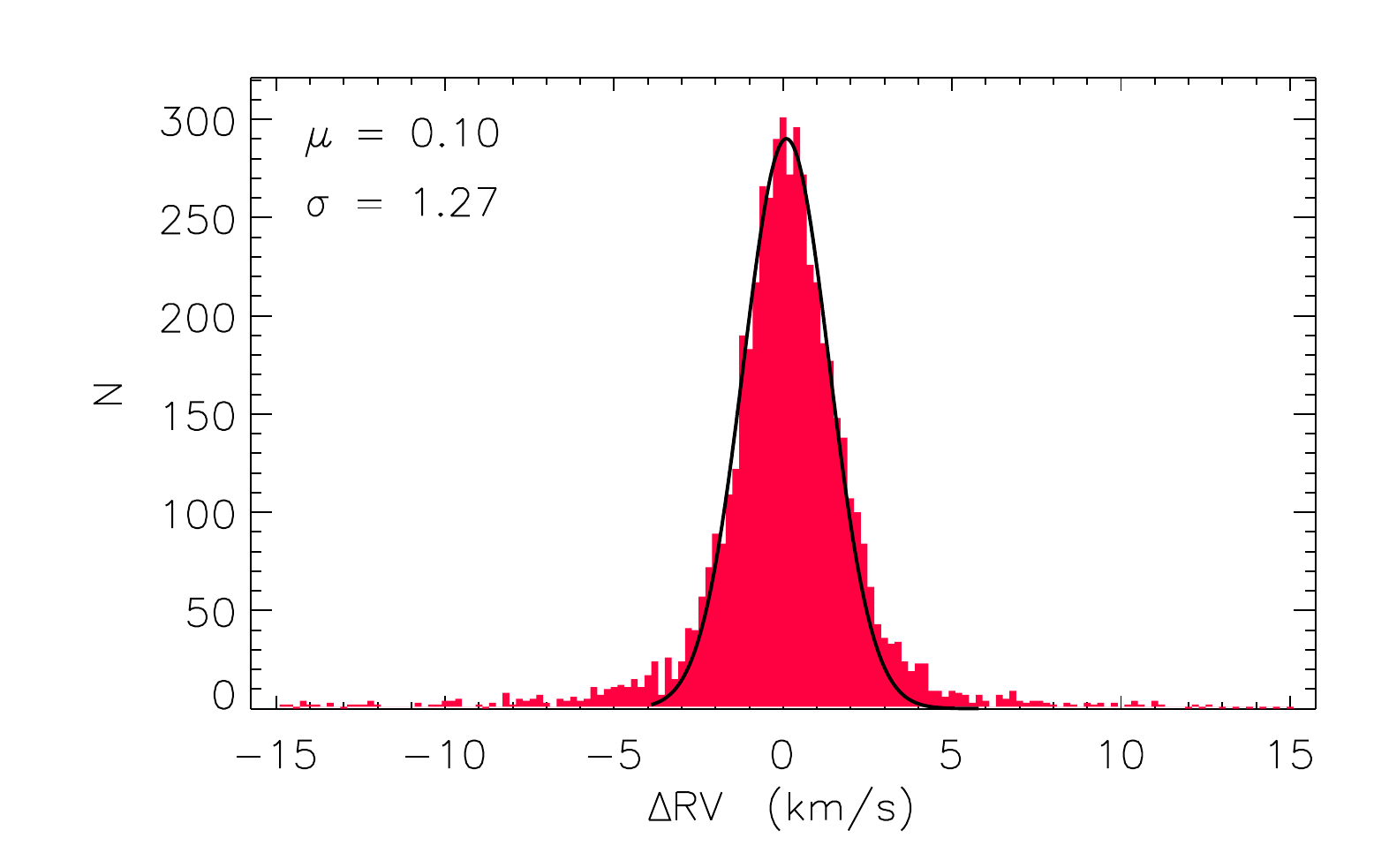}
\caption{Comparison with the APOGEE RVs. {\it Upper-left panel:} Comparison between the RV measured on the blue-arm \lamost\ MRS (Table~\ref{Tab:data_blue}) and APOGEE values \citep{2020AJ....160..120J}. Filled symbols represent the stars with multiple spectra. The one-to-one relation is shown by the continuous line. The RV differences between \lamost\ and APOGEE, $\Delta$RV, are displayed in the lower box and show an average value of $0.0$\,\kms\ (dashed line) and a standard deviation of 6.1\,\kms\ (dotted lines). {\it Upper-right panel:} Same as the upper-left panel but for the red-arm RVs listed in Table~\ref{Tab:data_red}. The purple squares in both panels enclose the more discrepant points (more than four times the rms). The distributions of the RV differences are shown by the histograms in the lower panels. The Gaussian fits are overplotted with black lines, and the center ($\mu$) and dispersion ($\sigma$) of the Gaussians are also marked in these boxes.}
\label{Fig:RV_comp}
\end{figure*}

The uncertainties evaluated by \rotfit\  (Sect.~\ref{Sec:APs})
are internal to the procedure and do not give an account of the real accuracy of the parameters. To this aim we compared the parameters that were derived in the present work with those available in the literature. 
As regards the RV, we compared the values measured in the blue-arm and red-arm spectra (Fig.~\ref{Fig:RV_comp}), corrected for the systematic offsets, with those measured by the Apache Point Observatory Galactic Evolution Experiment (APOGEE) and reported in the DR16 catalog \citep{2020AJ....160..120J}.
We note an almost zero average offset between our corrected RVs and APOGEE, both for the blue and red arm. The rms of the data dispersion around the mean is 6.1\,\kms\ for both arms, which can be considered as an upper limit for the data accuracy.
Indeed, some stars can have variable RVs because of pulsations or due to the presence of an unseen companion. 
The points that are more discrepant in Fig.~\ref{Fig:RV_comp} refer to 18 stars (KIC~5268955, 5527172, 5609753, 5688032, 6425135, 6777016, 6924881, 7119181, 7879399, 8022670, 8223328, 8687869, 9099927, 9651996, 10294429, 10987439, 11044668, and 11554998) that are known to be variable in RV in previous studies \citep[see][]{Frasca2016, 2016yCat.5149....0L, 2018yCat.5153....0L, Wang2019ApJS..244...27W, Wang2020ApJ...891...23W, 2020AJ....160..120J, 2020ApJS..246....4T} or show a significant RV spread in the MRS analyzed in this work.
If we discard these data the rms scatter decreases to 3.4\,\kms.
Another way to see the effect of stars with variable RV and to take it into account, is to investigate the distributions of the RV differences between LAMOST and APOGEE, which we have plotted in the lower panels of Fig.~\ref{Fig:RV_comp}.
The distributions, both for the red and the blue arm, are regular and symmetrical in their central part, which has been fitted with a Gaussian.
The excesses with respect to the Gaussian in the wings of the distributions are most likely due to objects with variable RV.
The Gaussian fits suggest an accuracy of the RV measurement of about 1.3\,\kms\ or better.

\begin{figure}[htb]
\includegraphics[width=8.cm]{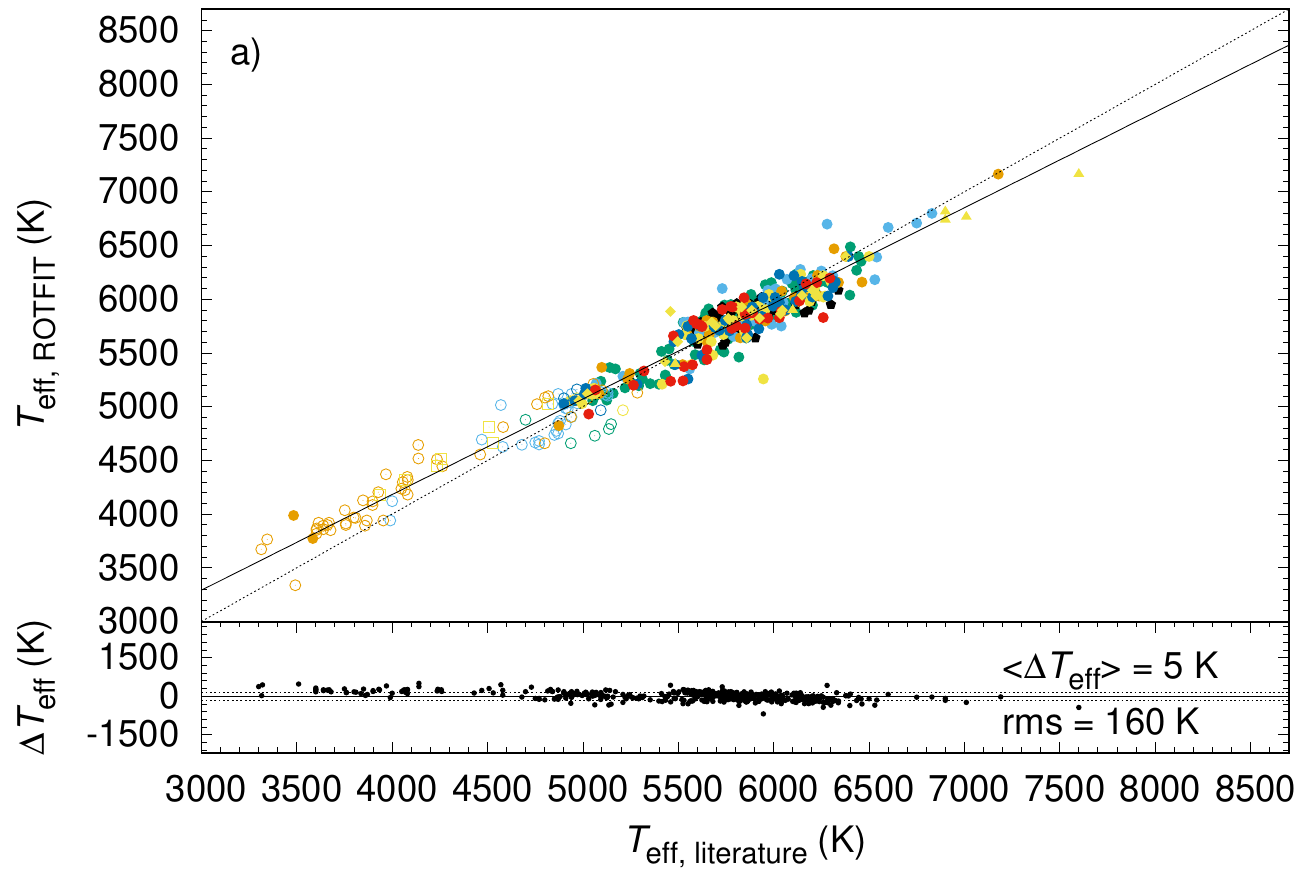}
\includegraphics[width=8.cm]{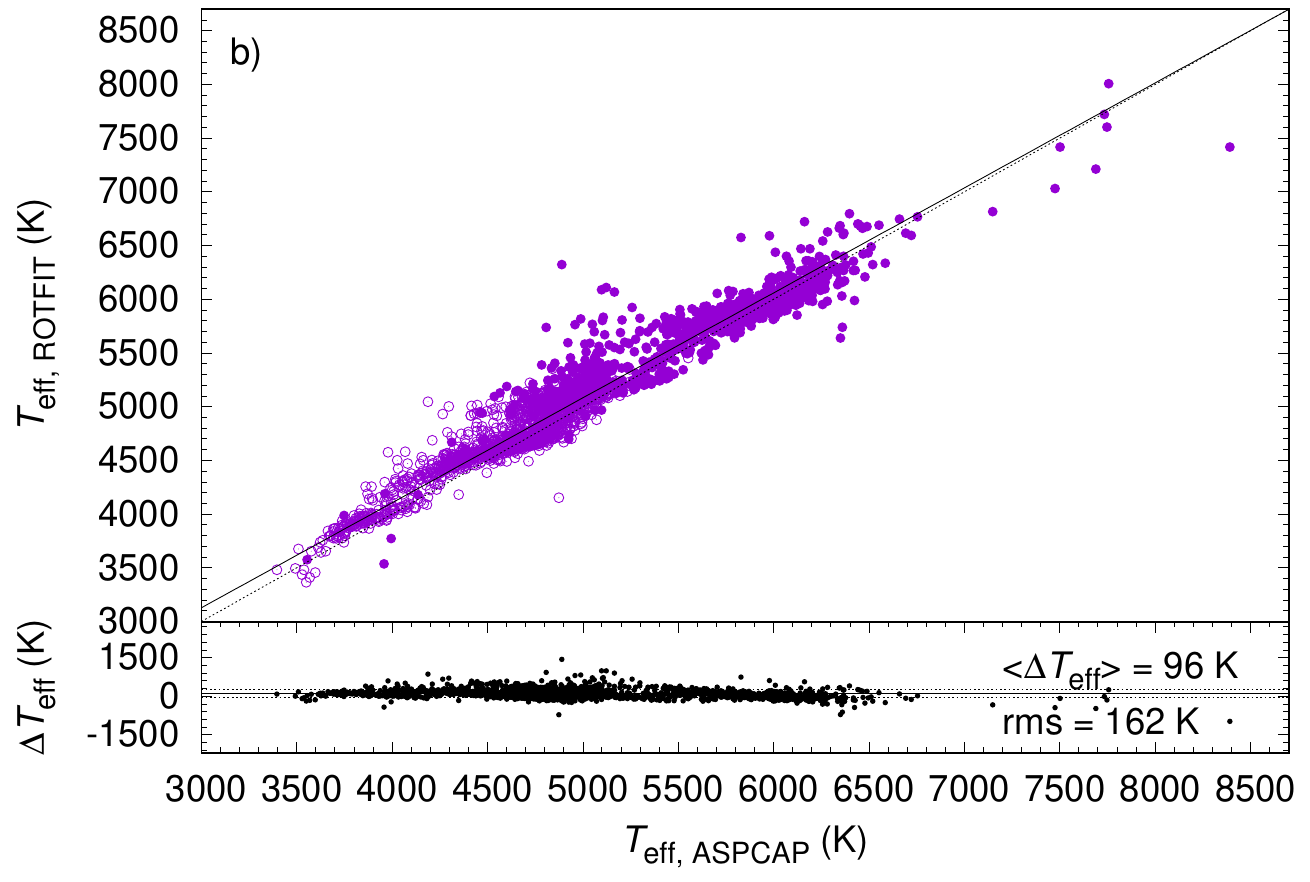}
\includegraphics[width=8.cm]{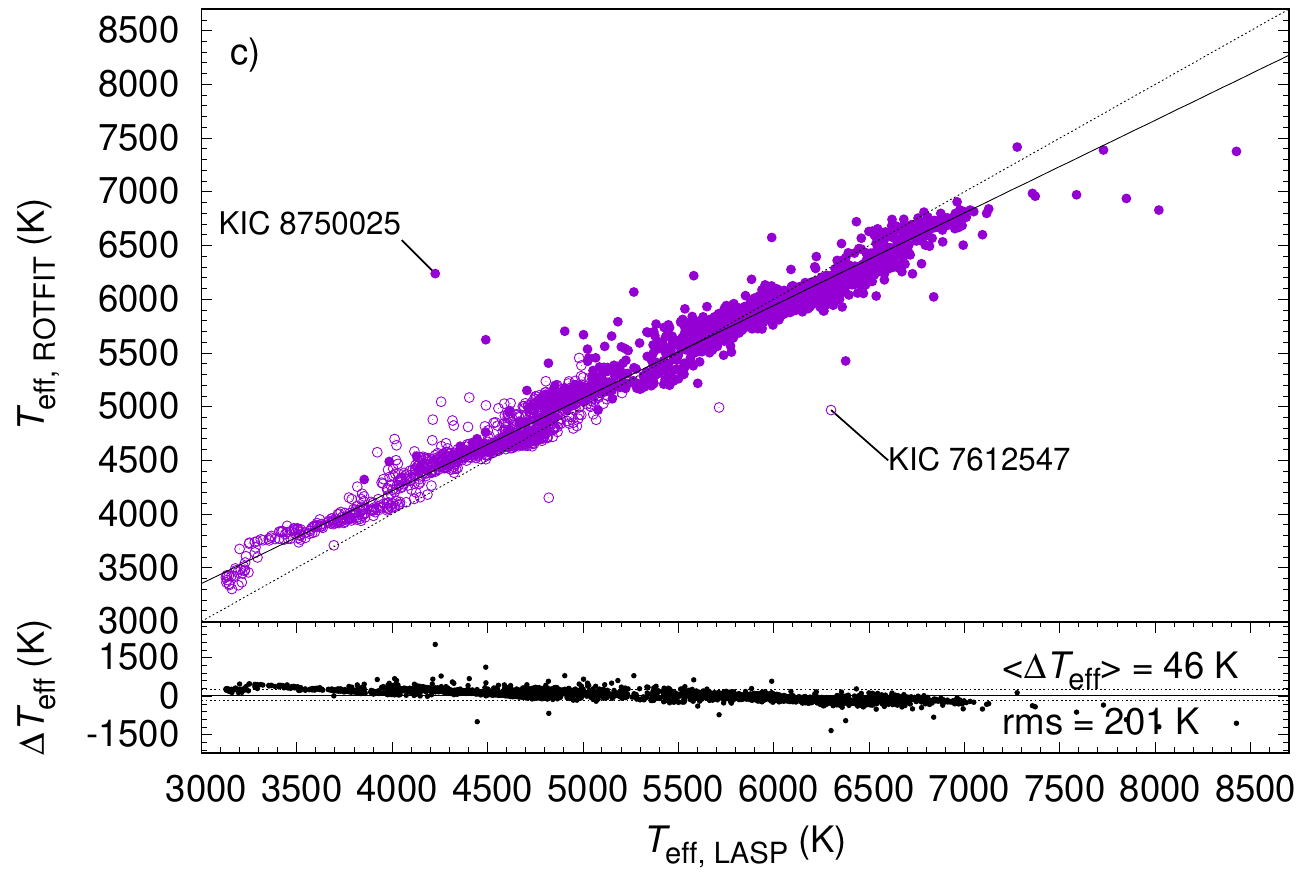}
\caption{Comparison of \teff\ values. (a) Comparison between the values of effective temperature in our database of \lamost\ spectra and in the literature. Filled symbols represent dwarfs ($\log g_{\rm ROTFIT} \ge 3.5$), open symbols represent giants ($\log g_{\rm ROTFIT} < 3.5$). Different colors have been used for different literature sources, as indicated in the main text. The comparison with the APOGEE and LASP
\teff\ is shown in the middle (b) and bottom (c) panels, respectively.
The dotted lines in the top box of each panel represent one-to-one relationships. The solid lines are linear fits to the data. The differences between ROTFIT and literature parameters are shown in the lower parts of each panel along with their average values and standard deviations. Stars indicated with arrows are discussed in the text.
}
\label{Fig:JMZ_comp_Teff}
\end{figure}

\begin{figure}[htb]
%\hspace{-0.2cm}
\includegraphics[width=8.cm]{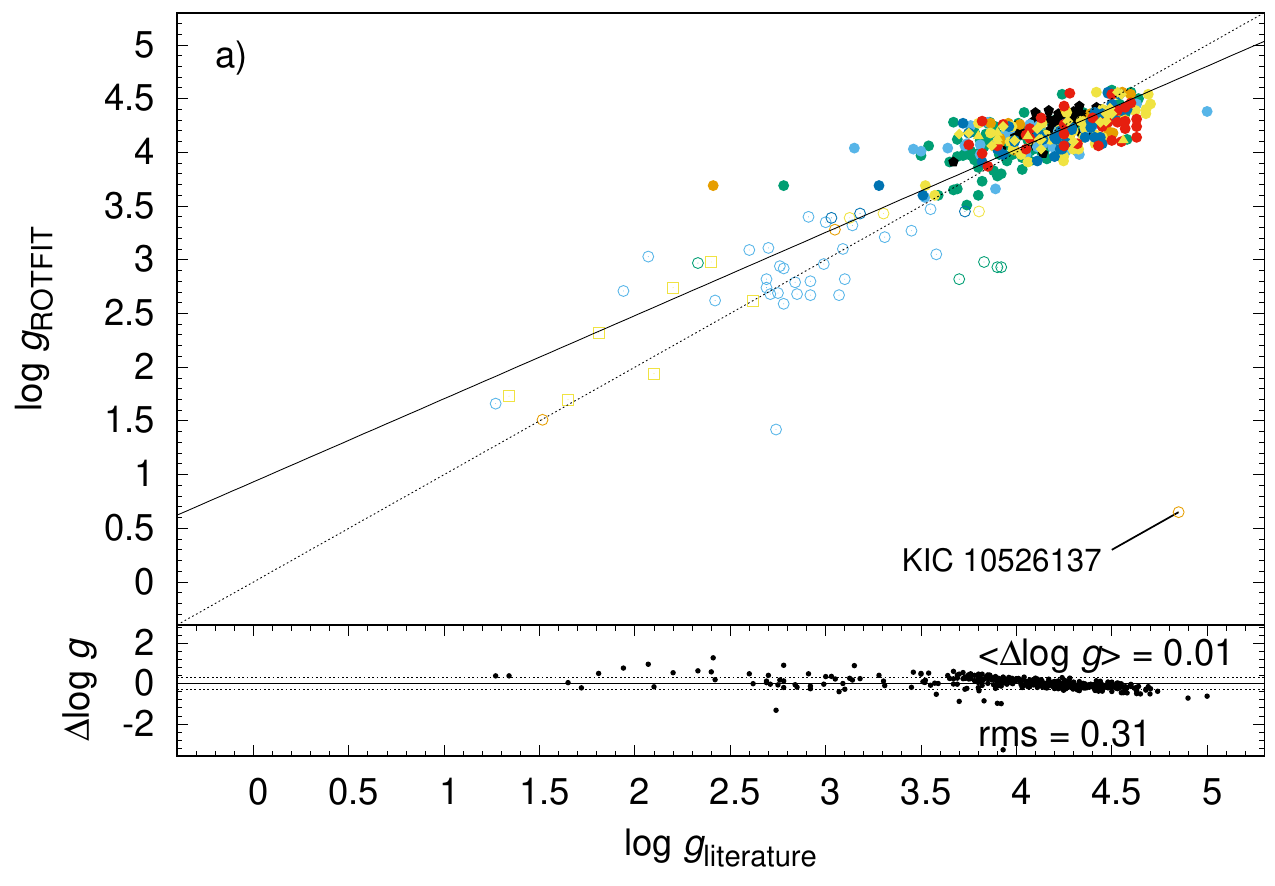}
\vspace{.2cm}
\includegraphics[width=8.cm]{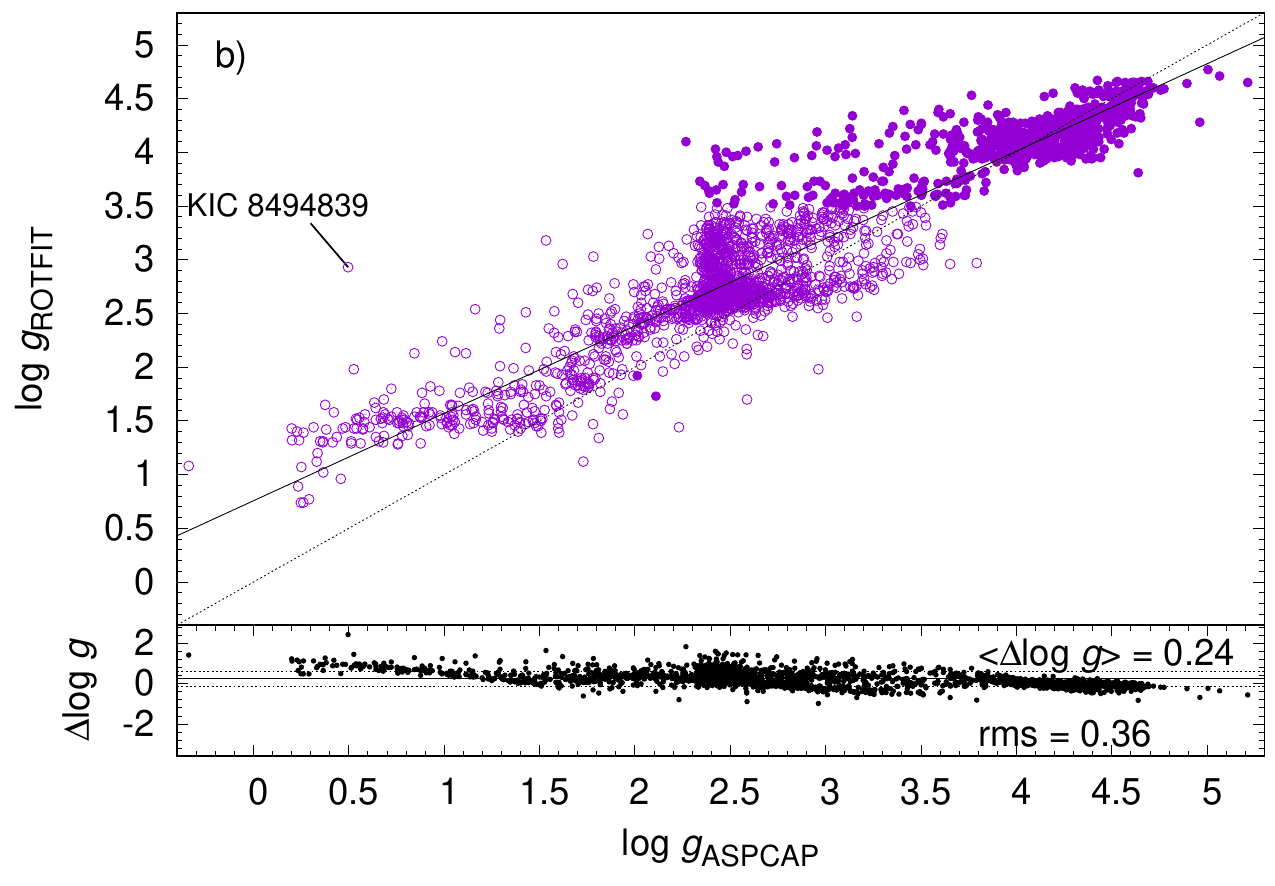}
\vspace{.1cm}
\includegraphics[width=8.cm]{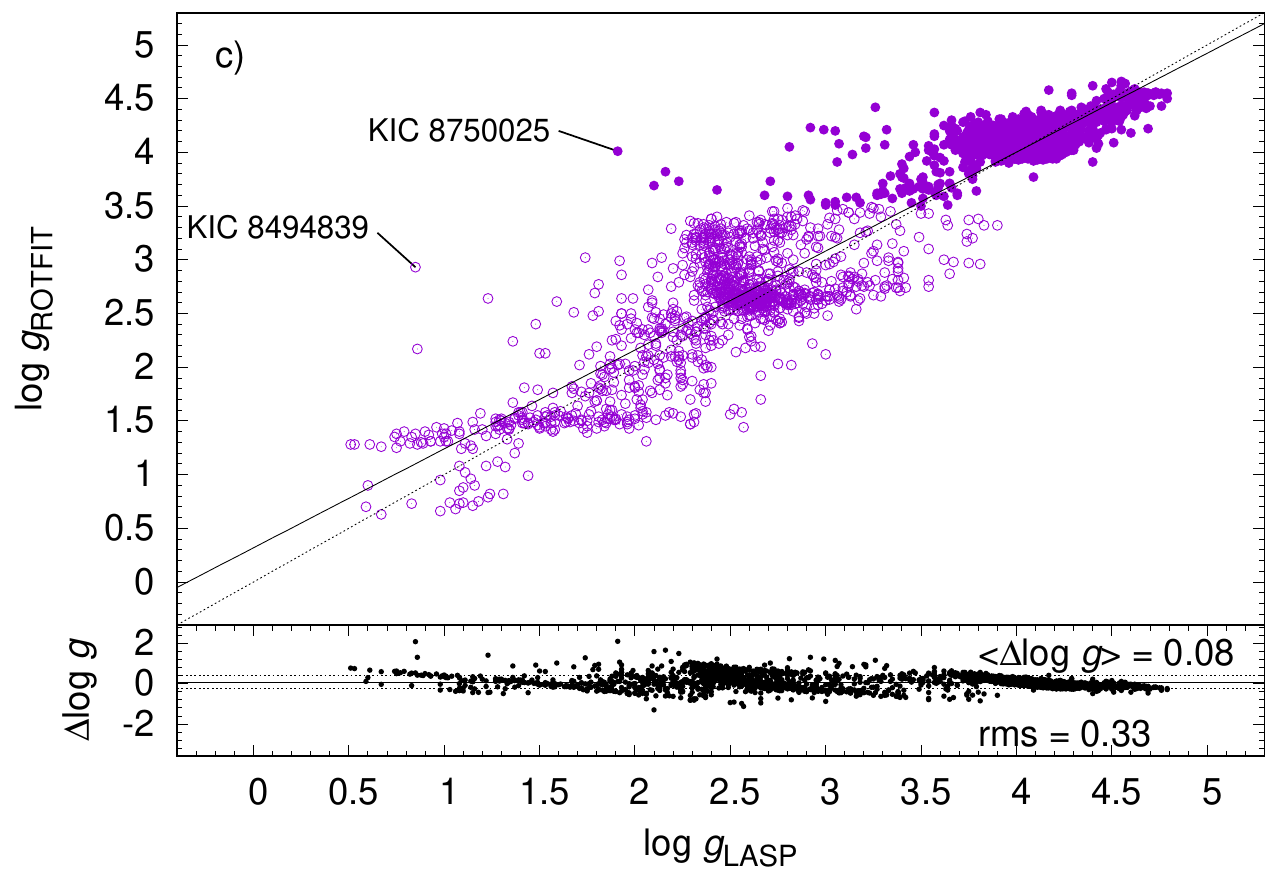}
\caption{Comparison of \logg\ values derived in the present work with those from the literature (a), APOGEE (b), and LASP (c). The meaning of lines and symbols is as in Fig.~\ref{Fig:JMZ_comp_Teff}.}
\label{Fig:JMZ_comp_logg}
\end{figure}

\begin{figure}[htb]
\includegraphics[width=8.cm]{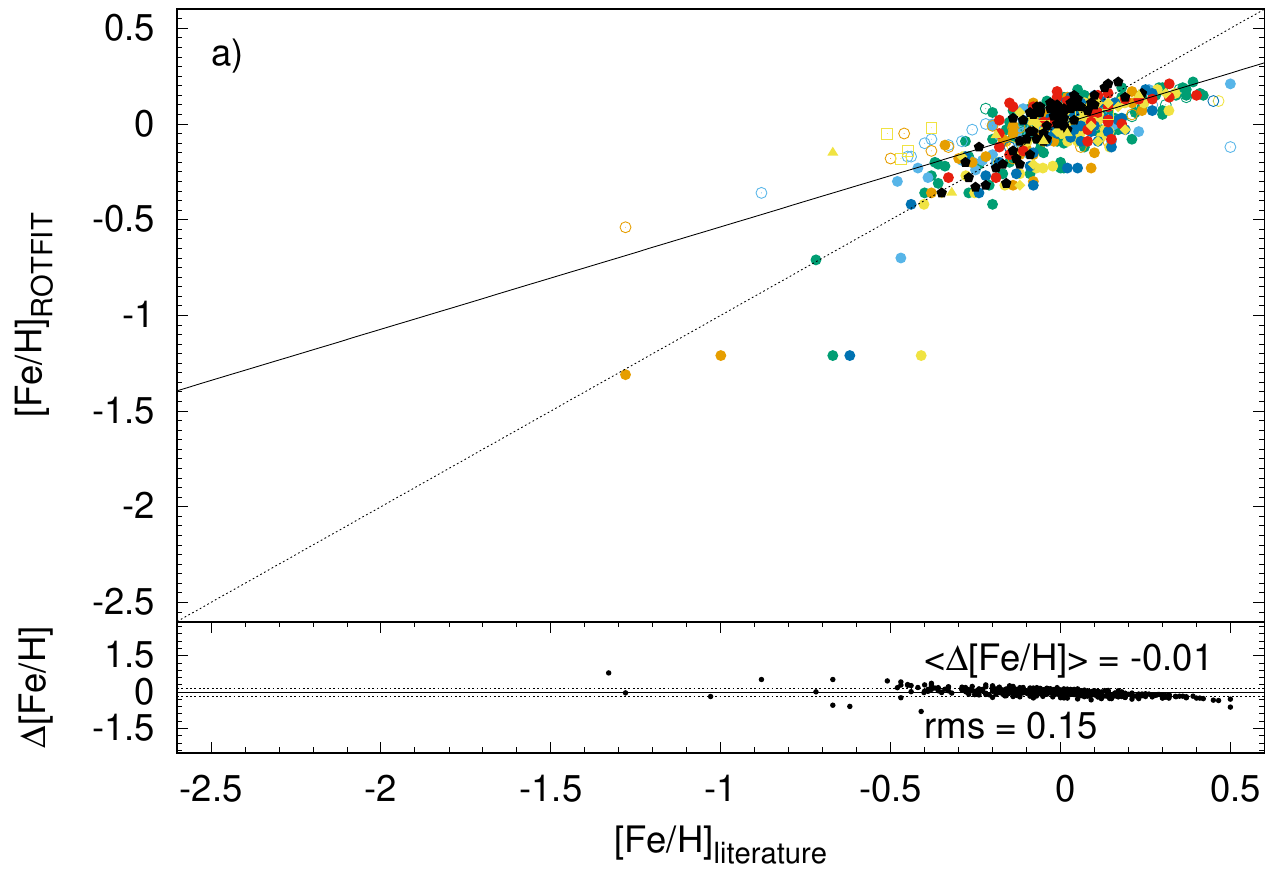}
\includegraphics[width=8.cm]{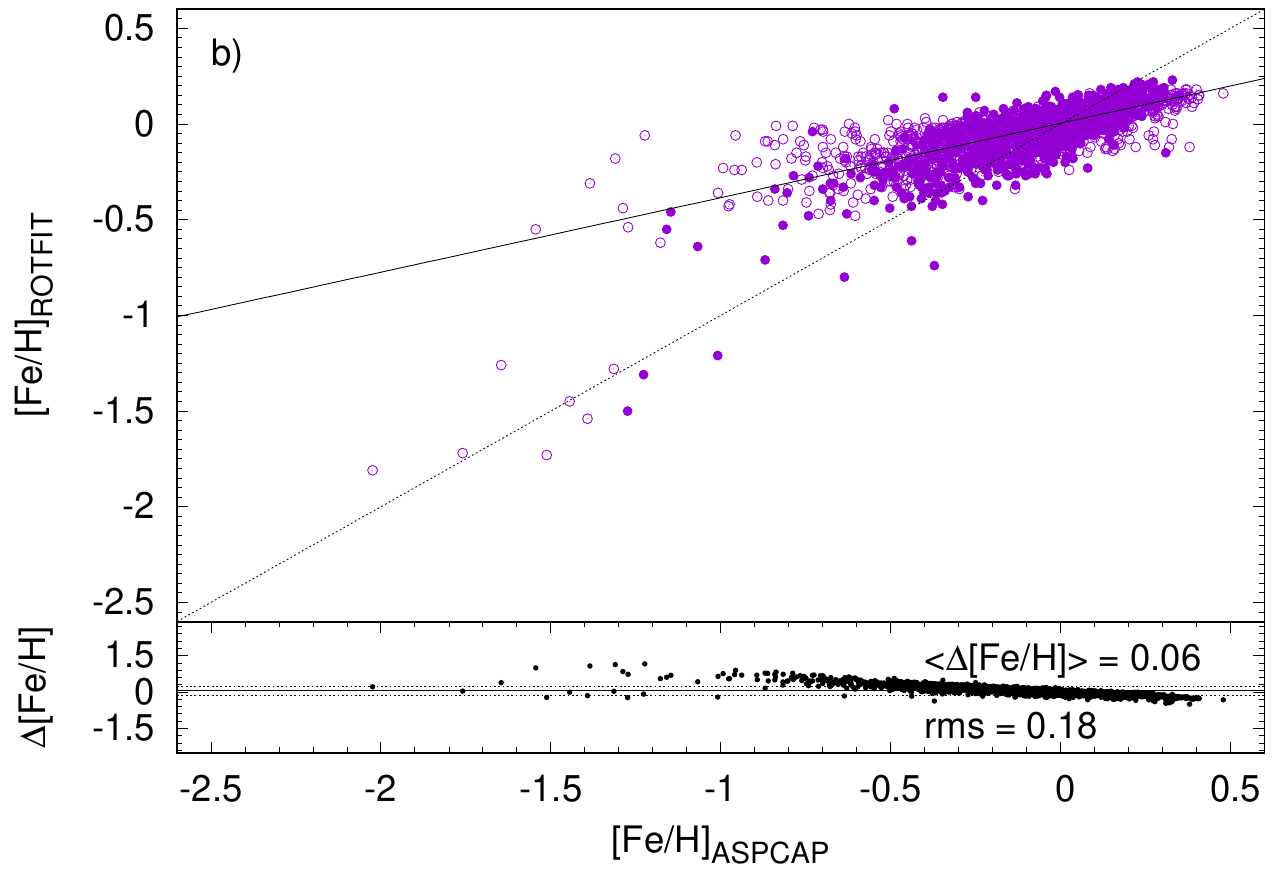}
\includegraphics[width=8.cm]{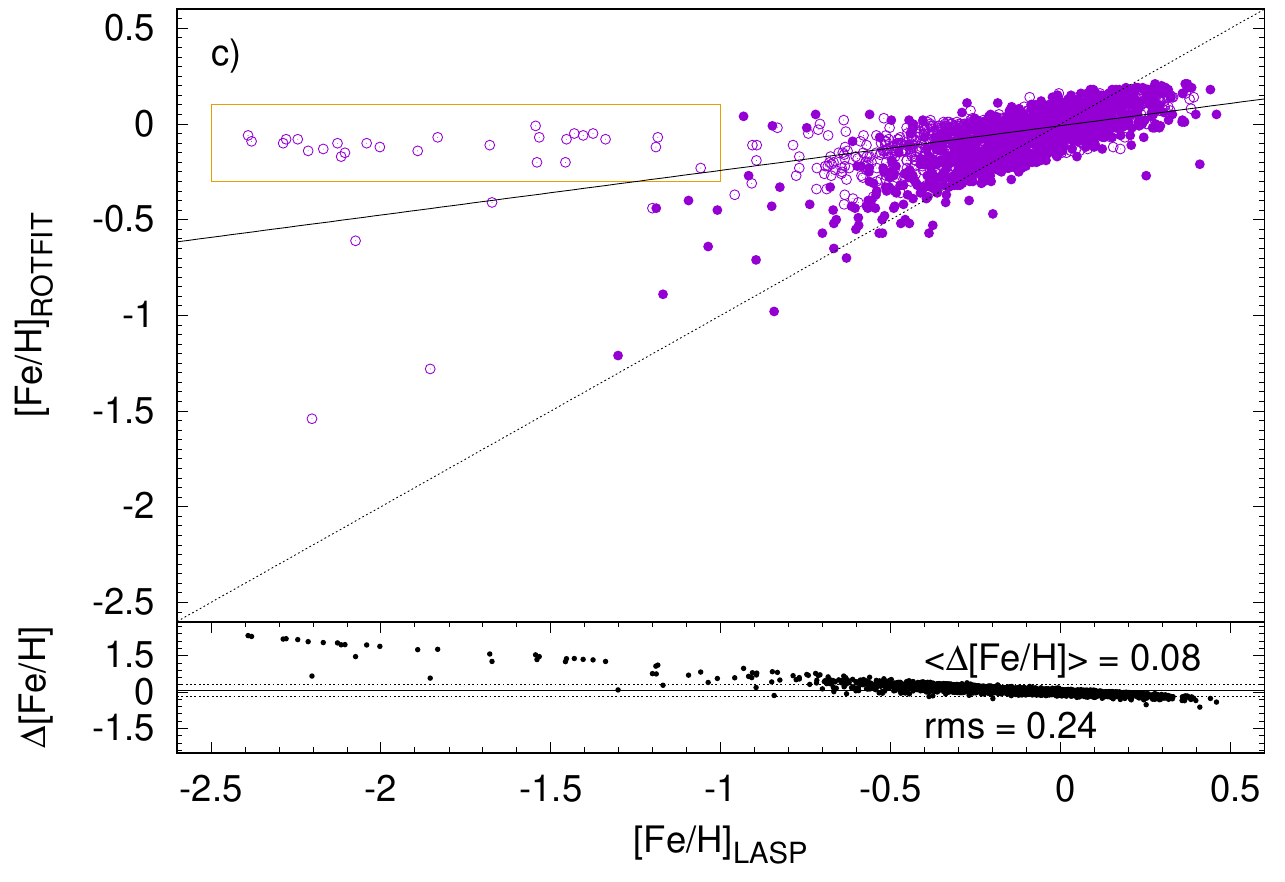}
\caption{Comparison of \feh\ values derived in the present work with those from the literature (a), APOGEE (b), and LASP (c).
The meaning of lines and symbols is as in Fig.~\ref{Fig:JMZ_comp_Teff}.
The sources enclosed in the yellow rectangle in box (c) are discussed in Appendix\,\ref{Appendix:Discrepant}.}
\label{Fig:JMZ_comp_FeH}
\end{figure}

The results of the comparison of the APs derived from MRS in the present work with those found in the literature are shown in Figs.~\ref{Fig:JMZ_comp_Teff}, \ref{Fig:JMZ_comp_logg}, and \ref{Fig:JMZ_comp_FeH} for \teff, \logg, and \feh, respectively. 
The majority of the literature parameters that are used for those comparisons were derived with the APOGEE Stellar Parameter and Chemical Abundance Pipeline (ASPCAP) for 2486 stars from the APOGEE DR16 catalog \citep{2020AJ....160..120J} and with \lasp\  adapted to the MRS \citep[see][and references therein]{Zong2020ApJS..251...15Z} for 2408 stars. 
The values of the APs determined by us are plotted versus the APOGEE ones in the middle panels of Figs.~\ref{Fig:JMZ_comp_Teff}, \ref{Fig:JMZ_comp_logg}, and \ref{Fig:JMZ_comp_FeH}, while the comparison with the \lasp\ ones is shown in the bottom panels of the same figures.
The remaining literature determinations, which are compared with our values in Figs.~\ref{Fig:JMZ_comp_Teff}a, \ref{Fig:JMZ_comp_logg}a, and \ref{Fig:JMZ_comp_FeH}a, were derived from high-resolution optical or, in two cases, low-resolution infrared spectra by 
\citet[][48 stars,  navy blue circles]{2018ApJS..237...38B}, 
\citet[][19 stars,  green diamonds]{2016ApJS..225...32B}, 
\citet[][127 stars, green circles]{2018ApJ...861..149F}, 
\citet[][34 stars,  red circles]{2018MNRAS.481.3244G}, 
\citet[][63 stars,  azure circles]{2017ApJ...838...25G}, 
\citet[][86 stars,  orange circles]{Huber2014_ApJS_211_2}, 
\citet[][7 stars,   yellow squares]{2019A&A...625A.141L}, 
\citet[][2 stars,   yellow triangles]{Niemczura2015}, 
\citet[][2 stars,   yellow triangles]{2017MNRAS.470.2870N}, 
\citet[][1 star,    yellow triangle]{2015PASJ...67...32N}, 
\citet[][51 stars,  yellow circles]{2017AJ....154..107P}, 
\citet[][62 stars,  black pentagons]{2020A&A...636A..85S}, or 
\citet[][1 star,    yellow triangle]{2013A&A...556A..52T}. 

Those 5397 determinations concern 4018 individual stars. For 2892 stars from that sample, the literature values of APs were provided only in one paper, for 980 stars
in two papers, for 90 stars in three papers, for 30 stars in four papers, for 12 stars in five papers, for seven stars in six papers, for three stars in seven papers, 
and for four stars in eight papers.
Since the above mentioned works do not provide information on all the three APs, the numbers of individual stars for which we found the literature values of \teff, \logg\, and \feh\, are, respectively, 5220, 3994, and 5179.

In order to find a linear relation between our determinations of the APs and the literature values, for each star we either used the literature values of the APs and their errors (stars with single literature determinations) or computed the weighted means of the APs and their errors (stars with two or more literature determinations). 
We note the overall good agreement between our \teff\ values and those from the literature, with an average offset of only $5$\,K and an rms of 160\,K. Indeed, a linear regression (full line in Fig.~\ref{Fig:JMZ_comp_Teff}a) gives a slope $b$=0.89, which is smaller than one-to-one relation (dotted line in the same plot). A similar slope is seen in Fig.~\ref{Fig:JMZ_comp_Teff}c in which we compare our \teff\ values with those derived with the \lasp\ pipeline. However, the agreement with the APOGEE temperatures, shown in the middle panel, looks better and the linear regression has a slope nearly equal to 1 ($b$=0.98). The dispersion of the \teff\ differences around the mean is of about 160\,K, which is an estimate of the accuracy of our \teff\ determinations.
In Fig.~\ref{Fig:JMZ_comp_Teff}c, there are two \teff\ outliers, which we discuss in more detail in Appendix\,\ref{Appendix:Discrepant}.

The values of \logg\ display also a good agreement with the literature with a small offset of only 0.01~dex and a scatter of about 0.31~dex (Fig.~\ref{Fig:JMZ_comp_logg}\,a). 
The comparison of our values of \logg\ with those from APOGEE and \lasp\ catalogs (Fig.~\ref{Fig:JMZ_comp_logg}\,b and c, respectively) displays an overall good agreement, with an rms dispersion of $\approx$\,0.3.
We note only very few discrepant sources that are labeled with their KIC identifiers and discussed in Appendix~\ref{Appendix:Discrepant}. 

The agreement of \logg\ with the literature is better than that found in \citet[][hereafter Paper\,II]{Frasca2016}, where we found most of the \logg\ values derived from low-resolution LAMOST spectra to be clustered around the gravity typical for red giants (\logg\,$\approx2.5$) and main-sequence (MS) stars (\logg\,$\approx4.5$). The clustering effect, which is mostly the result of the nonuniform density of templates in the space of parameters, is less pronounced in this case, likely due to the higher sensitivity to \logg\ of spectra with a higher resolution.  That effect is still visible for the data in the present work, especially for the stars with low gravity for which the \rotfit\ values tend to cluster around \logg\,=\,1.5, which is very likely caused by the small number of templates with a very low gravity. However, the comparison presented in this paper shows that the \logg\ values are accurate enough to distinguish between giant and MS stars, which, together with an accurate \teff\ determination unaffected by interstellar extinction, was one of the main purposes of our analysis. The knowledge of these parameters is, in fact, necessary for the spectral subtraction and flux calibration that we use to measure the chromospheric emission in the H$\alpha$ core (see Sects.~\ref{Sec:Activity} and \ref{Sec:Chromo}) and the lithium equivalent width and abundance (Sect.~\ref{Sec:Lithium}).

The agreement of \feh\ values with the literature is much poorer than of the previous parameters (Fig.~\ref{Fig:JMZ_comp_FeH}). We found a very small slope %($b$=0.43)
for the linear regression of the data ($b$=0.54 in Fig.~\ref{Fig:JMZ_comp_FeH}a or $b$=0.39 in Fig.~\ref{Fig:JMZ_comp_FeH}b), which indicates that we are getting correct \feh\ values only around the solar one (\feh=0) and we are systematically overestimating the metallicity for metal-poor stars (\feh$<-0.3$) and underestimating it for  metal-rich stars (\feh$>+0.2$).
We think that this effect, already seen in the results based on the low-resolution data of Paper\,II, is due to the relative scarcity of metal-poor and super metal-rich stars among our templates that generates a sort of ``smoothing'' of the final \feh\ values. Indeed, the handful of stars with the lowest metallicity in Fig.~\ref{Fig:JMZ_comp_FeH}\,a and b lie close to the one-to-one relation. For these stars, the templates with a higher metallicity have a spectrum so different that they have not played any role in the \feh\ determination.

We thus propose a correction relation for the \lamost\ metallicity, based on the linear fit shown in Fig.~\ref{Fig:JMZ_comp_FeH}\,b, which can be expressed as
\begin{equation}
{\rm [Fe/H}]_{\rm corr}  =  2.57\cdot{\rm [Fe/H]} -0.01,
\label{Eq:FeH} 
\end{equation} 
{\noindent applicable in the range}\\
~\\
$ {\rm [Fe/H]}>-1.0.$\\

Additionally, in Fig.~\ref{Fig:JMZ_comp_FeH}c we find a group of 27 stars, which have been enclosed in a yellow rectangle. All those stars are cool (\teff $\lesssim$ 4500\,K) giants for which the \lasp\ \feh\ values are very low while the ROTFIT \feh\ values are close to zero. Those stars we discuss in more detail in Appendix~\ref{Appendix:Discrepant}.

\subsection{Balmer \halpha\ and lithium equivalent width}             
\label{Sec:Activity}

\begin{figure}[th]
\hspace{-.7cm}
\includegraphics[width=9.5cm]{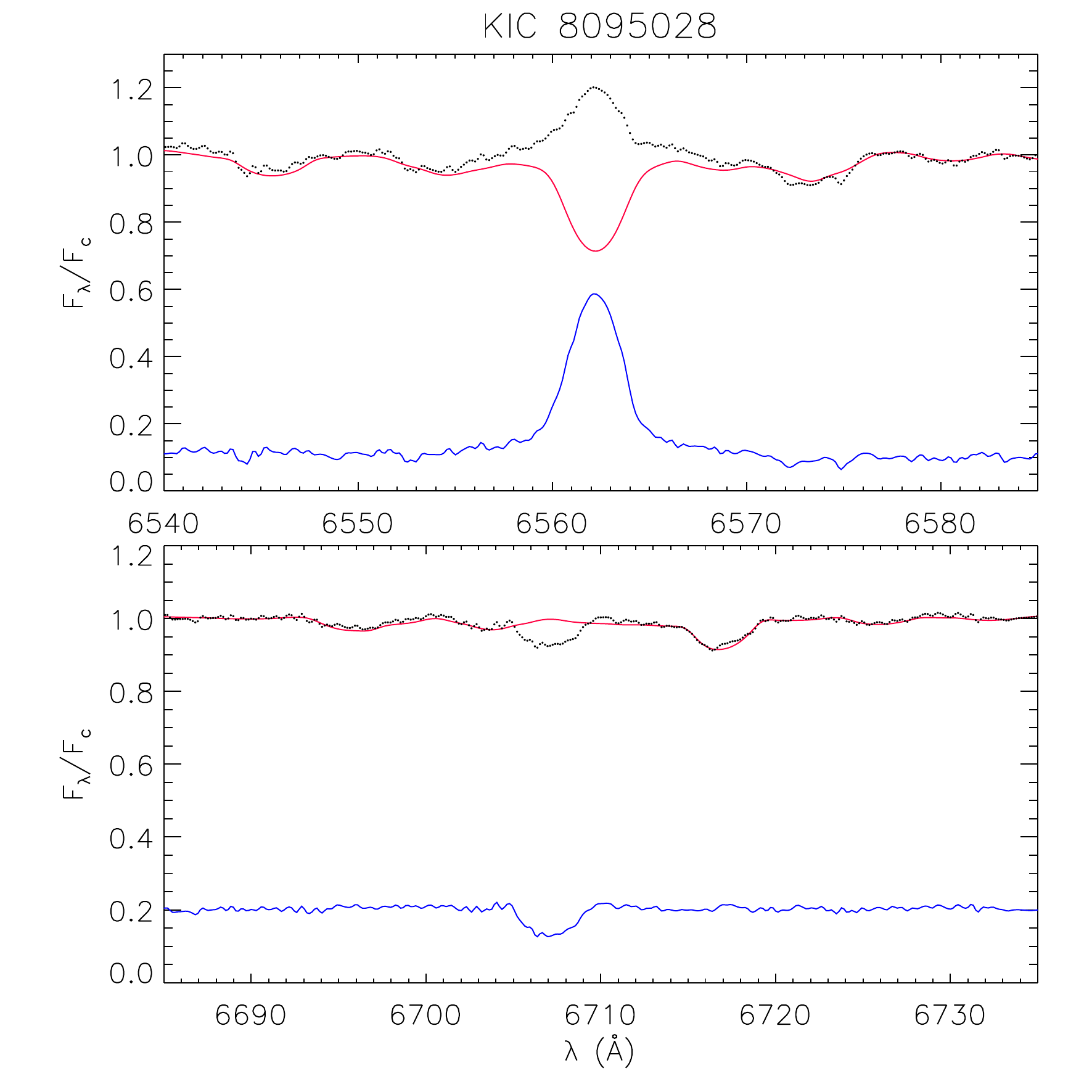}
\caption{Example of the subtraction of the best non-active, lithium-poor template (red line) from
the spectrum of KIC~8095028 (black dots), which reveals the chromospheric emission in the \halpha\ core (blue line in the {\it top panel}) and emphasizes the \ion{Li}{i} $\lambda$6708\,\AA\ absorption line, with the nearby blended lines removed ({\it bottom panel}).} 
\label{Fig:Ha_Li}
\end{figure}

The most sensitive diagnostics of magnetic activity in the range covered by the \lamost\  MRS is the Balmer \halpha\ line.
We therefore identified objects with \halpha\ emission, which can be produced by various physical mechanisms in addition to the presence of an active chromosphere, such as magnetospheric accretion in the youngest evolutionary phases of low- and intermediate-mass stars, or circumstellar (or circumbinary) matter. 
As the chromospheric emission can only show up as a small to moderate filling of the line core, depending on the activity level and on the photospheric flux of the star, the removal of the photospheric spectrum is crucial to emphasize the \halpha\ core emission.
To this aim, we subtracted the non-active template that best matches the final APs from each \lamost\ red-arm spectrum. 
This template has been aligned to the target RV, rotationally broadened at the \vsini\ of the target and resampled on its spectral points. 
The ``emission'' \halpha\ equivalent width, \Whalpha, was interactively measured by integrating the residual emission profile (see Fig.~\ref{Fig:Ha_Li}, upper panel). 

To speed up this procedure and to reduce spurious detections, we firstly made the measurements of \Whalpha\ only on the spectra with a S/N\,$\geq$\,20. 
Moreover, among these spectra, we selected a subsample of likely active stars as those for which an automatic procedure, which measure the \halpha\ equivalent width in a fixed wavelength range of 3\,\AA\ around the line center, gave rise to a value larger than 0.1\,\AA. 
Then, to pick up low-signal (6\,$<$\,S/N\,$<$\,20) spectra with a relevant filling or a pure \halpha\ emission above the continuum, we added the spectra with S/N\,$<$\,20 for which the ``initial'' automatic measure of \Whalpha\ was larger than 0.5\,\AA\ in the same integration range. 
For all these spectra, the \Whalpha\ was measured interactively, as described above, and the error was calculated as the product of the integration range times the error in the placement of the continuum, which was evaluated as the rms of the values of the subtracted spectra in two regions at the two sides of the \halpha\ line.
We end up with a total of 546+31 (cool,+\,unclass) spectra of 334 stars displaying \halpha\ in emission or filled in by a minimum amount as defined above. 
The maximum value, \Whalpha\,$=10.98$\,\AA, was found for KIC\,8749284, an active star already discovered by us in Paper\,II as an object with \halpha\ emission above the continuum. 
%\citep[][hereafter Paper\,II]{Frasca2016}.   

The values of \Whalpha, along with their errors, are quoted in Table\,\ref{Tab:active}. 
We also report whether the line is observed as a pure emission feature and whether the measure is uncertain as a result of the low S/N or other possible spectral issues.

The subtraction of the photospheric template also allowed us to measure the equivalent width of the \ion{Li}{i}\,$\lambda$\,6707.8\,\AA\ line by removing the nearby lines (basically \ion{Fe}{i}\,$\lambda$\,6707.4\,\AA), which are blended with the lithium line in the observed spectrum.
We used a similar approach to that used for \halpha\ to select the spectra on which we measure the lithium equivalent width, \WLi.
We fixed a threshold of 0.05 \AA\ (50\,m\AA) from an automatic measurement of the lithium absorption in the residual spectrum integrating it in a range of 6 \AA\ centered on the \ion{Li}{i}\,$\lambda$\,6707.8\,\AA\ line for the spectra with an S/N larger than 20 and 150\,m\AA\ for the low-signal spectra. 
We detected a \WLi\ above the given thresholds for 
2763 spectra, corresponding to 1657 different stars.
For stars with multiple visits, we calculated the weighted average of the values of \WLi\ measured in different epochs, adopting a weight $w=1/\sigma_{W_{\rm Li}}^2$, where 
$\sigma_{W_{\rm Li}}^2$ is the error on the individual measure.
We took the maximum of the weighted standard deviation and the standard error of the weighted mean as the final uncertainty. 

The highest value of lithium equivalent width, $W_{\rm Li}$ = 626\,m\AA, was found for KIC\,11657857 (= IRAS\,19170+4937), which is also one of the coldest lithium-rich giants ever discovered.

\section{Results}
\label{Sec:Results}

\subsection{Chromospheric activity}
\label{Sec:Chromo}

Late-type stars, which have convective envelopes, rotation and differential rotation able to produce strong magnetic fields by a dynamo action, display a complex of phenomena known as magnetic activity. These include, radio and/or X-ray coronal emission, UV and optical  emission lines produced in a chromosphere, fast energy releases (in the continuum and/or in spectral lines) known as flares, and rotational modulation of brightness produced by cool spots. 
For stars with a SpT later than mid-F (\teff$\leq 6500$\,K) the \halpha\ line is an efficient diagnostic of magnetic activity, especially for the cooler ones with a moderate to high level of activity, if we subtract the underlying photospheric spectrum, as we have explained in Sect.\,\ref{Sec:Activity}.

The equivalent width of an emission line formed in a chromospheric layer can be used to quantify the activity level, when it is converted into energetic units. 
Therefore, more accurate indicators of chromospheric activity are the line flux in units of stellar surface, $F$, and the ratio between the line luminosity and bolometric luminosity, $R'$, which can be calculated, for the \halpha , as
\begin{eqnarray}
F_{\rm H\alpha} & = & F_{6563}W^{\rm res}_{\rm H\alpha} \\  
R'_{\rm H\alpha}&  = & L_{\rm H\alpha}/L_{\rm bol} = F_{\rm H\alpha}/(\sigma T_{\rm eff}^4),
\end{eqnarray}
{\noindent where $F_{6563}$ is the continuum surface flux at the \halpha\ center, which we evaluated from the BT-Settl synthetic spectra \citep{Allard2012} at the stellar temperature and surface gravity of the target.}
The flux error includes both the error of the equivalent width and the uncertainty in the continuum flux at the line center, which is obtained by propagating the \teff\  and \logg\  errors. 

%\begin{longtable}{ccccccccccccc}
\begin{table*}[tb]
\caption[ ]{Activity indicators.}
\label{Tab:active}
\scriptsize
\begin{tabular}{ccccccccccccc}
\hline
\hline
\noalign{\smallskip}
 DESIG & KIC & RA & Dec. & \teff &  err & \logg & err & \Whalpha & err & $F_{\rm H\alpha}$ & err & $R'_{\rm H\alpha}$ \\ 
        &       & $\degr$ & $\degr$ &  \multicolumn{2}{c}{(K)} & \multicolumn{2}{c}{(dex)} & \multicolumn{2}{c}{(\AA)}  & \multicolumn{2}{c}{(erg\,cm$^{-2}$s$^{-1}$)} &   \\ 
\noalign{\smallskip} 
\hline
\noalign{\smallskip} 
 J184805.69+415751.8 & KIC06497820 & 282.023712 & 41.964409 & 5164 & 101 & 4.44 &    0.26 & 0.779 & 0.163 & 3.632e+06 & 8.214e+05 & -4.045 \\ 
 J184847.43+421316.2 & KIC06752578 & 282.197632 & 42.221169 & 3536 & 90 & 4.77 &  0.13 & 1.478 & 0.152 & 7.400e+05 & 1.557e+05 & -4.078 \\ 
 J184942.18+435303.5 & KIC08008128 & 282.425781 & 43.884331 & 5920 & 177 & 4.02 & 0.15 & 0.095 & 0.061 & 7.637e+05 & 4.982e+05 & -4.960 \\
 J184951.21+431802.6 & KIC07661598 & 282.463379 & 43.300739 & 5834 & 136 & 4.29 &   0.16 & 0.036 & 0.034 & 2.736e+05 & 2.596e+05 & -5.380 \\ 
 J185114.02+451313.2 & KIC08932950 & 282.808441 & 45.220352 & 5829 & 99 & 4.33 &   0.12 & 0.107 & 0.032 & 8.102e+05 & 2.482e+05 & -4.907 \\ 
 J185236.87+451902.5 & KIC09002183 & 283.153656 & 45.317371 & 5309 & 135 & 4.49 &   0.14 & 0.113 & 0.069 &  5.934e+05 & 3.677e+05 & -4.880 \\ 
 J185336.19+444900.1 & KIC08671812 & 283.400818 & 44.816711 & 4980 & 103 & 3.09 &   0.27 & 0.375 & 0.057 & 1.436e+06 & 2.569e+05 & -4.385 \\ 
 J185342.86+450847.1 & KIC08866716 & 283.428589 & 45.146439 & 4927 & 132 & 2.93 &   0.27 & 0.197 & 0.080 & 7.184e+05 & 3.052e+05 & -4.668 \\ 
 J185441.45+451133.4 & KIC08867218 & 283.672729 & 45.192612 & 4944 & 132 & 4.55 & 0.11 & 0.121 & 0.028 & 4.640e+05 & 1.215e+05 & -4.863 \\
 J185450.75+401409.3 & KIC05077994 & 283.711487 & 40.235920 & 5608 & 97 & 4.19 & 0.27 & 0.244 & 0.091 & 1.584e+06 & 6.004e+05 & -4.549 \\ 
 J185501.45+445803.9 & KIC08736810 & 283.756073 & 4.967758 & 5874 & 166 & 4.08 &  0.15 & 0.672 & 0.245 & 5.241e+06 & 1.995e+06 & -4.110 \\ 
 J185519.56+445212.3 & KIC08672632 & 283.831512 & 44.870090 & 4730 & 124 & 4.58 &  0.11 & 0.268 & 0.170 & 8.305e+05 & 5.382e+05 & -4.534 \\ 
 J185633.42+451348.1 & KIC08935655 & 284.139252 & 45.230042 & 3575 & 106 & 4.71 &  0.12 & 4.820 & 0.276 & 2.359e+06 & 5.429e+05 & -3.594 \\ 
 J185637.96+443001.1 & KIC08479165 & 284.158173 & 44.500309 & 6076 & 130 & 4.09 &  0.17 & 0.073 & 0.044 & 6.490e+05 & 3.944e+05 & -5.076 \\ 
 \dots & \dots & \dots & \dots & \dots & \dots & \dots & \dots & \dots & \dots & \dots & \dots & \dots \\
\noalign{\smallskip} 
\hline       
\noalign{\smallskip} 
\end{tabular}
\\
{\bf Notes.} The full table is only available in electronic form at the CDS.
\end{table*}

\begin{figure*}  
\begin{center}
\includegraphics[width=8.8cm]{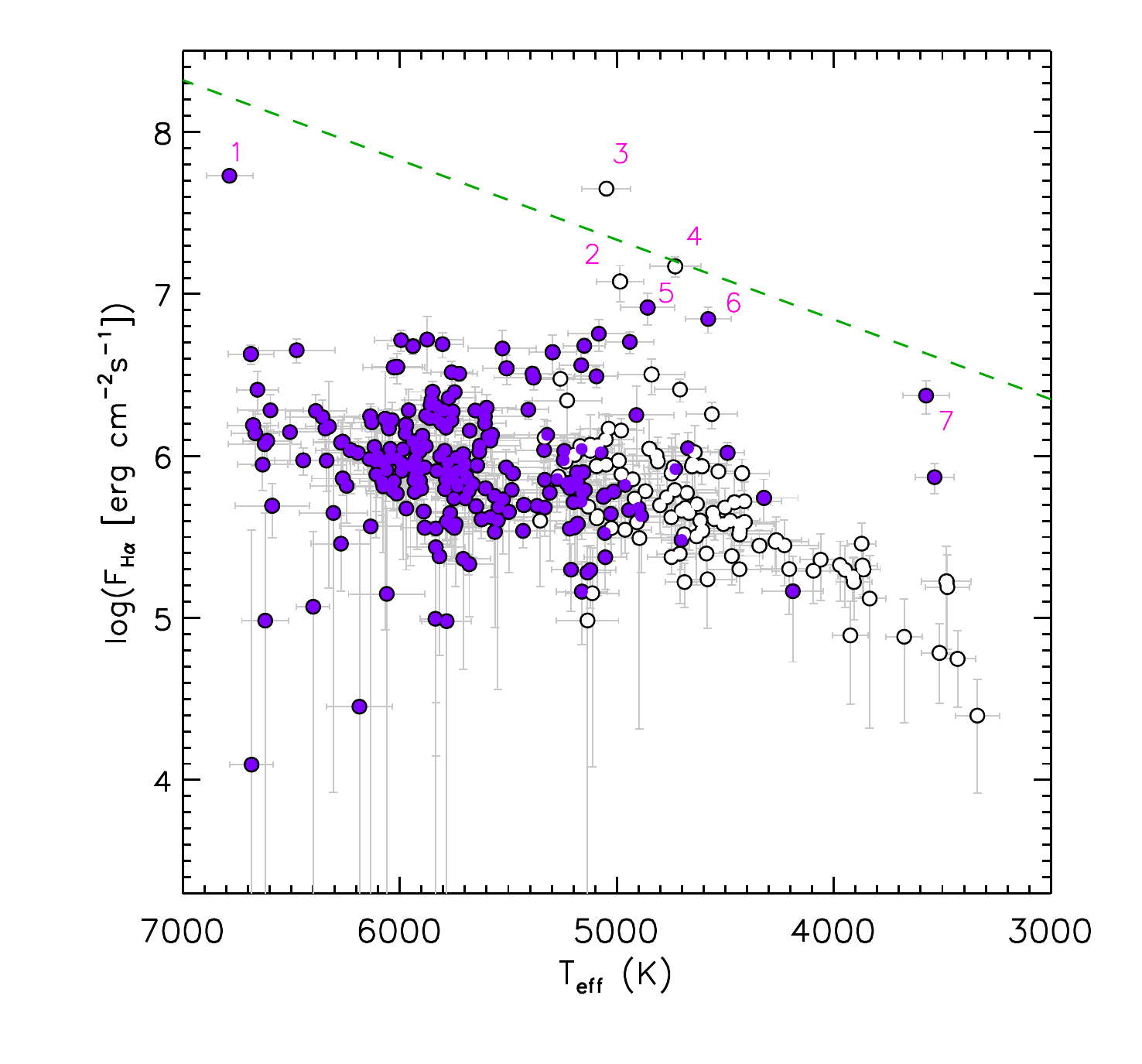} 
\includegraphics[width=8.8cm]{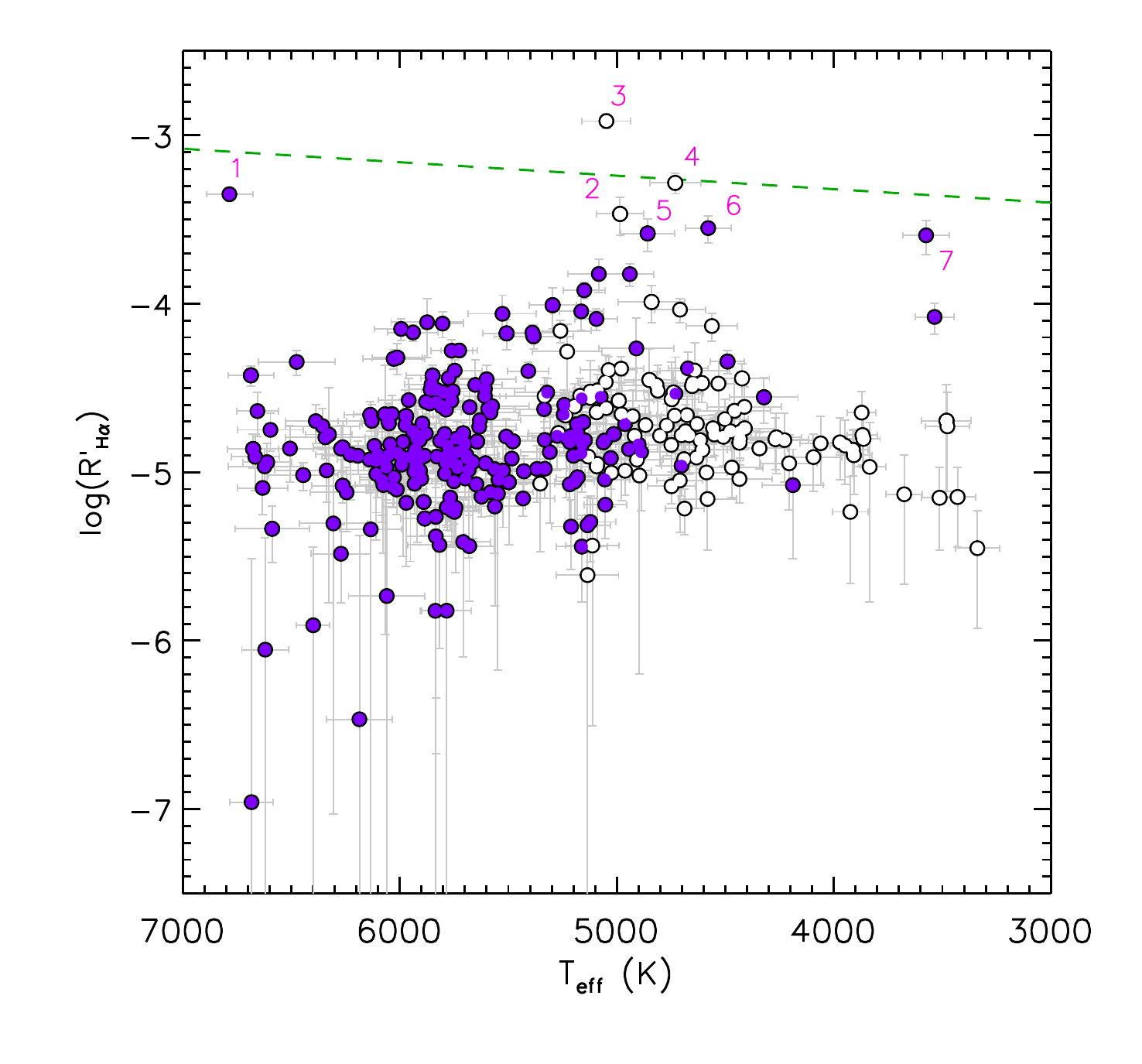}  
\caption{Activity indicators. {\it Left panel:} \halpha\ flux versus \teff\ (open circles for giants, purple dots for MS stars). {\it Right panel:} $R'_{\rm H\alpha}$ versus \teff\ (same symbols).
The straight dashed line in each panel is the boundary between chromospheric emission (below) and accretion as derived by \citet{Frasca2015}. The objects with a flux close to this boundary are labeled with integer numbers and are discussed briefly in the text.}
\label{Fig:Flux_Teff}
\end{center}
\end{figure*}

The \halpha\ fluxes and $R'_{\rm H\alpha}$ of our targets are plotted as a function of the effective temperature in Fig.~\ref{Fig:Flux_Teff} using different symbols for giant (open circles) and MS stars (dots). This plot also shows the boundary between young stars with mass accretion and chromospherically active stars. The latter lie below this line, which is also close to a ``saturated'' activity regime for MS stars \citep[see, e.g.,][]{Frasca2015}. 

We note that only two stars lie higher (or right over) the dividing line (3 and 4 in Fig.\,\ref{Fig:Flux_Teff}), but we classified them with \rotfit\ as giant stars. Their high \halpha\ flux is likely the fingerprint of an extremely high activity level due to an enhanced rotation rate, which could be the result of a particular evolutionary phase or of spin-orbit synchronization for an unresolved binary or an SB1 system in which the companion of the cool giant is overwhelmed by its flux (similar to many long-period RS\,CVn binaries). We cannot exclude the contribution of other processes, such as circumstellar matter or shells, as the cause for the strong \halpha\ emission (see also their \halpha\ profiles in Fig.\,\ref{Fig:Halpha_prof}).

The stars with the highest \halpha\ flux, which lie above or close to the dividing line between accretors and chromospherically active stars, are indicated with progressive numbers (with decreasing \teff) in Fig.\,\ref{Fig:Flux_Teff}. In the following we report a few notes on these sources.

KIC 9720659 (\#1) is the star with the 
earliest SpT (F0\,V, \teff\,=\,6785\,K) in our sample of H$\alpha$ emitters. According to the H$\alpha$ profile, it could be a Herbig star, with broad wings typical of an A- or F-type star with an emission core, similar to \object{CQ~Tau} \citep[e.g.,][and references therein]{Alcala2021}. We did not detect any other emission line in its spectrum.
However, a moderate infrared excess, indicative of circumstellar matter or a protoplanetary disk is visible in its spectral energy distribution (SED; Fig.\,\ref{Fig:SED_KIC9720659}).

KIC 8022670 = V2279 Cyg (\#2) is mentioned in the SIMBAD astronomical database as an ``eruptive variable star.'' It has been reported as a periodic variable (Cepheid?) by \citet{Schmidt2007} and \citet{Pigulski2009} with a period of about 4.12 d and an amplitude of 0.30 mag.  \citet{Schmidt2011} included this star in their study of type\,II Cepheid candidates and related stars from the Robotic Optical Transient Search Experiment 1 (ROTSE-1), which was based on LRS. They report \teff\ in the range 4905--4957\,K, \logg\,=\,4.1, and a solar metallicity. They also report H$\alpha$ emission (c.f. their Fig. 1) with a Lick-IDS (Lick Observatory Image Dissector Scanner) index of about $-2.29$.  We find nearly the same effective temperature, \teff\,=\,4985$\pm$110\,K, but a different \logg=\,2.64$\pm$0.41\,dex. The latter is in better agreement with other literature values, such as \logg\,=\,2.68 \citep{Christiansen2012} or  \logg\,=\,2.82 \citep{Ho2017} that indicate this star as a giant.
It was subsequently reported as an active flaring star by \citet{2016ApJ...829...23D} and \citet{Olah2021} on the basis of \kepler\ photometry. In particular, \citet{Olah2021} classify it as a flaring giant with a radius of 3.3 $R_{\sun}$ for which 55 flares have been detected by \kepler.

KIC 8749284 (\#3) is reported in SIMBAD as a ``rotationally variable star'' with a period $P_{\rm rot}\simeq 3.22$\,d \citep{Debosscher2011}. It is also classified as an active giant, based on \kepler\ light photometry (see \citealt{2020A&A...639A..63G}), which displays intense flare activity \citep{Olah2021}, with 88 flares detected during \kepler\ observations. We find \teff=5048$\pm$112\,K and \logg\,=\,3.20$\pm$0.30, which indicates an evolved star, as only a weak lithium absorption line (\WLi\,$\simeq 40$\,m\AA), which is not compatible with a pre-main-sequence (PMS) object, has been detected. This is in agreement with the \teff=5089\,K and the radius of 4.6\,$R_{\sun}$ reported by \citet{Olah2021}. The SED of this object reveals a strong infrared excess that begins from the $H$ band, which is likely due to cold circumstellar matter. We note that this star was already included in the \projectlrs\ and we already detected the H$\alpha$ emission in the low-resolution \lamost\ spectra as well as the infrared excess in its SED \citep{Frasca2016}. 

KIC~6201369 (\#4)  %(2MASS\,J19223143+4133342) 
is not present in SIMBAD. In the {\it Gaia} early third data release (EDR3) there is a nearby source at about 6\arcsec\ that is about 6 mag fainter. No spectroscopic observations can be found in the literature for this object. We find \teff\,=\,4731$\pm$118\,K and \logg\,=\,2.76$\pm$0.27, meaning that it is likely another active giant with a very strong and redshifted \halpha\ emission. The \ion{Li}{i}$\lambda$6708 line is not detected in the MRS spectrum of this source.
There is no information in the literature about brightness variations; this source, although in its field of view, was not observed by \kepler.

KIC~8095028 (\#5) is a young (strong lithium) fast-rotating ($P_{\rm rot}=0.397$\,d, \citealt{McQuillan2014}) and flaring star (see \citealt{2016ApJ...829...23D} and \citealt{2019ApJS..244...37G}). An MRS spectrum of this target is displayed in Fig.\ref{Fig:Ha_Li}. We classified it as a K3.5\,V star with \teff=4858$\pm$126\,K and \logg=4.58$\pm$0.15, to be compared with the values of \teff\,=\,4464\,K and \logg\,=\,4.49 reported by \citet{McQuillan2014}. The lithium equivalent width of \WLi\,=\,272 m\AA\ places this star just below the Pleiades upper envelope
in Fig.~\ref{Fig:Li} (i.e., it is the MS star with the highest \WLi\ in our sample; see Sect.~\ref{Sec:Lithium}).

KIC 10063343 (\#6) is an ultrafast rotator ($P_{\rm rot}=0.337$\,d, \citealt{McQuillan2014}),
which shows several flare episodes in the \kepler\ light curves \citep[see, e.g.,][]{2016ApJ...829...23D,Yang2019}. We confirm it as a K5\,V star (\teff\,=\,4579$\pm$105, \logg\,=\,4.65$\pm$0.11) rotating at \vsini\,=\,104\,\kms.
The \ion{Li}{i}$\lambda$6708 line is not visible in the low-signal MRS spectrum of this source.

KIC 8935655 (\#7) is a dMe flaring star (see \citealt{2017ApJ...849...36Y, 2016ApJ...829...23D}). \citet{2019AJ....158...75H} report an M4\,V SpT and \teff\,=\,3241\,K; we find this star slightly earlier (M1.5\,V) and hotter (\teff\,=\,3575$\pm$102\,K) and confirm it as a slowing rotating (\vsini$<8$\,\kms) dMe star. Our temperature is instead in very good agreement with that based on APOGEE spectra, \teff\,=\,3554$\pm$69\,K \citep{2020AJ....160..120J}. The \ion{Li}{i}$\lambda$6708 line is not visible in the MRS spectrum of this source.

\subsection{Lithium abundance and age}
\label{Sec:Lithium}

\begin{table*}[tb]
\caption[ ]{Lithium equivalent widths and abundances. }
\label{Tab:Lithium}
\scriptsize
\begin{tabular}{ccccccccc}
\hline
\hline
\noalign{\smallskip}
 DESIG & KIC & RA      & Dec.   & \teff & \logg  & $A$(Li) & \WLi  & Class \\ 
       &     & $\degr$ & $\degr$ &  (K) & (dex) & (dex) & (m\AA)  &     \\ 
\noalign{\smallskip} 
\hline
\noalign{\smallskip} 
 J184441.98+422834.4 & KIC06922059 & 281.174930 & 42.476238 & 5829$\pm$ 70 & 4.31$\pm$0.12 & 2.32$^{+0.20}_{-0.27}$ &  45$\pm$17  &  Hya   \\
 J184456.49+423715.0 & KIC07091292 & 281.235410 & 42.620834 & 5812$\pm$141 & 4.09$\pm$0.18 & 2.52$^{+0.16}_{-0.18}$ &  68$\pm$11  &  UMa   \\
 J184502.55+422843.5 & KIC06922204 & 281.260650 & 42.478771 & 5843$\pm$140 & 4.21$\pm$0.20  & 2.50$^{+0.22}_{-0.24}$ &  64$\pm$18  &  Hya   \\
 J184529.95+422421.2 & KIC06922382 & 281.374820 & 42.405910 & 5894$\pm$ 97 & 4.04$\pm$0.15  & 2.14$^{+0.40}_{-2.44}$ &  28$\pm$28  &  Hya   \\
 J184613.19+421535.2 & KIC06751420 & 281.554960 & 42.259800 & 4688$\pm$112 & 2.56$\pm$0.25  & 1.20$^{+0.45}_{-1.30}$ &  26$\pm$26  &  \dots \\
 J184700.25+440107.3 & KIC08144149 & 281.751070 & 44.018719 & 5728$\pm$117 & 4.34$\pm$0.19  & 2.05$^{+0.41}_{-2.32}$ &  29$\pm$29  &  Hya   \\
 J184713.75+435409.1 & KIC08075941 & 281.807310 & 43.902538 & 5400$\pm$241 & 3.93$\pm$0.43  & 2.09$^{+0.36}_{-0.49}$ &  51$\pm$24  &  UMa   \\
 J184715.22+421233.3 & KIC06751838 & 281.813420 & 42.209251 & 4625$\pm$121 & 2.51$\pm$0.22  & 1.01$^{+0.46}_{-1.10}$ &  20$\pm$20  &  \dots \\
 J184732.37+435138.3 & KIC08007262 & 281.884890 & 43.860657 & 6821$\pm$ 73 & 4.18$\pm$0.11  & 3.03$^{+0.16}_{-0.20}$ &  56$\pm$16  &  \dots \\
 J184741.36+442038.4 & KIC08344972 & 281.922360 & 44.344002 & 5739$\pm$147 & 4.28$\pm$0.18  & 2.08$^{+0.20}_{-0.23}$ &  30$\pm$ 7  &  Hya   \\
 J184753.87+435727.6 & KIC08076240 & 281.974460 & 43.957680 & 5660$\pm$132 & 4.20$\pm$0.21  & 2.11$^{+0.29}_{-0.43}$ &  37$\pm$19  &  Hya   \\
 J184805.84+440407.9 & KIC08144580 & 282.024350 & 44.068871 & 4883$\pm$126 & 3.03$\pm$0.32  & 1.69$^{+0.39}_{-0.73}$ &  48$\pm$35  &  \dots \\
 J184805.93+424448.3 & KIC07175400 & 282.024720 & 42.746750 & 6015$\pm$156 & 4.08$\pm$0.14  & 2.51$^{+0.29}_{-0.40}$ &  52$\pm$25  &  \dots \\
 J184814.26+423130.9 & KIC07009054 & 282.059450 & 42.525253 & 6633$\pm$130 & 4.09$\pm$0.16  & 3.10$^{+0.15}_{-0.17}$ &  81$\pm$14  &  \dots \\
 J184826.05+434740.2 & KIC07938762 & 282.108550 & 43.794521 & 5571$\pm$147 & 4.43$\pm$0.17  & 2.56$^{+0.28}_{-0.34}$ & 102$\pm$38  &  UMa   \\
 \dots & \dots & \dots & \dots & \dots & \dots & \dots & \dots & \dots   \\
\noalign{\smallskip} 
\hline       
\noalign{\smallskip} 
\end{tabular}
\\
{\bf Notes.} The full table is only available in electronic form at the CDS.
\end{table*}

\begin{figure}[th]
\includegraphics[width=9.0cm]{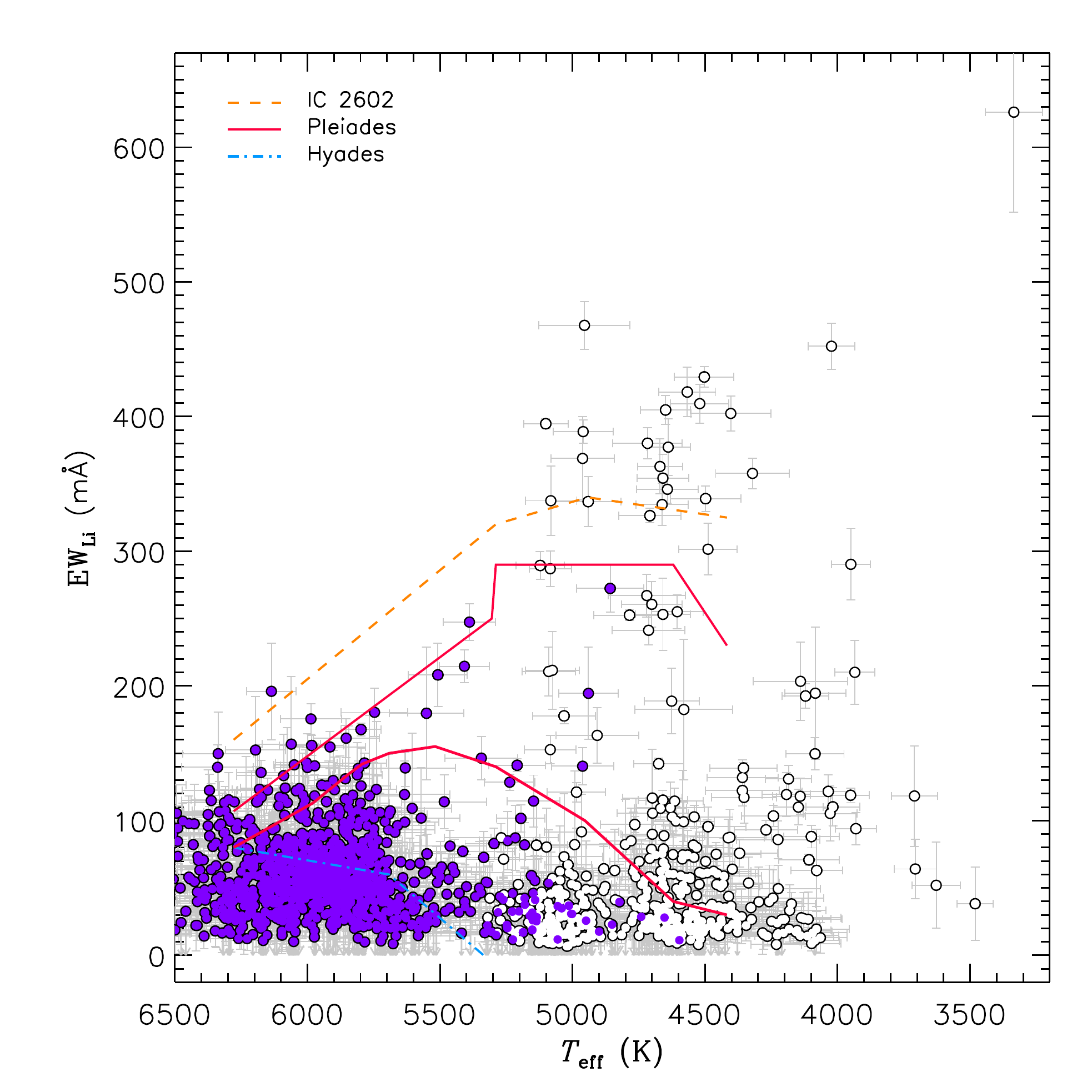} 
\caption{Equivalent width of the \ion{Li}{i}\,$\lambda$6707.8 line (\WLi) plotted as a function of \teff\ (open circles for giants, purple dots for MS stars). 
The lines show the upper boundaries for Hyades (dash-dotted blue), Pleiades (solid red for both the lower and upper boundary), and IC~2602 (dashed orange) clusters. 
%The black dashed line indicates the lower limit for positive detection of the  \ion{Li}{i}\,$\lambda$6707.8 line estimated by us (10\,m\AA). 
Most of the stars with a strong lithium absorption are lithium-rich giants.
}
\label{Fig:Li}
\end{figure}

Lithium is burned in stellar interiors at temperatures of about $2.5\,\times\,10^6$\,K. Therefore, it is progressively depleted from 
the stellar atmospheres of late-type stars with deep enough convective envelopes.  The degree of depletion depends on the internal structure of the star and therefore, for the MS stars, on the mass. 
Thus, its abundance can be used as an empirical indicator of age for MS stars cooler than about 6500\,K. A simple and effective way to get
an age estimate is a diagram showing the equivalent width of lithium as a function 
of effective temperature (or a color index) together with the upper envelopes of 
young open clusters that can be used as boundaries to separate the star sample in age classes (Fig.\,\ref{Fig:Li}). 

We adopted the upper envelopes of the Hyades \citep{Soderblom1990},
the Pleiades \citep{Soderblom1993c,Neuhauser1997}, and IC~2602 \citep{Montes2001}, whose ages are of about 650 \citep{White2007AJ133},  125 \citep{White2007AJ133}, and 30~Myr \citep{Stauffer1997}, respectively. 
The lower envelope of the Pleiades corresponds to the upper boundary for the members of the Ursa Major (UMa) cluster ($age\approx$\,300 Myr; \citealt{Soderblom1993b}).
The boundaries are remarkably close to each other for temperatures higher than 6000\,K and therefore, also taking into account the errors on \WLi, we preferred to restrict the age classification to objects with \teff\,$\leq$\,6000\,K. In line with what done by \citet{Guillout2009} and \cite{Frasca2018}, we have defined four age classes for the MS stars: (i) ``PMS-like'' encompasses objects that lie above the Pleiades upper boundary; (ii) ``Pleaides-like'' comprises stars that fall between the upper and lower Pleiades envelopes; (iii) ``UMa-like'' stars are those located between the lower Pleiades boundary (corresponding to the UMa upper envelope) and the Hyades envelope; and (iv) ``Hyades-like'' stars are those with \WLi$>20$\,m\AA\ that lie below the Hyades upper envelope.

We adopted a lower threshold of 20\,m\AA\ for the Hyades-like stars, because the typical error of \WLi\ is 20\,m\AA, and lower values of \WLi\ are normally associated with stars older than the Hyades. Therefore, the stars with \WLi\,$<$\,20\,m\AA\ and those for which no \ion{Li}{i} absorption has been detected are not classified.
The stars that belong to the four age classes defined above are labeled as PMS, Ple, UMa, and Hya, respectively, in Table~\ref{Tab:Lithium}.   
The MS star with the highest value of \WLi\ is KIC 8095028, which is close to the upper envelope of the Pleiades stars.
In the end, only three MS stars were classified as PMS-like, but all of them lie very close to the Pleiades upper envelope and therefore their PMS nature is questionable; they should be broadly considered as very young stars. We find 19 Pleiades-like (1.6\%), 247 UMa-like (21\%), and 343 Hyades-like stars (30\%) in our sample of MS stars (\logg$>$3.7) with a detection of the lithium line (1184 stars).

\begin{figure}[htb]
\includegraphics[width=9.0cm]{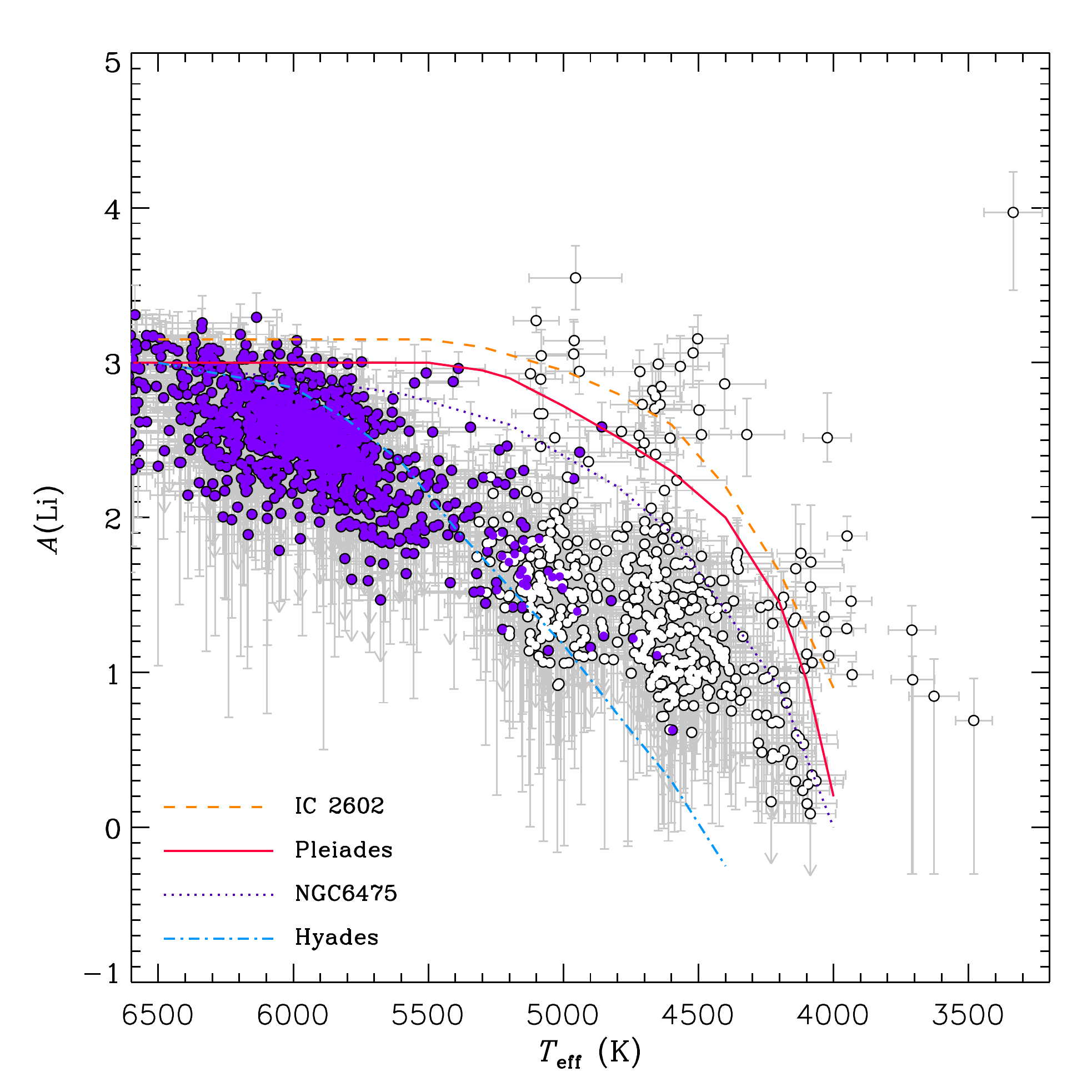}           
\caption{Lithium abundance as a function of \teff. 
The upper envelopes of $A$(Li) for IC~2602, Pleiades,  NGC\,6475 ($age\approx$\,300 Myr), and Hyades clusters adapted from \citet{Sestito2005} are overplotted.
The meaning of the symbols is the same as in Fig.\,\ref{Fig:Li}. 
}
\label{Fig:NLi}
\end{figure}

We calculated the lithium abundance, $A$(Li), from our values of \teff, \logg, and \WLi\ by interpolating the curves of growth of \citet{Lind2009}, which span the \teff\ range 4000--8000\,K and log from 1.0 to 5.0 and include nonlocal thermal equilibrium corrections. We note that for the few stars with \teff$<4000$\,K the lithium abundance is extrapolated and could be less accurate. 
The errors of $A$(Li) take into account both the \teff\ and $W_{\rm Li}$ errors\footnote{We did not consider the errors of \logg\ for the evaluation of $A$(Li) errors because an error $\sigma_{\log g}$\,=\,0.5 dex translates into an uncertainty of only a 0.1 or 0.2 dex, at most, in $A$(Li), in the \teff\ range of our targets. }.
In Fig.\,\ref{Fig:NLi} we show the lithium abundance as a function of \teff\ along with the upper envelopes of the distributions of some young open clusters shown by \citet{Sestito2005}.
We note that the age of NGC\,6475 of about 300\,Myr is close to the UMa cluster, and its upper envelope corresponds to the lower envelope of the Pleiades in Fig.\,\ref{Fig:Li}.
This plot shows that the age classification made on the basis of the $W_{\rm Li}$ diagram and upper envelopes of open clusters is confirmed by the $A(Li)$ diagram.

The object with the highest lithium abundance, $A$(Li)\,=\,3.97$\pm$0.26, is KIC\,11657857 (= IRAS\,19170+4937), which is also the coolest lithium-rich giant. 
The SED (see Sect.\,\ref{Sec:Lithium-rich_giants}) of this source (Fig.~\ref{Fig:SED_KIC11657857}) displays infrared excess starting from about 10\,$\mu$m, which is clearly revealed by IRAS \citep{IRAS2015}, AKARI \citep{AKARI}, and WISE \citep{WISE} data.
Another object with a very large value of lithium abundance ($A$(Li)=2.98$\pm$0.15) is KIC\,8363443, which is a lithium-rich giant already known from the literature. No infrared excess is visible in the SED of this source (Fig.~\ref{Fig:SED_KIC8363443}).

%%%%%%%%%%%%%%%%%%%%%%%%%%%%%%%%%%%%%%%%%%%%%%%%%%%%%%%%%%%%%%%%%%%%%%%%%%%%%%%%%%%%%%
% Li-rich giants
%%%%%%%%%%%%%%%%%%%%%%%%%%%%%%%%%%%%%%%%%%%%%%%%%%%%%%%%%%%%%%%%%%%%%%%%%%%%%%%%%%%%%%

\subsection{Lithium-rich giants}
\label{Sec:Lithium-rich_giants}

As mentioned in Sect.~\ref{Sec:Lithium}, lithium is a fragile element, which is progressively destroyed by nuclear reactions at temperatures much lower than those of H-burning cores.
As such, it is very sensitive to stellar evolution. Canonical models \citep{ib67_2,ib67,Soderblom1993c} predict Li depletion as a direct consequence of the first dredge-up (FDU) once 
stars reach the red giant branch (RGB). According to this, we do not expect to find abundances of Li above 1.5 dex in red giants \citep{Charbonnel2000}. Observations confirm this 
theoretical scenario \citep{Bon59,Gra04}. However, a few giants with Li abundances larger than this value, the so-called Li-rich giants, have been discovered since several decades 
\citep[e.g.,][]{Wallerstein82,Luck82}. A large number of such stars have recently been  found in the field \citep{Casey2016,Smiljanic2018,Gao19, Martell21}, globular clusters \citep{Kirby16}
as well as open clusters \citep{Monaco14, 6067, 2345, Magrini21} representing a fraction around 1$\%$ of all known giants.
These stars have from early-F to A-type progenitors on the MS ($M_{\ast}=1.5-2.5\,M_{\sun}$).

\begin{figure}[h!]
  \centering         
  \hspace{-0.2cm} 
  \includegraphics[width=9.2cm]{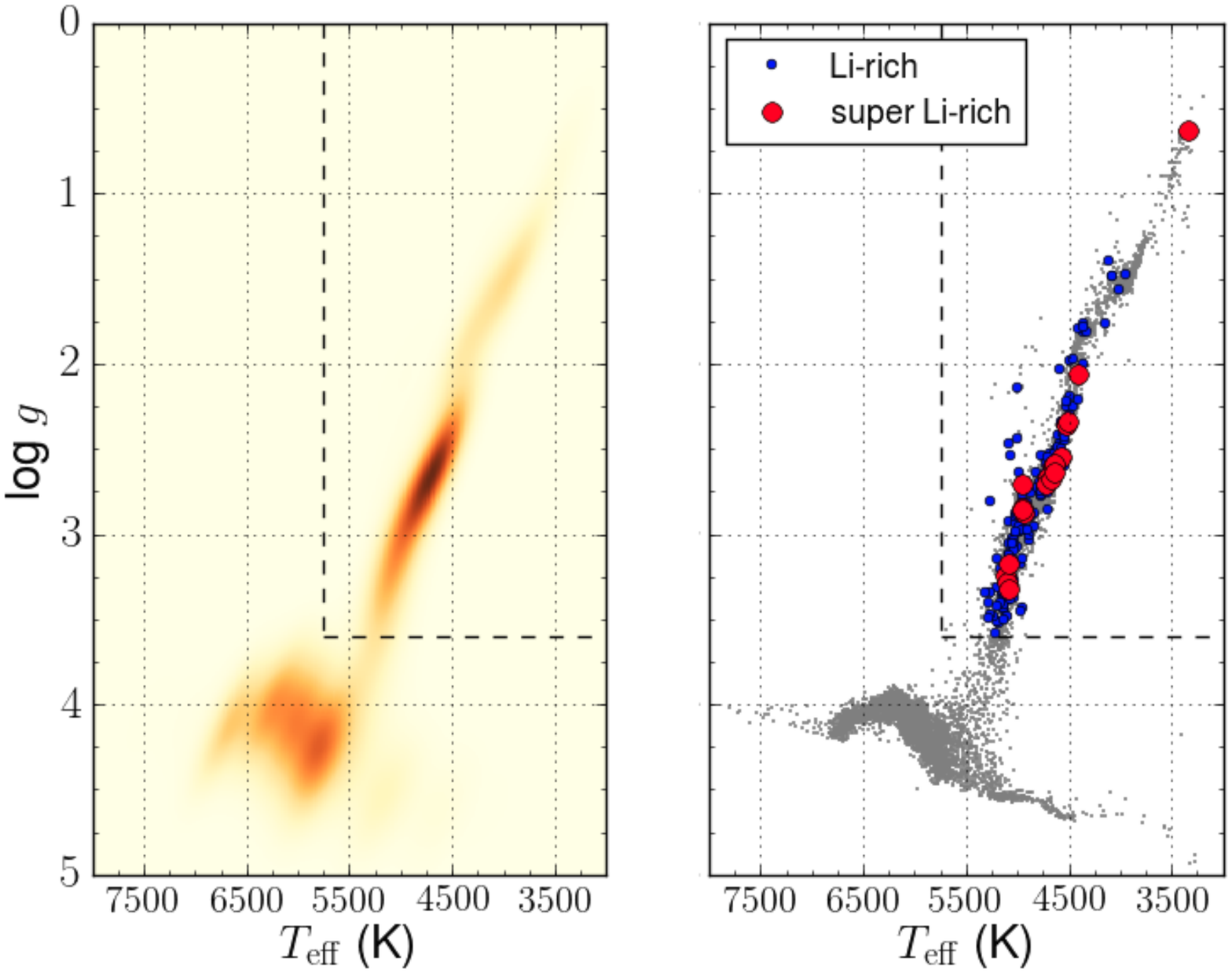}  
  \caption{Kiel (\teff--\logg) diagram for the final sample of the stars observed in this work. The left panel shows the stellar density and the right one highlights the  position of the Li-rich 
  (1.5\,$\leq$\,$A$(Li)\,$<$\,2.7) and super Li-rich giants ($A$(Li)\,$\geq$\,2.7). The dashed lines delimit the area populated by giants according to our selection criteria, as explained in the text.}
  \label{fig_pHR_li} 
\end{figure}

As noted above, we find 1657 stars with a detectable amount of lithium in our sample of 7443 stars, representing about 21\% of the total.
Excluding 1184 stars with \logg$>$3.7, which we classified as MS stars, we are left with 473 evolved stars with a detectable \ion{Li}{i}\,$\lambda$6708 line. Among them there are the Li-rich giants
($A$(Li)$\geq$1.5), including those that exhibit an abundance above the primordial value (i.e., $A$(Li)$\geq$2.7). We refer to this subset as super Li-rich stars. In order to determine 
their fraction, we first need to know the total number of giants in our sample. To identify them, we plotted the Kiel (\teff--\logg) diagram and examined the distribution of the stars in it.
In the left panel of Fig.~\ref{fig_pHR_li} giants clearly stand out from MS stars. As seen, their maximum density is centered at \teff\,$\approx$\,4800\,K and \logg$\,\approx$\,2.6. 
Moving down in the diagram toward the point of maximum density of MS stars, we set the giants boundary at a place approximately equidistant from both points. In this way, stars with 
\teff\,$\leq$\,5750\,K and \logg\,$\leq$\,3.6 will be considered as giants. By taking this criterion into account, we found 3276 giants in our sample, out of which 195 have $A$(Li)$\geq$1.5,
which implies a fraction of Li-rich giants of 6.0\%. This number is practically not affected by minor changes to the selection criterion. Surprisingly, this fraction is much higher than $\approx$1.3\%, 
which is the result obtained by the most recent surveys \citep{Gao19,Martell21} after analyzing stellar samples much larger than ours.
Among the 195 Li-rich giants detected, 34 were already observed by \citet{Gao19} while the remaining 161 are reported here for the first time.

\begin{figure}[htb]
\hspace{-.3cm}
\includegraphics[width=9.2cm]{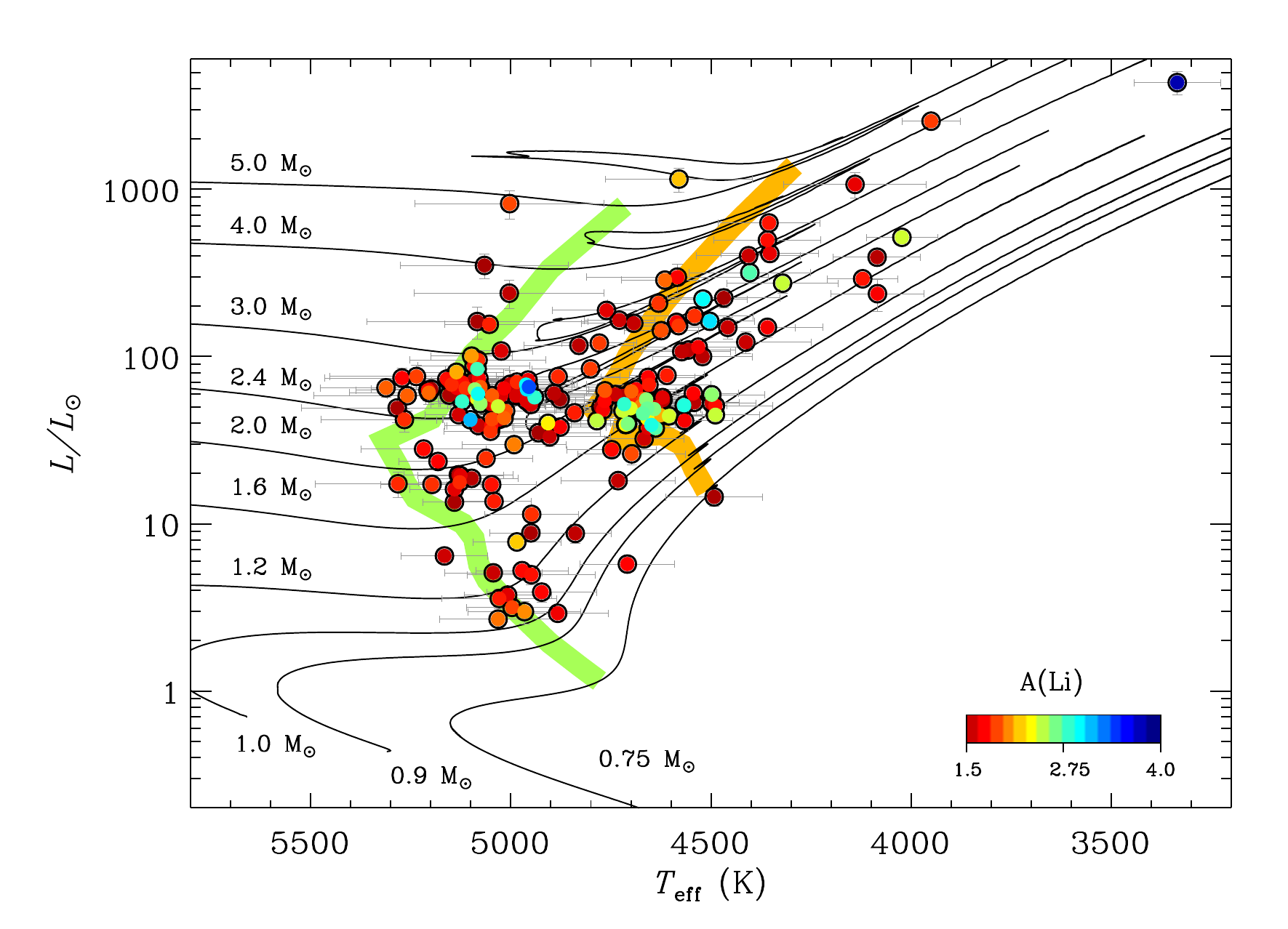}           
\caption{Hertzsprung-Russell diagram for all the Li-rich giants. The symbols are color coded by the lithium abundance $A$(Li). The PARSEC post-MS evolutionary tracks  
\citep{Bressan2012} for a metallicity Z\,=\,0.017 are plotted with black lines. The green and orange strips mark the base of the RGB and the RGB bump, respectively. 
}
\label{Fig:HR_Li-rich}
\end{figure}

In order to understand the evolutionary status of these Li-rich giants, we placed them on the Hertzsprung-Russell (HR) diagram. For this task, we first calculated their luminosity from the analysis of
the SED, as explained in Appendix~\ref{Appendix:SED}. Then, we added the PARSEC evolutionary tracks \citep{Bressan2012} for a solar metallicity (Z\,=\,0.017) and finally, we marked the base of the RGB and the 
RGB bump on them. The resulting HR diagram of these sources is shown in Fig.~\ref{Fig:HR_Li-rich}. As inferred from their position on the diagram, the vast majority of our sources are low-mass stars with
masses in the range 1$-$3\,M$_{\sun}$.

According to their position in the HR diagram and the loci indicated in Fig.~\ref{Fig:HR_Li-rich}, three evolutionary stages can be distinguished. The first one corresponds to the stars located to the left of the green strip. These objects could have not yet reached the RGB\footnote{Although these stars were selected as giants, they could actually be at the end of the subgiant phase according to the HR diagram in Fig.~\ref{Fig:HR_Li-rich}. Regardless of this group, the fraction of Li-rich giants is still very high, 5.5\%. Even when considering the stars over the strip or those that have already passed it but whose positions, within the errors, might be indicating that they are still over or slightly before the strip ($\approx$\,45 stars), the fraction of Li-rich giants would be around 4.2\%, still much higher than the 1.3\% found by other authors.} and their initial lithium is still observable in their atmospheres. Those stars in the diagram placed between both strips are  classified as RGBs. Approximately in the middle of this phase, the FDU takes place and drastically depletes the amount of the photospheric Li. As commented before, $A$(Li)$>$1.5 should not be observed in these stars. Some of them, the closest to the green strip, could not yet have undergone the FDU, and therefore, their lithium abundance is explained as in the previous case. Nevertheless, with regard to the remaining stars, the existence of these Li-rich giants implies the additional contribution of a mechanism capable of enhancing  the atmospheric lithium content. This is also valid for those stars found after the RGB bump, to the right of the orange strip.

However, a single universal mechanism that can explain this phenomenon as a whole does not seem to exist; there is more likely a mixture of various processes at work.
External pollution is one of the scenarios most often claimed to explain it. The extra contribution of Li could be due to both the engulfment of a substellar companion \citep{Si99,Ag16} and the enrichment of the local interstellar medium caused by the explosion of a nearby supernova \citep{Woo95}.
On the other hand, the internal production via the Cameron-Fowler mechanism \citep[CFM;][]{CFM} would allow these stars to create new Li. However, an extra-mixing process after the FDU is absolutely required in this case.
Binarity has also been invoked to solve the Li puzzle, either as an external or internal mechanism. In the first case, the Li-rich giant would be the consequence of the merge of a white dwarf with a red giant in a binary system \citep{Zhang20} while in the second one, the tidal spin-up from a binary companion would provide the internal extra-mixing 
necessary to start the CFM \citep{Casey2019}.

\begin{figure} 
\centering         
\includegraphics[width=9cm]{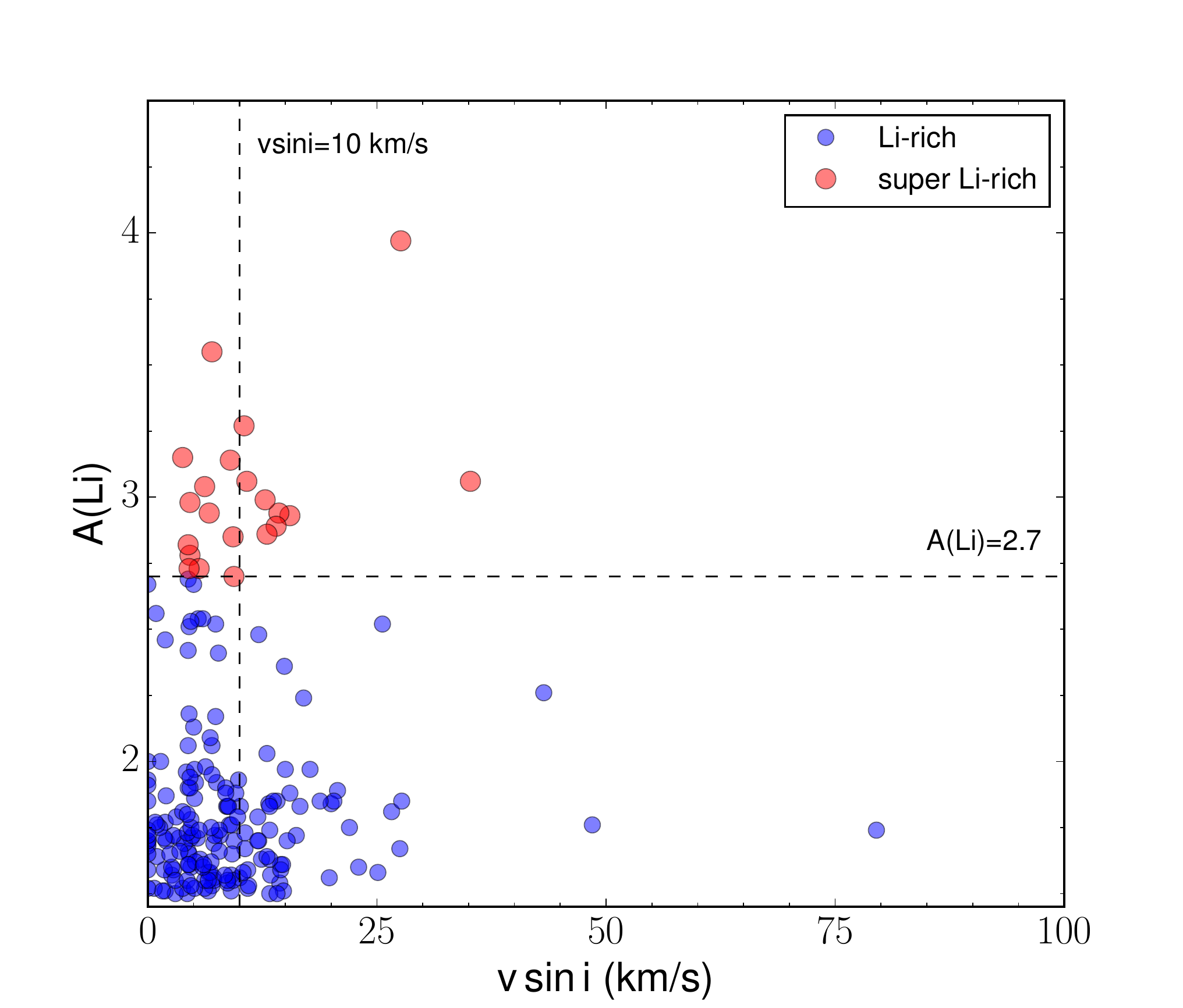} 
\caption{Abundance of Li as a function of the \vsini\ among our Li-rich giant sample. The vertical line shows the limit between slow and fast rotators, while the horizontal one 
represents the boundary between Li-rich and super Li-rich giants.}
  \label{fig_li_vsini} 
\end{figure}

We tested some of these hypotheses by taking advantage of our observations. Firstly, binarity is unlikely to explain the Li enhancement, as no binaries have been found among our sample of Li-rich giants. 
The scenario of engulfment of a substellar companion or a white dwarf does not match what is observed either. According to it, we would expect enhanced rotational velocities among the Li-rich stars, as a natural
consequence of the transfer of angular momentum \citep{Ag16,Privitera16}. However, in our sample, no significant differences are found with regards to the fraction of fast rotators (as well as the metallicity distribution) between Li-rich and Li-normal giants (Table~\ref{tab_normal_rich}). Additionally, fast rotators\footnote{In line with other works \citep[e.g.,][]{Carlberg12,Martell21}, we considered 10\,km\,s$^{-1}$ an adequate value to separate fast from slow rotators since the typical $v$\,sin$i$ among giants is around 3--5\,km\,s$^{-1}$ and the macroturbulence velocity $\approx$7\,km\,s$^{-1}$ \citep{Carney08}.} ($v$\,sin$i$\,$\geq$10\,km\,s$^{-1}$) show an average
$A$(Li)=2.0$\pm$0.5, a value fully consistent with that of stars with $v$\,sin$i<$10\,\kms, $A$(Li)=1.9$\pm$0.4, and a correlation between lithium abundance and stellar rotation is not observed
(see Fig.~\ref{fig_li_vsini}), as also found in previous works \citep[e.g.,][]{Martell21}. However, the occurrence of fast rotators seems to be
higher among the super Li-rich giants (42.9\%) than in the Li-rich sample (29.9\%), although the sample is not so large as to have good statistical significance.

\begin{table}[ht]
\caption{Percentages of fast rotators (\vsini$\geq$10\,\kms) and stars with a solar metallicity (i.e., $-$0.2$\leq$[Fe/H]$\leq$+0.2) among the Li-normal and Li-rich giants. N is the number of stars in each sample.}
    
\begin{center}
\begin{tabular}{lccc}
\hline\hline
Sample &  N  & Fast rotators & Solar stars\\
\hline
Li-normal &  3081  &  31.1  &  91.4  \\
Li-rich   &   195  &  31.3  &  90.3  \\
\hline
\end{tabular}
\end{center}

\label{tab_normal_rich}
\end{table}

Despite the fact that Li-rich giants are observed throughout the entire RGB phase, recent studies \citep{Smiljanic2018,Deepak19,Singh19,Casey2019,Kumar2020,Martell21}
show that their frequency is significantly higher at the red clump. A higher occurrence in a particular evolutionary phase (i.e., during the He-core burning) would 
therefore imply an internal origin \citep[likely related to the He flash,][]{Kumar2020} as the main responsible for the Li observed.
The accurate classification of the Li-rich giants in the different evolutionary stages is beyond the scope of this work. 
Only for a merely indicative purpose, we assume
that the region in the Kiel diagram (right panel in Fig.~\ref{fig_pHR_li}) with the highest density of stars (the sequence with 2.3$\leq$\logg$\leq$3.0) corresponds to the red clump.
Then, in line with other works \citep[e.g.,][]{Martell21} super Li-rich giants are more concentrated in this part of the diagram than Li-rich giants (71.4\% versus 51.1\%, respectively).  

\begin{figure} 
\centering         
\includegraphics[width=\columnwidth]{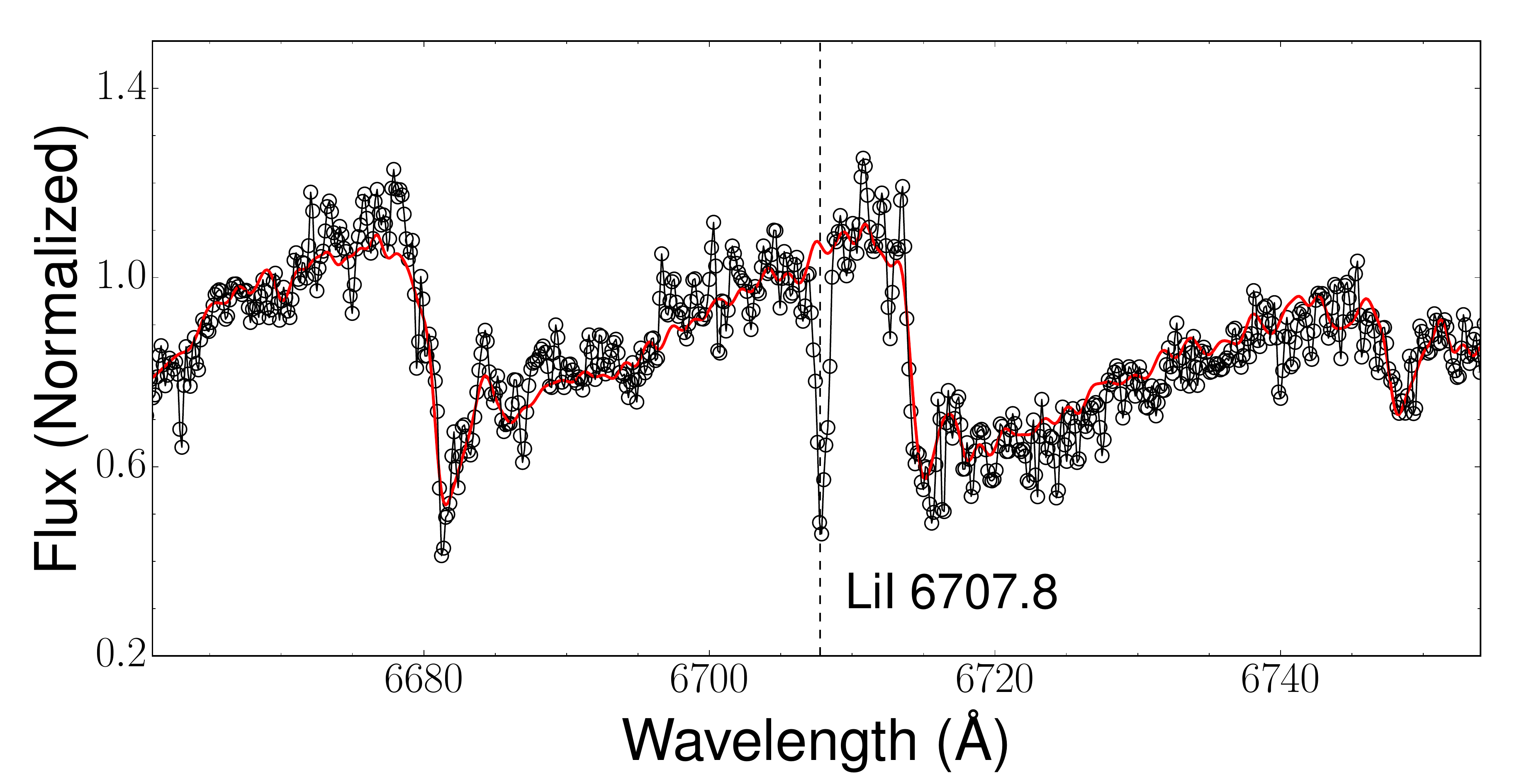} 
\caption{Spectrum of KIC\,11657857 (black line and dots), the star with the highest abundance of Li in our sample, versus that of a Li-normal giant with similar stellar parameters (red line) around the Li line at 6708\,\AA, whose center is marked with a vertical dashed line.}
  \label{fig_very_Li} 
\end{figure}

Finally, as commented above, KIC\,11657857 is the object that exhibits the highest Li abundance, $A$(Li)=3.97, in our sample (see Fig.\ref{fig_very_Li}). It is a super Li-rich M-type star \citep[see, e.g.,][for another example]{Alcala11}, which also shows an infrared excess. According to its location in the HR diagram (the brightest and the coolest), KIC\,11657857 is most likely to be an asymptotic giant branch (AGB) star, the only one in our Li-rich sample. High-resolution spectroscopic observations would be necessary to properly characterize this interesting object.

%%%%%%%%%%%%%%%%%%%%%%%%%%%%%%%%%%%%%%%%%%%%%%%%%%%%%%%%%%%%%%%%%%%%%%%%%%%%%%%%%%%%%%%%%%%%%
% SB2-3
%%%%%%%%%%%%%%%%%%%%%%%%%%%%%%%%%%%%%%%%%%%%%%%%%%%%%%%%%%%%%%%%%%%%%%%%%%%%%%%%%%%%%%%%%%%%%

\subsection{Spectroscopic binaries}
\label{Sec:SB2}

The resolution of the \lamost\ MRS is sufficient to detect binary and multiple systems and to study their RV curves for deriving the orbital elements and the physical parameters of the components of these systems \citep[see, e.g.,][]{Pan2020,Wang2021}. 

As anticipated in Sect.\,\ref{Sec:Analysis}, in some spectra (more than 250) we noted two or three peaks in the CCF that resulted to be significant in comparison with the CCF noise. These objects are SB2 or SB3 systems, respectively. In these cases, the analysis with \rotfit\ is meaningless, so we do not report any parameter in Tables\,\ref{Tab:data_red} and \ref{Tab:data_blue} and flag them as SB2 or SB3. In some cases the binary nature is questionable, because the CCF displays only an asymmetric peak (i.e., no clear valley between the peaks is visible) that can be due to both an unresolved SB2 system or starspots. For these spectra we kept the APs in Tables~\ref{Tab:data_blue} and \ref{Tab:data_red}, whenever we considered them as reliable, and flagged them as ``SB2?.'' 
In total, we found 27 spectra corresponding to 7 different sources, which display three peaks in the CCF (SB3) and more than 230 spectra corresponding to 98 sources with two peaks in the CCF (SB2). The SB3 and SB2 binaries in our sample of MRS are listed in Table~\ref{Tab:Binaries}. 

\begin{figure}[th]
\begin{center}
\includegraphics[width=8.5cm]{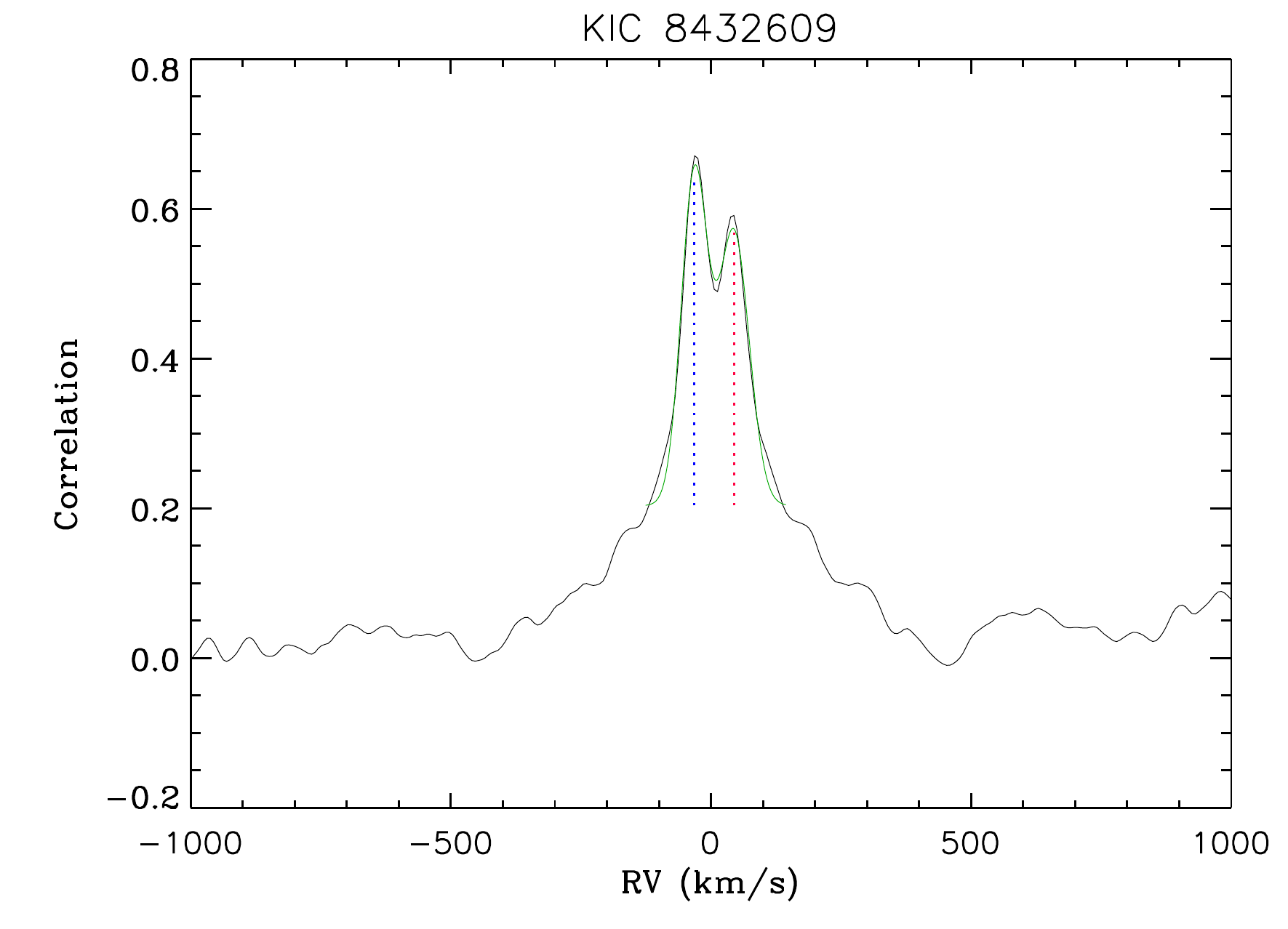}\\        
\includegraphics[width=8.5cm]{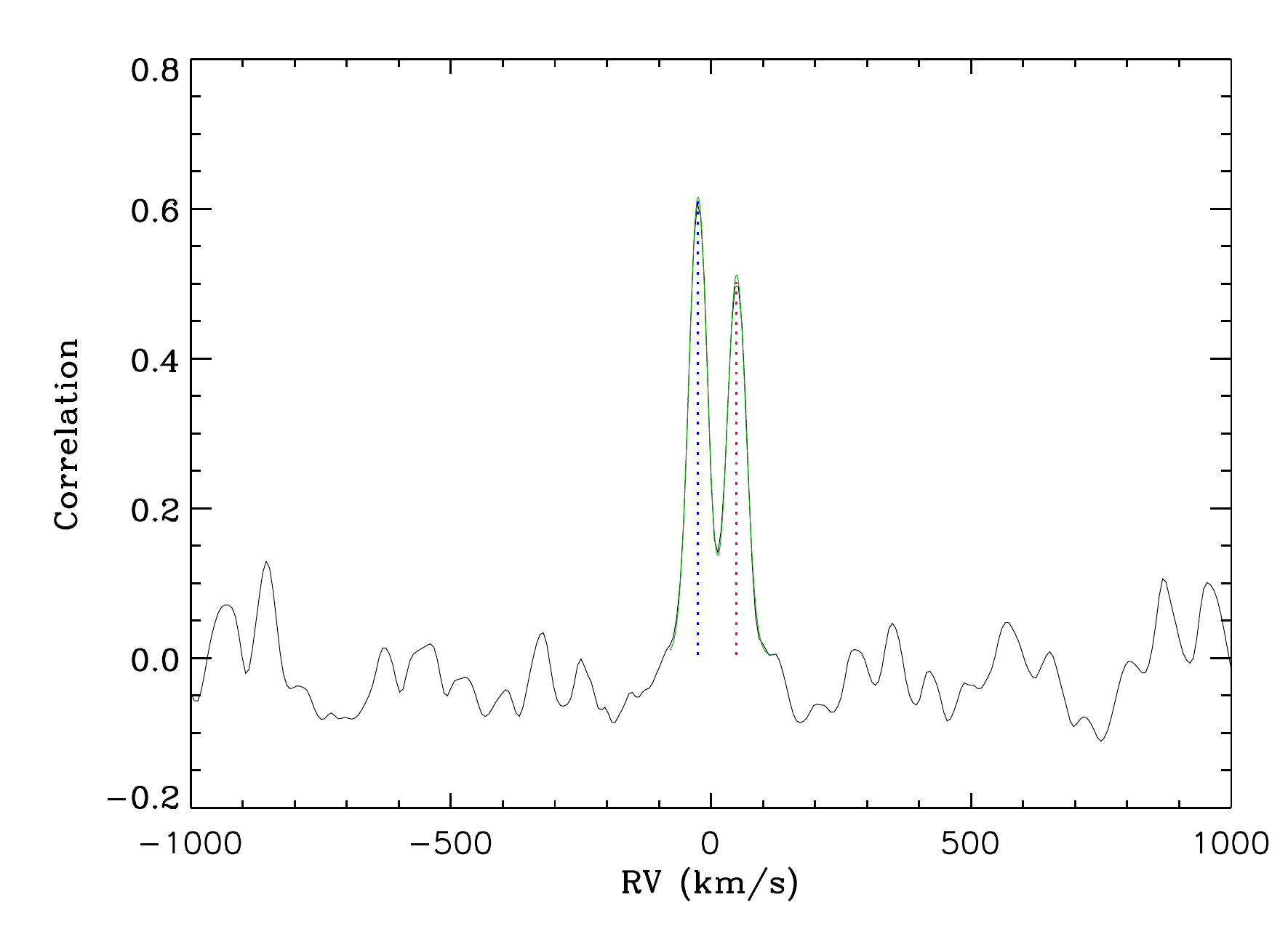}           
\caption{Cross-correlation function (black line) for a double-lined spectroscopic binary (SB2) in the blue arm ({\it upper panel}) and in the red arm ({\it lower panel}). The two-Gaussian fit of CCF is displayed with a green line, and the  centers of the primary (more luminous) and secondary components are indicated with vertical dotted blue and red lines, respectively.
}
%The cyan regions at the two sides of the CCF peak show the region used for evaluating the CCF noise that is used for estimating the RV errors.}
\label{Fig:CCF_SB2}
\end{center}
\end{figure}

For the spectra with multiple peaks in the CCF that are sufficiently resolved, we derived the RV of the individual components by fitting the CCF with multiple Gaussians with an approach similar to that used for single-lined objects. An example of the fitting procedure for an SB2 system is shown in Fig.\,\ref{Fig:CCF_SB2}.
The RV values for the individual components of the SB3 and SB2 systems listed in Table~\ref{Tab:Binaries} are not reported in this paper and will be published in forthcoming works dedicated to the double and multiple systems in our sample.

Whenever an  SB2 system is observed near the conjunctions, only one peak is visible in the CCF and the RV, measured with a single Gaussian fit, has to be considered as a ``blended'' value of the
RVs of the two components, which is close to the velocity of the barycenter of the system.

In the following, we show the case of a few systems for which we could build an RV curve, including also data from a few spectra acquired in 2019 (DR7).
We used the weighted average of the corrected RVs measured in the red and the blue arm, with a weight inversely proportional to the squares of their errors. 
We used the stacked spectra to maximize the S/N for binary systems with a period longer than a few days, while the individual spectra acquired during each night have been used for binaries with a short or unknown period.  
We mention here the cases of the eclipsing binaries KIC\,8301013 and KIC\,5359678, whose RV curves based on \lamost\ MRS data have been studied by \citet{Pan2020} and \citet{Wang2021}, respectively. They also presented an in-depth analysis of their \kepler\ light curves. 

\begin{figure}[th]
\begin{center}
\includegraphics[width=9cm]{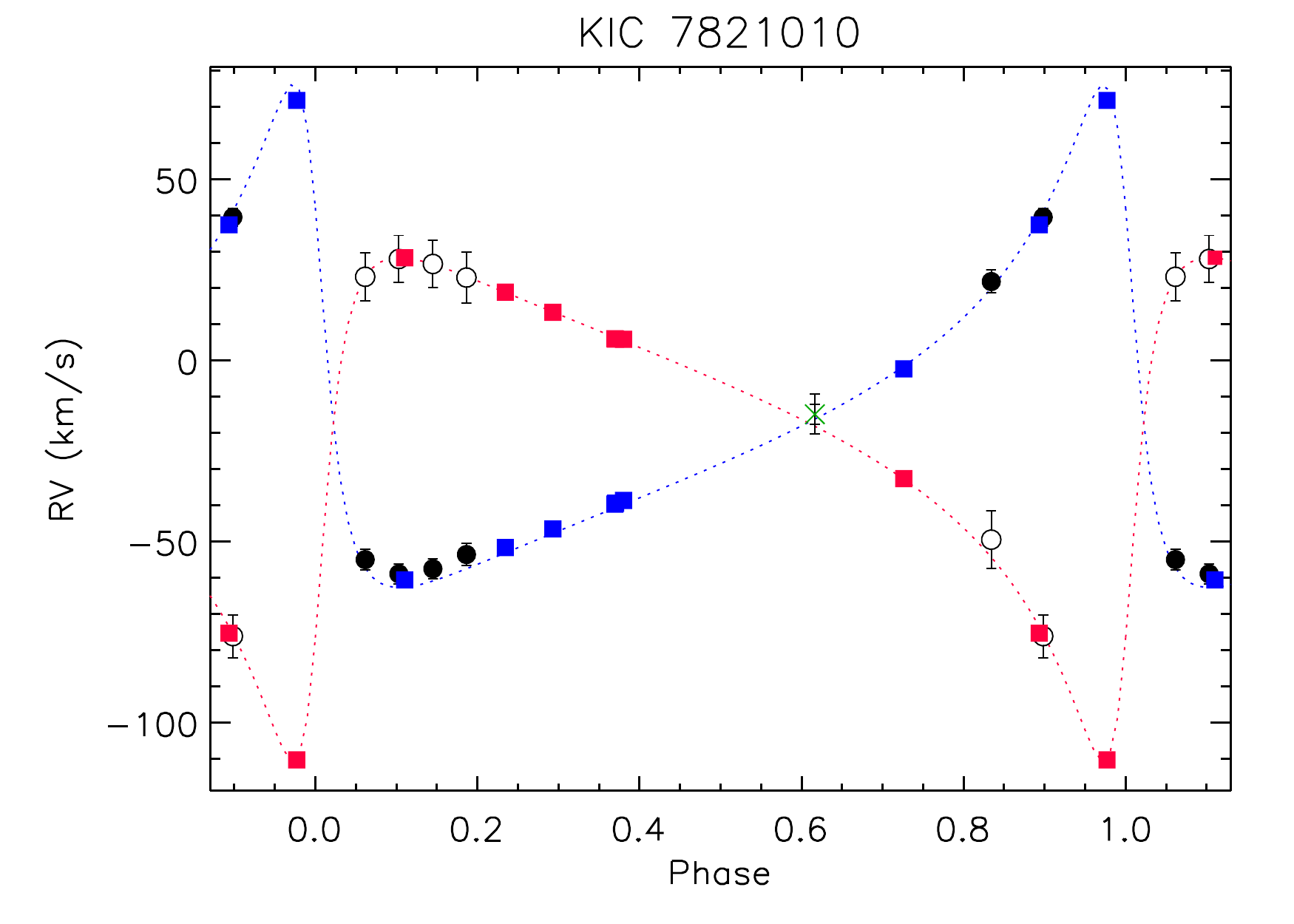}           
\caption{Radial velocity curve of KIC~7821010. The RV data from \citet{Helminiak2019} are shown with blue and red squares and the orbital solution with dotted blue and red lines for the primary and secondary component, respectively.
The \lamost\ MRS RVs derived in this work are overplotted with filled black circles for the primary and open circles for the secondary component, respectively. The green x symbols display the RV values measured by us near the conjunctions, where a single peak is visible in the CCF.
}
\label{Fig:RV_KIC7821010}
\end{center}
\end{figure}

Another eclipsing binary with a period $P_{\rm orb}\simeq$\,24.24 days and a very eccentric orbit ($e\simeq$0.680) is KIC\,7821010. An orbital solution, based on the \kepler\ light curve and RVs derived from high-resolution HIDES spectra, has been presented by \citet{Helminiak2019}.
Our RVs of KIC~7821010 are displayed in Fig.~\ref{Fig:RV_KIC7821010}. We note the very good agreement with the RVs and the orbital solution of \citet{Helminiak2019}, which are also displayed in Fig.~\ref{Fig:RV_KIC7821010}. 

\begin{figure}[th]
\begin{center}
\includegraphics[width=9cm]{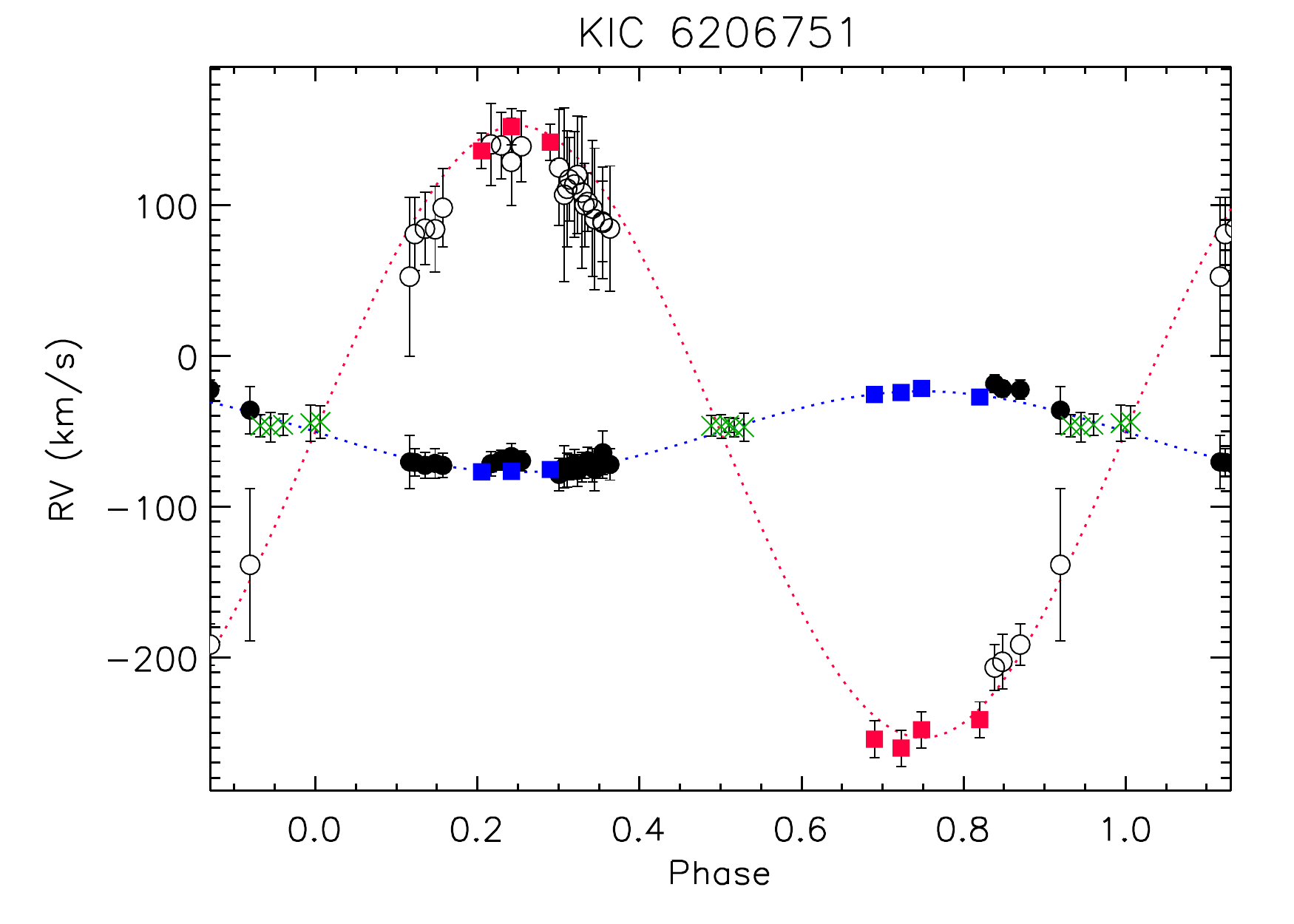}           
\caption{Radial velocity curve of KIC~6206751. The RV data from \citet{Matson2017} are displayed with blue and red squares and the orbital solution with dotted blue and red lines for the primary and secondary component, respectively.
The \lamost\ MRS RVs derived in this work are overplotted with filled black circles for the primary and open black circles for the secondary component.  The green x symbols display the RV values measured by us near the conjunctions, where a single peak is visible in the CCF.
}
\label{Fig:RV_KIC6206751}
\end{center}
\end{figure}

We also measured the RVs of the eclipsing binary system KIC~6206751, which has a much smaller orbital period, $P_{\rm orb}=1.2453$ days, and a circular orbit \citep{Lee2018}. We note that the RVs of the primary component are in very good agreement with those from \citet{Matson2017}, while for the secondary component, which is much less luminous than the primary, the agreement is less good (Fig.~\ref{Fig:RV_KIC6206751}), but our values are still compatible within the errors.  

\begin{figure}[th]
\begin{center}
\includegraphics[width=9cm]{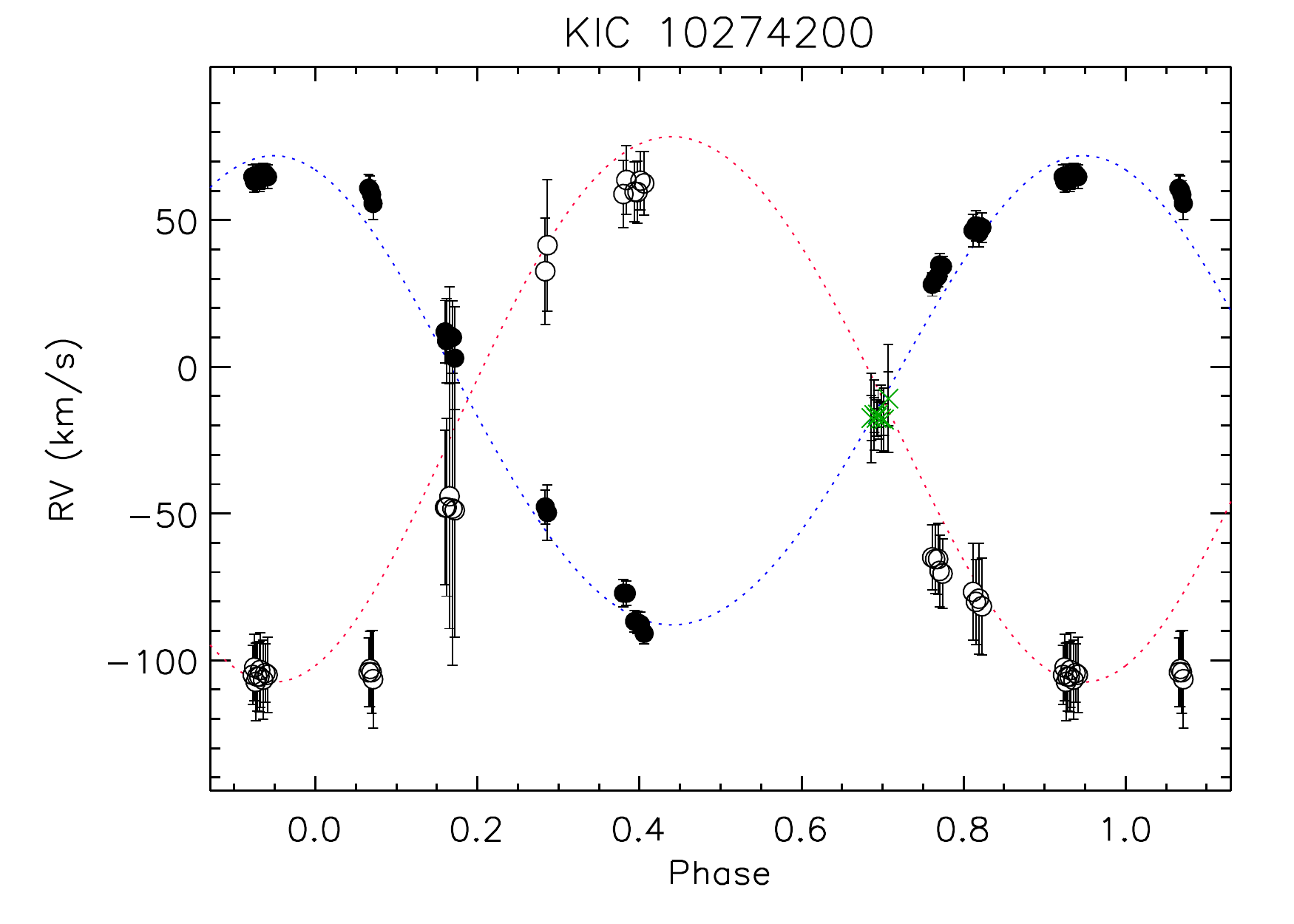}           
\caption{Radial velocity curve of KIC~10274200. 
The \lamost\ MRS RVs derived in this work are plotted with filled circles for the primary and open circles for the secondary component.  The green x symbols display the RV values measured near the conjunctions, where a single peak is visible in the CCF. The orbital solution is displayed with a dotted blue and red line for the primary and secondary component, respectively.
}
\label{Fig:RV_KIC10274200}
\end{center}
\end{figure}

As a further example, we show the case of KIC~10274200, a non-eclipsing SB2, for which we determined the period and the orbital parameters  from our data only  (see Fig.~\ref{Fig:RV_KIC10274200}).
We searched for the best orbital period by applying the periodogram analysis \citep{Scargle1982} to the RVs of the primary and secondary components.  
Then, we fitted the observed RV curve with the {\sc curvefit} routine \citep{Bevington} to determine the orbital parameters and their standard errors, which are listed in Table~\ref{Tab:Param}. We remark that these are very preliminary estimates and new data are needed to refine the orbital solution. As mentioned above, a more in-depth study of the RVs of binary and multiple systems will be the subject of future works. 

\begin{table}   %[ht]
\caption{Orbital parameters of KIC~10274200.}
\begin{tabular}{lr}
\hline
\hline
\noalign{\smallskip}
Parameter &  Value   \\ 
\hline
\noalign{\smallskip}
 HJD0$^a$      &  58020.45$\pm$0.05  \\
 $P_{\rm orb}$ ($d$) & 4.278$\pm$0.001    \\
 $e$            &  0.04$\pm$0.04  \\
 $\omega$ (\degree) &  20.0$\pm$0.5  \\
 $\gamma$ (\kms)    &  $-11\pm$3 \\
 $K_1$ (\kms)       &  80$\pm$1  \\
 $K_2$ (\kms)       &  93$\pm$3  \\
 $M_1\sin^3i$ ($M_{\sun}$) & 1.23$\pm$0.08 \\
 $M_2\sin^3i$ ($M_{\sun}$) & 1.06$\pm$0.05 \\
 $M_2$/$M_1$  &  0.86$\pm$0.03  \\
 $a\sin i$ ($R_{\sun}$) & 14.6$\pm$0.2  \\
\noalign{\smallskip}
\hline \\
\end{tabular}
\label{Tab:Param}
\begin{list}{}{}
\item[$^{\rm a}$] Heliocentric Julian date (HJD-2,400,000) of the periastron passage.
\end{list}
\end{table}

\section{Summary}
\label{Sec:Concl}

We have presented the results of the analysis of about 16,300 \lamost\ MRS of cool (FGKM-type) stars in the field of the \kepler\ space telescope performed with the code \rotfit. 
We were able to determine the APs (\teff, \logg, and \feh), the RV, and the projected rotational velocity (\vsini) for about 14\,300 spectra corresponding to 7443 different stars. The average uncertainties of these parameters are %of 
about 0.7\,\kms, 2.5\,\%, 0.25\,dex, and 0.15 dex for the RV, \teff, \logg, and \feh, respectively. We were able to measure \vsini\ values larger than 8\,\kms, which is the lower threshold for
this measure resulting from the resolution and sampling of the MRS, as found from Monte Carlo simulations. The typical uncertainty for \vsini\ measures is 3--4\,\kms. The comparison of our determinations of \teff\ and \logg\  with literature values derived from spectroscopic analyses shows an overall good agreement with very few outliers. The case is different for \feh, which is clearly overestimated for metal-poor stars by our analysis method. This is likely the result of the nonuniform distribution of the grid of templates in the space of parameters, which is more important in the low-metallicity domain. We propose a calibration relation to correct the metallicity values. 

The resolution $R\sim 7500$ and the spectral coverage of the MRS were adequate to discover and study very interesting objects, such as spectroscopic binaries, chromospherically active stars, young stars, and lithium-rich giants. Based on the emission or filling of the H$\alpha$ line core, we discovered 327 active stars that are young MS stars or more evolved stars that rotate faster than usual due to tidal spin-orbit synchronization or to evolutionary effects. We detected the \ion{Li}{i}$\lambda$6708\,\AA\ line and measured its equivalent width, \WLi, in 1657 stars, both giants and MS stars. 
We also calculated the lithium abundance $A$(Li) from \WLi\ and the APs, thanks to suitable curves of growth.
Regarding the MS stars, we carried out a discrete age classification based on the \WLi\ and $A$(Li) and the upper envelopes of the \WLi--\teff\ diagrams of young open clusters. We found, among the 1184 MS stars where the lithium line was detected, only 3 stars above (but very close to) the Pleiades upper envelope, 19 stars with an age comparable to the Pleiades ($age \approx$ 125 Myr), 247 UMa-like stars whose age should be similar to that of the UMa cluster ($age \approx 300$\,Myr), and 343 Hyades-like stars ($age \approx 600$\,Myr).

We have also found 195 Li-rich giants, out of which 161 are reported here for the first time. The super Li-rich giant KIC\,11657857, with $A$(LI)$\approx$4, stands out among them. It is an IRAS source that shows infrared excess and, according to its position in the HR diagram, is an AGB star. The fraction of Li-rich giants found in this work, around 4\%, is considerably higher than that observed in other works. The lack of enhanced rotational velocities among the Li-rich giants allows us to exclude external mechanisms (mergers) as the most likely explanation for the observed enrichment of Li in our sample. We find that fast rotators are more recurrent among the super Li-rich giants than among the Li-rich giants. Additionally, the former seem to occur more frequently at the red clump with respect to the latter, which appear to be more evenly distributed over the entire RGB phase.

Moreover, we have found 98 double-lined spectroscopic binaries (SB2) and 7 triple systems (SB3) from the shape of the CCF, which displays two and three significant peaks for SB2 and SB3 systems, respectively. For three SB2 systems with MRS taken at different epochs, we have also presented the RV curves.

\begin{acknowledgements}
We are grateful to the anonymous referee for a careful reading of the manuscript and very useful suggestions that helped us to improve our work.
Guoshoujing Telescope (the Large Sky Area Multi-Object Fibre Spectroscopic Telescope \lamost) is a National Major Scientific Project built by the Chinese Academy of Sciences. Funding for the project has been provided by the National Development and Reform Commission. \lamost is operated and managed by the National Astronomical Observatories, Chinese Academy of Sciences.
Support from the Italian {\it Ministero dell'Istruzione, Universit\`a e Ricerca} (MIUR) is also acknowledged. 
JM-\.Z acknowledges the Wroc{\l}aw Centre for Networking and Supercomputing grant no.~224.
JNF and WZ acknowledge the support of the Joint Fund of Astronomy of National Natural Science Foundation of China (NSFC) and Chinese Academy of Sciences through the Grants 11833002, 12090040, 12090042 and 11903005.
This research made use of SIMBAD and VIZIER databases, operated at the CDS, Strasbourg, France.
This publication makes use of data products from the Two Micron All Sky Survey, which is a joint project of the University of Massachusetts and the Infrared Processing and Analysis Center/California Institute of Technology, funded by the National Aeronautics and Space Administration and the National Science Foundation.
This publication makes use of data products from the Wide-field Infrared Survey Explorer, which is a joint project of the University of California, Los Angeles, and the Jet Propulsion Laboratory/California Institute of Technology, funded by the National Aeronautics and Space Administration.
\end{acknowledgements}

\nocite{*}  % to test all bib entrys
\bibliographystyle{aa}
%\bibliography{lamostmidres}
\bibliography{43268_ArXiv.bib}

%\Online

\newpage
\begin{appendix}
\section{Monte Carlo simulations for \vsini}
\label{Appendix:MonteCarlo}

We run a Monte Carlo simulation to evaluate the resolution in \vsini\  that can be reached with the \lamost\ MRS data and the \rotfit\ code. 
To this aim, we used synthetic BT-Settl spectra \citep{Allard2012} with \logg=4.5 and \teff\,=\,4000, 5000, and 6000\,K. We brought these spectra
to the resolution of \lamost\ MRS by the convolution with a Gaussian kernel of the proper width (see Sect.\,\ref{Sec:APs}) and resampled them on 
the \lamost\ MRS points.
Then we rotationally broadened these spectra from 0 to 30\,\kms\  in steps of 1\,\kms\  and made 200 simulations per each \vsini,  adding a random noise corresponding to signal-to-noise ratios S/N=50 and 100. 
We applied the code ROTFIT to each simulated spectrum, analogously to what we did with the target spectra, to determine the \vsini\  both for the red- and the blue-arm spectral ranges.

\begin{figure}[th]
\includegraphics[width=8cm]{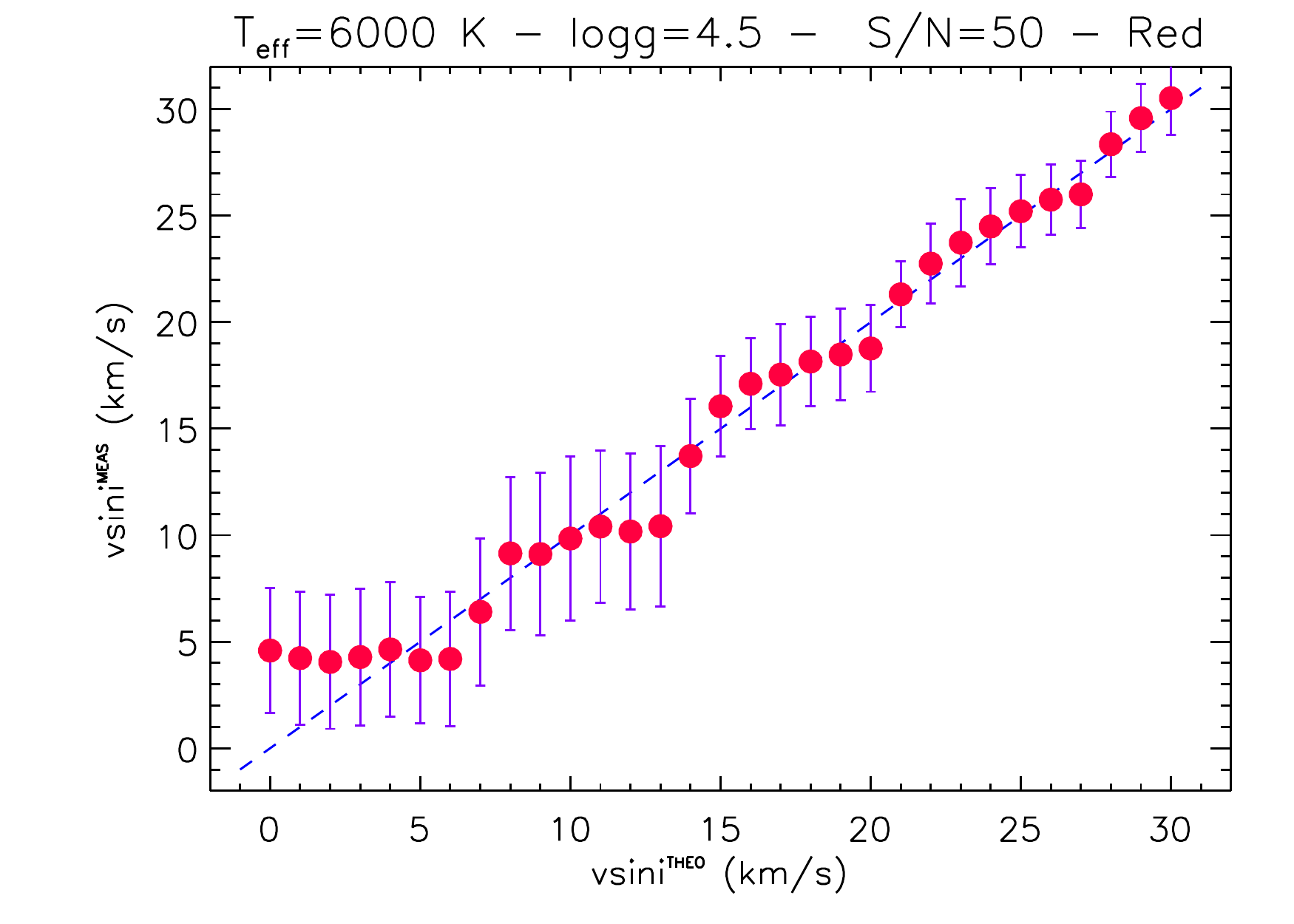}
\vspace{.4cm}
\includegraphics[width=8cm]{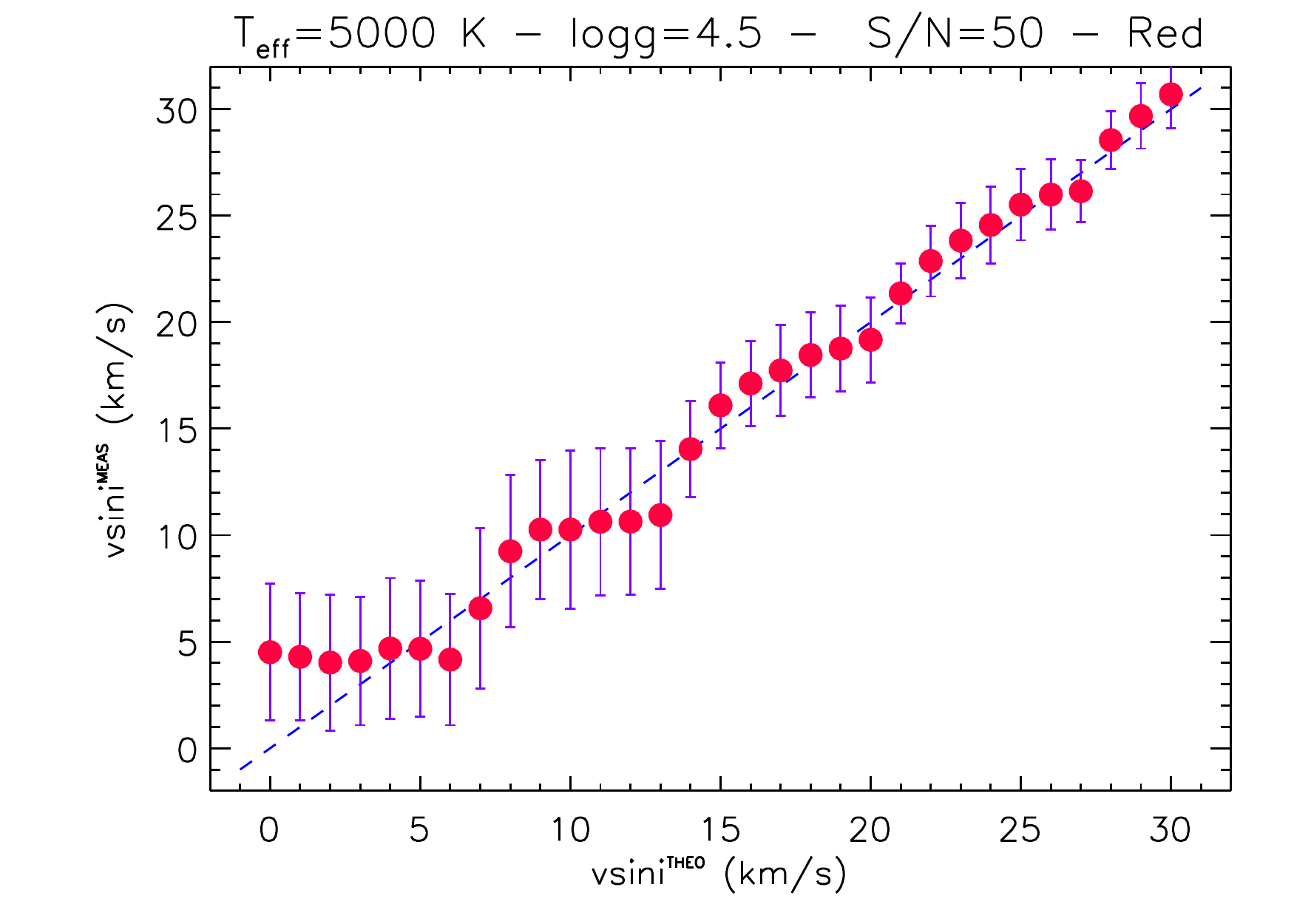}   
\includegraphics[width=8cm]{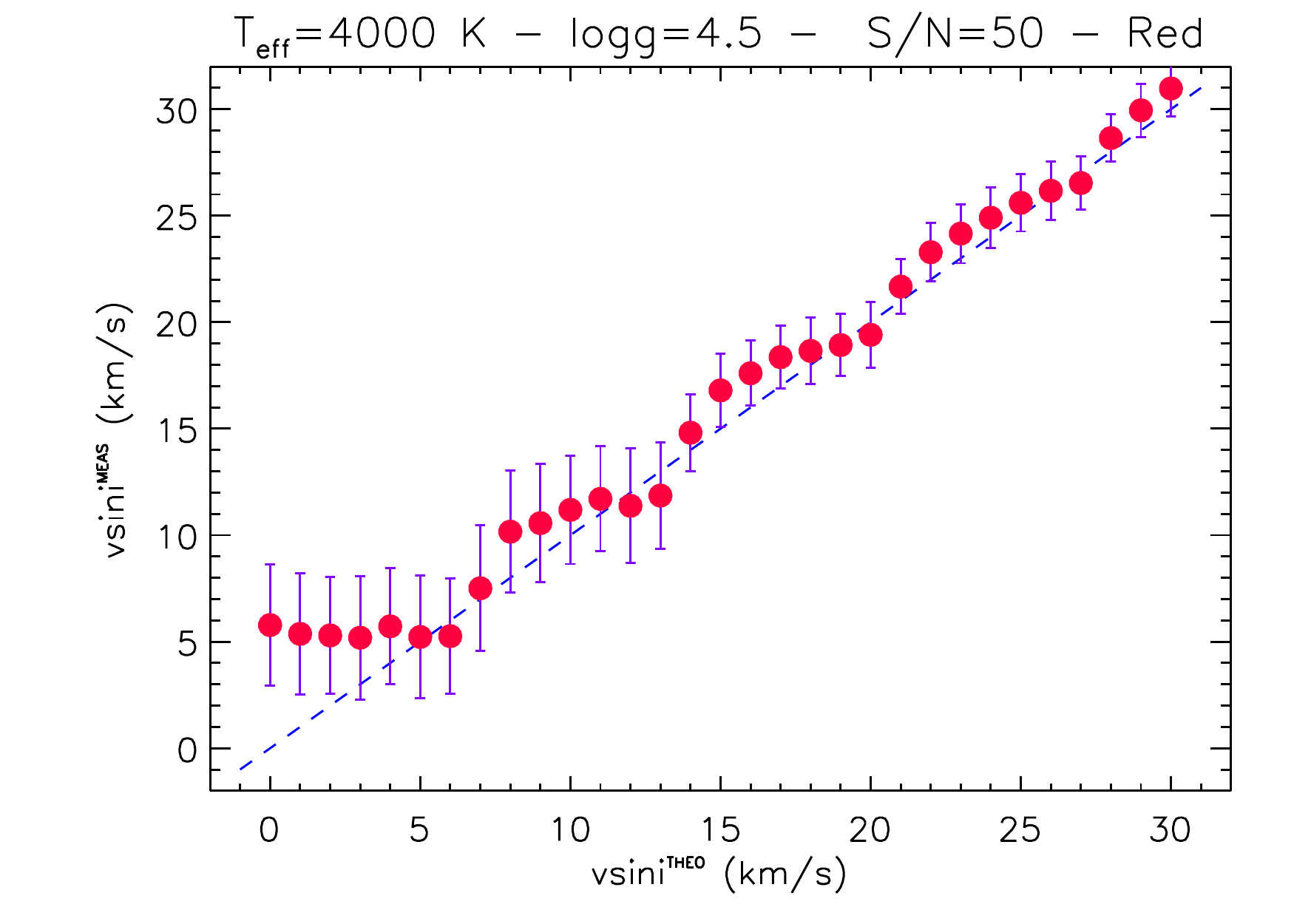} 
\caption{Results of the Monte Carlo simulations on $v\sin i$  for red-arm \lamost\ MRS. BT-Settl synthetic spectra with \logg=4.5 and three different \teff\ values, as indicated in the title of each box, have been adopted. The S/N=50.
The average $v\sin i$ values measured with our procedure (dots) are plotted against the ``theoretical'' $v\sin i$ to which the spectra have been broadened. The error bars represent the standard deviations of the 200 simulations at each \vsini.
The 1:1 relation is plotted with a dashed blue line.}
\label{Fig:MonteCarlo}
\end{figure}

\begin{figure}[th]
\includegraphics[width=8cm]{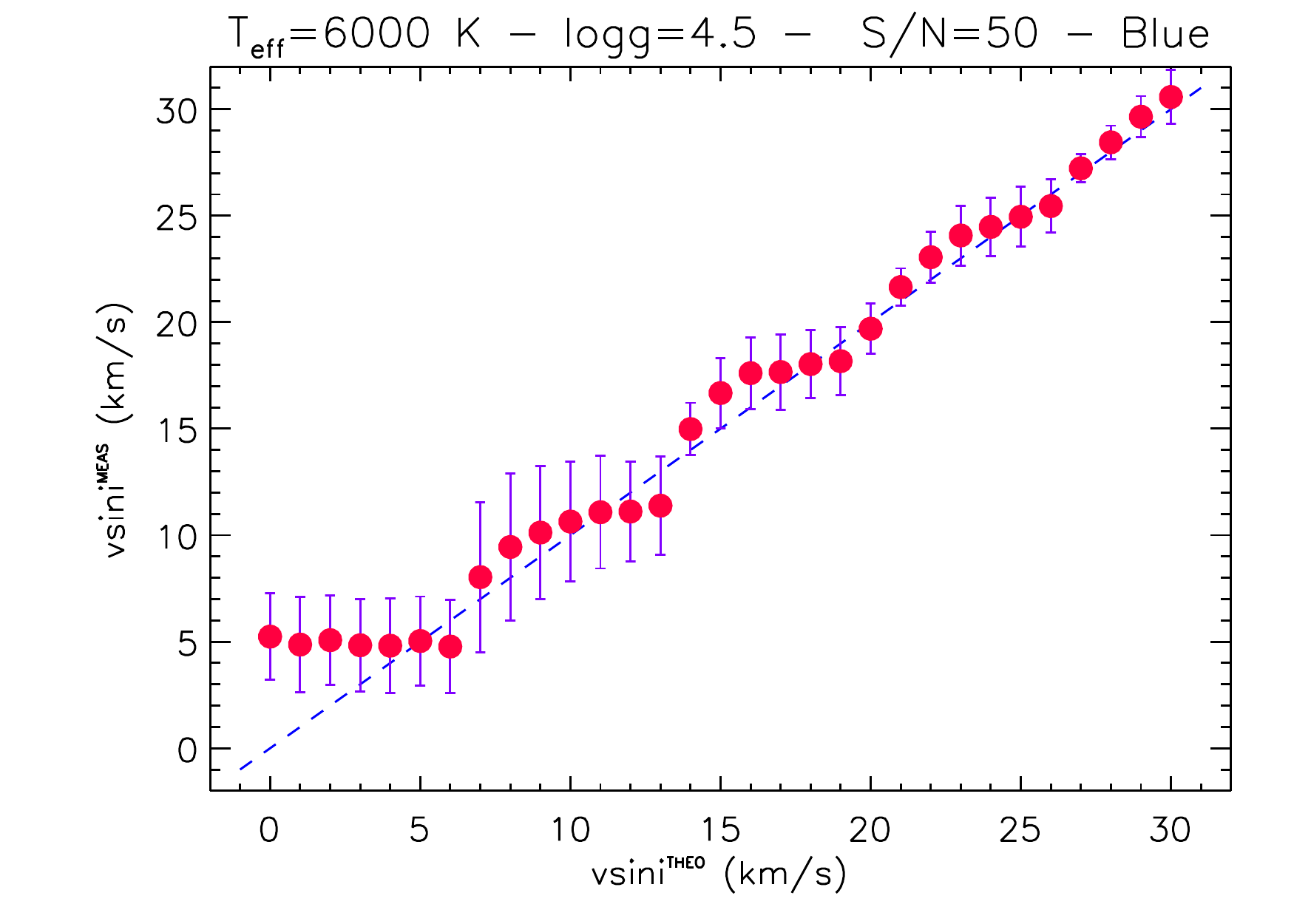}    
\includegraphics[width=8cm]{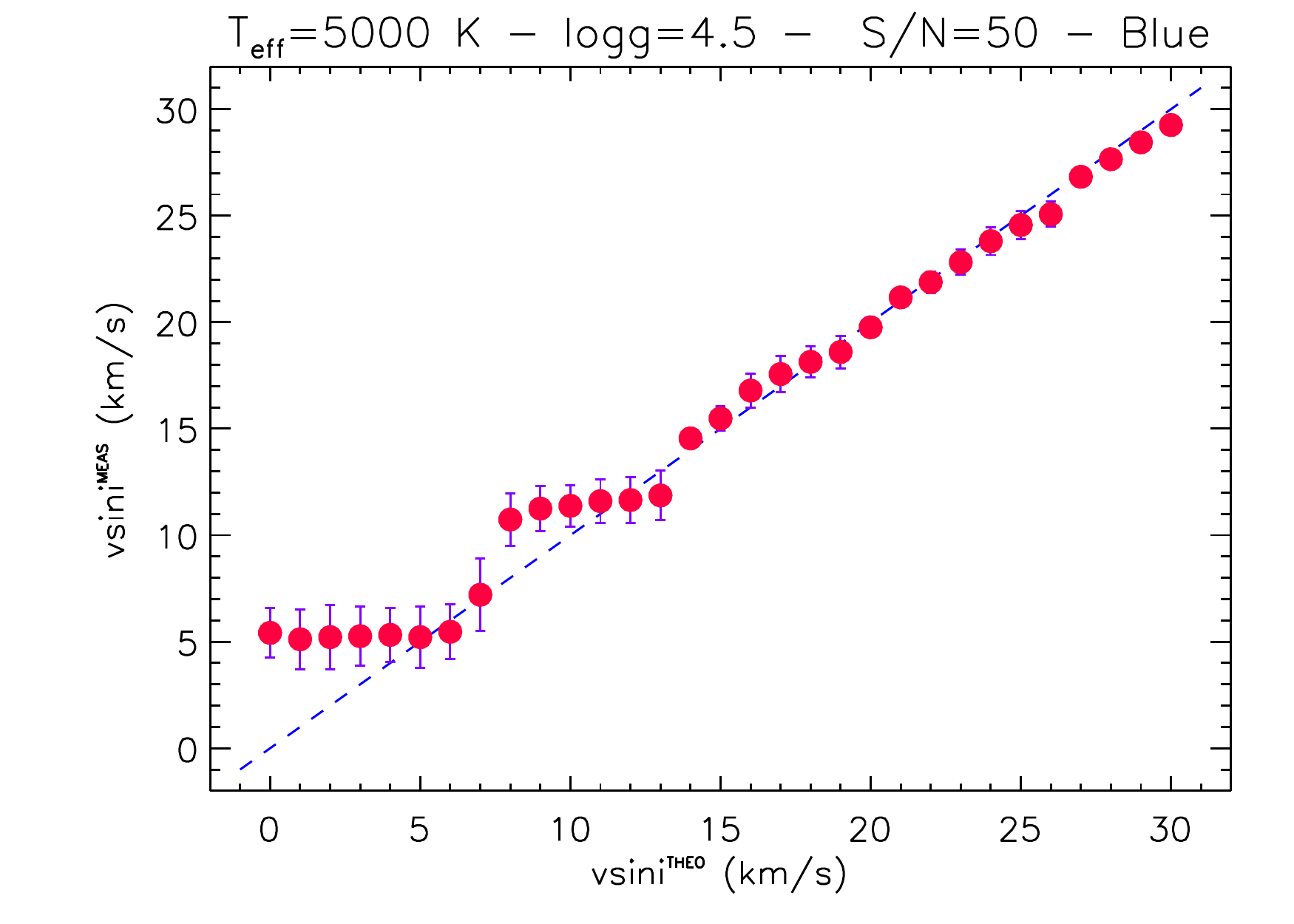}    
\includegraphics[width=8cm]{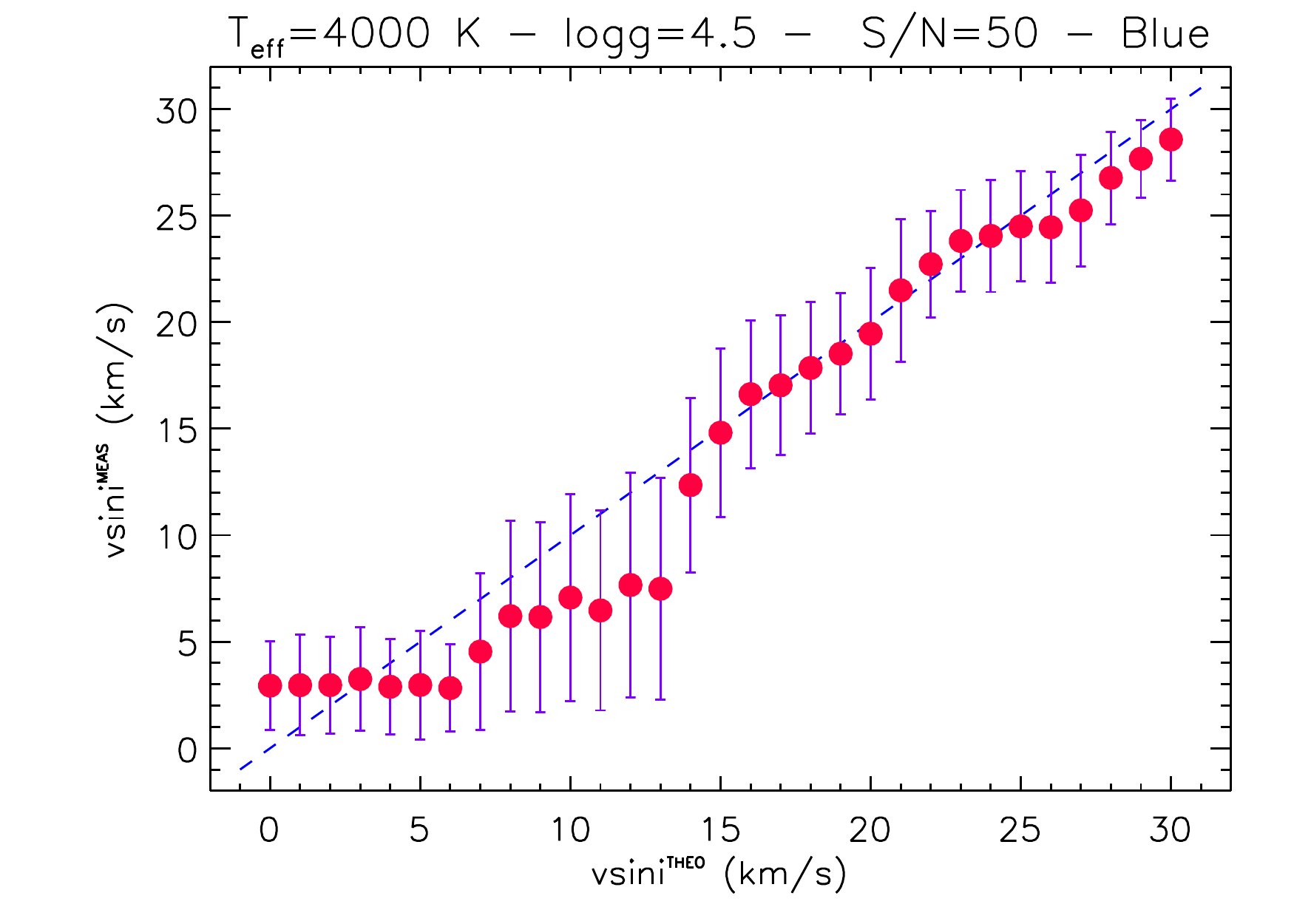}    
\caption{Results of the Monte Carlo simulations on \vsini\ for blue-arm  \lamost\  MRS. The \teff, \logg, and S/N are indicated in the top title of each box. The meaning of the symbols is the same as in Fig.~\ref{Fig:MonteCarlo}.}
\label{Fig:MonteCarlo_blue}
\end{figure}

In Figs.~\ref{Fig:MonteCarlo} and \ref{Fig:MonteCarlo_blue}, we show some examples of the results of these simulations for the red- and blue-arm spectra, respectively.
As apparent from these plots, we are not able to resolve \vsini's lower than about 8\,\kms, for which the measured values, \vsini$^{\rm MEAS}$, stay at a nearly constant level. Thereafter, the measured  \vsini\  starts to grow with the simulated \vsini\  and smoothly follows the one-to-one line
(dashed blue line in the plots).
Therefore, we adopted 8\,\kms\  as the upper limit for the \vsini\ measures on the  \lamost\ MRS.  
We point out that this limit is related to the sampling and resolution of the \lamost\ MRS and it is not the result of inaccuracies in the analysis code. 
These simulations provide also us with another estimate of the \vsini\ accuracy.
The \vsini\  errors measured with \rotfit\ on the target spectra were in most cases comparable with those inferred by the aforementioned Monte Carlo simulations, which are in the range 3--4 \kms\ for spectra with S/N=50.

\section{Stars with discrepant values of stellar parameters compared to the literature}
\label{Appendix:Discrepant}

We discuss in this section the stars whose \teff, \logg, or \feh\ values appear as outliers in Figs.~\ref{Fig:JMZ_comp_Teff}, \ref{Fig:JMZ_comp_logg}, or \ref{Fig:JMZ_comp_FeH} where we compare our APs determinations with those from the literature.

As regards the effective temperature, we note two stars, KIC\,7612547 and KIC\,8750025, with very discrepant values ($\Delta$\teff$>$1300\,K) that are marked with their KIC identifiers in Fig.~\ref{Fig:JMZ_comp_Teff}c. Furthermore, we measure a lower temperature for the very few objects with \teff$>$\,7500\,K in the APOGEE or LASP samples. This is obviously due to the template grid used by \rotfit, which is optimized for FGKM stars and does not contain a sufficient number of hot stars. 

As regards the gravity, we found three stars with discrepant \logg\ values ($\Delta$\logg$>$2.0\,dex): KIC\,8494839, KIC\,8750025, and KIC~\,10526137. Those stars are marked in Figs.~\ref{Fig:JMZ_comp_logg}a, b, or c with their KIC identifiers.

Finally, as regards the metallicity, we notice a group of 27 stars (KIC 5949422, 5957774, 6127083, 6685477, 6849670, 6852771, 6862553, 6864450, 6938848, 6949847, 7039688, 7186646, 7287513, 7360068, 7532435, 7603295, 7691709, 7820467, 7880048, 7880839, 7949214, 8036035, 8169032, 8686000, 8950928, 9026866, and 9099050) for which the LASP pipeline produces very low values of \feh\ but which in the same time are found in this paper to be of solar metallicity. Those stars fall into a yellow rectangle in Fig.~\ref{Fig:JMZ_comp_FeH}c and are discussed at the bottom of this section.

In Table~\ref{Tab:discrepant}, we provide the APs derived from the \lamost\ MRS spectroscopy in this paper for four stars indicated with their KIC numbers in Figs.~\ref{Fig:JMZ_comp_Teff} or \ref{Fig:JMZ_comp_logg}. That table provides also the APs derived from the \lamost\ MRS data by \cite{Zong2020ApJS..251...15Z} who used the LASP pipeline (two stars) or by \cite{Wang2020ApJ...891...23W} who used the Stellar Parameters and Chemical Abundances Network (SPCANet) neural network (two stars, we provide the mean values and their standard errors calculated from individual AP measurements), high-resolution spectroscopy by \cite{2020AJ....160..120J} (two stars), Dartmouth stellar isochrones by \cite{Huber2014_ApJS_211_2} (one star), the \lamost\ low-resolution spectroscopy by \cite{2019ApJS..245...34X} (two stars, for KIC\,10526137 we provide the mean values and their standard errors calculated from two individual AP measurements), or the Bayesian stellar parameters derived from the Gaia DR2 parallaxes and optical photometry by \cite{2019AaA...628A..94A} (four stars). 
%For \textcolor{red}{one} star, we provide the \logg\ values derived from the granulation-driven light curve flicker by \cite{2016ApJ...818...43B} who used the \teff\ values adopted from \cite{Huber2014_ApJS_211_2}. 
We note that not for all the literature values of APs listed in Table~\ref{Tab:discrepant} the $1\sigma$ uncertainties have been provided.

Since the APs of stars listed in Table~\ref{Tab:discrepant} show noticeable scatter, our general conclusion is that the APs derived by means of methods other than analysis of high-resolution spectroscopy should be taken with care, and the respective stars should be observed again, possibly with high-resolution spectroscopy.

\begin{table}   %[ht]
\begin{center}
\caption{Atmospheric parameters of stars marked in Figs.~\ref{Fig:JMZ_comp_Teff} or \ref{Fig:JMZ_comp_logg}.}
\begin{tabular}{llll}
\hline
\hline
\noalign{\smallskip}
\teff\ & \logg\ & \feh\ & Reference \\
(K)    & (dex)  & (dex) &  \\
\hline
\noalign{\smallskip}
\multicolumn{4}{c}{KIC~7612547}\\ \hline
4969$\pm$246 & 2.02$\pm$0.39 & $-$1.54$\pm$0.24 & this paper\\
5029$\pm$110 & 2.03$\pm$0.09 & $-$1.39$\pm$0.02 & J20       \\ % APOGEE
6301$\pm$705 & 2.82$\pm$0.99 & $-$2.20$\pm$0.26 & Z20       \\ % MRS LAMOST Zong
5056$\pm$53  & 2.53$\pm$0.12 & $-$1.45$\pm$0.05 & W20 (mean)\\ % MRS LAMOST SPCANet - mean values
5038$\pm$33  & 1.88$\pm$0.06 & $-$1.42$\pm$0.05 & X19       \\ % LRS LAMOST                     
5093         & 2.47          & $-$0.65          & A19       \\ % StarHorse
\hline
\noalign{\smallskip}
\multicolumn{4}{c}{KIC~8494839} \\ \hline
5046$\pm$162 & 2.93$\pm$0.47 & $-$0.44$\pm$0.17 & this paper\\
4188$\pm$80  & 0.50$\pm$0.07 & $-$1.23$\pm$0.01 & J20       \\ % APOGEE % Teff & log g: Parameter in possibly unreliable range of calibration determination 
4278         & 1.29          & $-$0.69          & A19       \\ % StarHorse
\hline
\noalign{\smallskip}
\multicolumn{4}{c}{KIC~8750025} \\ \hline
6239$\pm$129 & 4.01$\pm$0.14 & $-$0.02$\pm$0.14 & this paper\\
4226$\pm$634 & 1.91$\pm$0.51 & $-$0.75$\pm$0.20 & Z20       \\ % MRS LAMOST Zong
4791$\pm$673 & 2.38$\pm$0.89 & $-$0.56$\pm$0.24 & W20 (mean)\\ % MRS LAMOST SPCANet - mean values
4009         & 1.13          & $-$0.33          & A19       \\ % StarHorse
\hline
\noalign{\smallskip}
\multicolumn{4}{c}{KIC~10526137} \\ \hline
3337$\pm$79  & 0.65$\pm$0.34 & $-$0.12$\pm$0.11 & this paper    \\
3161         & 0.19          & \hspace{7.5pt}0.25 & A19         \\ % StarHorse
5857$\pm$404 & 5.30$\pm$0.45 & $-$1.15$\pm$0.22 & X19 (mean)    \\ % LRS LAMOST - mean values
3492$\pm$92  & 4.85$\pm$0.09 & \hspace{7.5pt}0.06$\pm$0.18 & H14\\ % spectroscopy Huber
\hline
\\
\end{tabular}
\begin{minipage}{7.5cm}
A19 = \citet{2019AaA...628A..94A}; H14 = \citet{Huber2014_ApJS_211_2}; J20 = \citet{2020AJ....160..120J}; W20 =  \citet{Wang2020ApJ...891...23W}; X19 = \citet{2019ApJS..245...34X}; Z20 = \citet{Zong2020ApJS..251...15Z}. % B16 = \citet{2016ApJ...818...43B}; 
\end{minipage}
\label{Tab:discrepant}
\end{center}
\end{table}  

\begin{figure}[ht]
\begin{center}
\includegraphics[width=8.5cm]{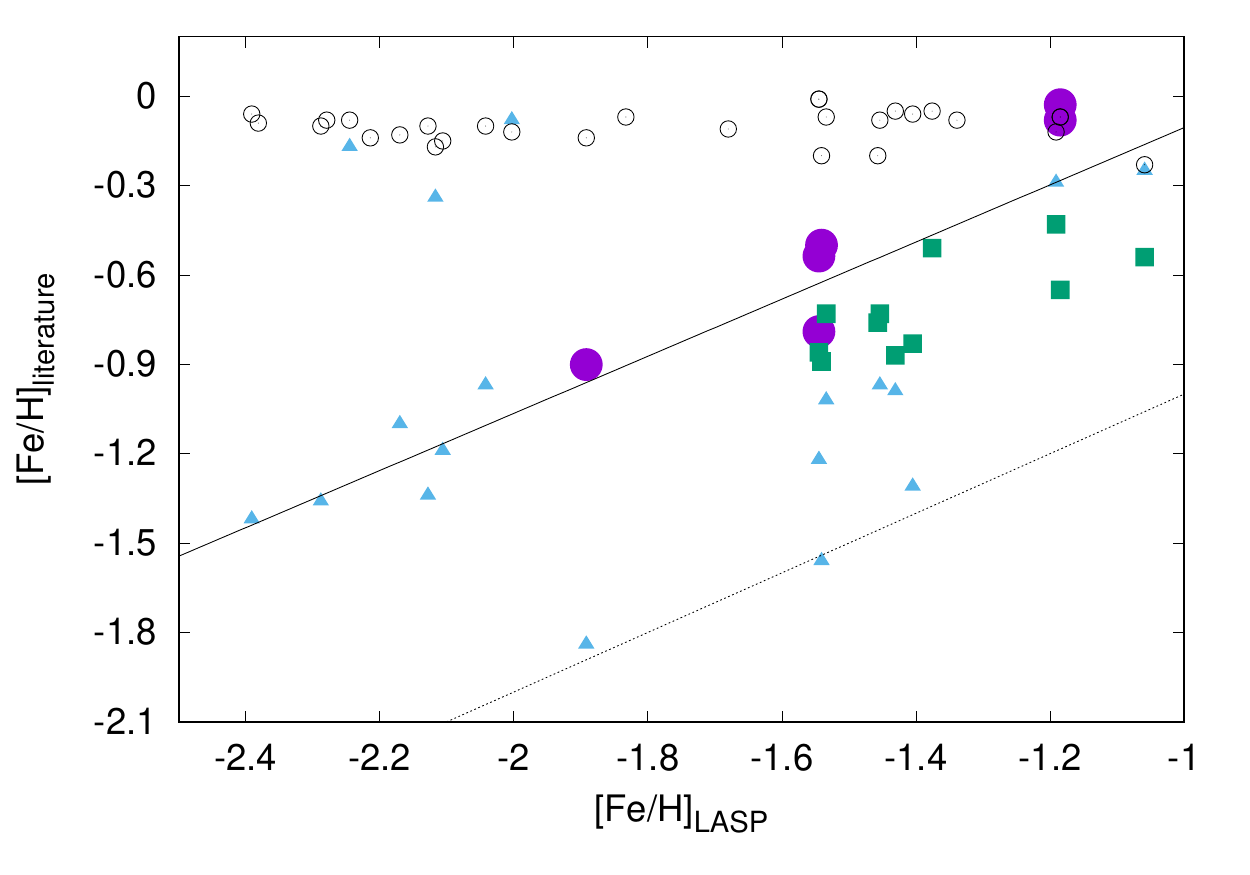}
\caption{Comparison of the [Fe/H] values derived by \citet{Zong2020ApJS..251...15Z} with the values reported in the literature. Different colors and symbols are used for different literature sources, as indicated in the text. The solid line is a linear fit to the data. The dotted line represents the one-to-one relationship.}
\label{Fig:FeH_Zong_discrepant}
\end{center}
\end{figure}

For KIC\,7612547, which is marked in Fig.~\ref{Fig:JMZ_comp_Teff}c, we found a very large discrepancy ($\simeq 1330$\,K) between our \teff\ value and the one derived with the \lasp\ pipeline on the same spectra. However, the APs reported in the present work are in excellent agreement with those from the literature, as can be seen in Table~\ref{Tab:discrepant}. Therefore, we conclude that our APs are fully reliable and the \lasp\ values are affected by issues, as also apparent from their very large errors.

For KIC\,8494839, which is marked in Figs.~\ref{Fig:JMZ_comp_logg}b and c,
    the \logg\ value derived in this paper is significantly higher than the \logg\ reported by 
    \cite{2020AJ....160..120J}. We note, however, that this value was flagged by \cite{2020AJ....160..120J} as possibly falling in an unreliable range of calibration determination. The \logg\ value derived by \cite{2019AaA...628A..94A} falls in between the determinations reported in this paper and by \cite{2020AJ....160..120J}.
    \vspace{5pt}

The APs of KIC\,8750025, which is marked in Figs.~\ref{Fig:JMZ_comp_Teff}c and \ref{Fig:JMZ_comp_logg}c, are very different from the literature determination listed in Table~\ref{Tab:discrepant}. Also, \cite{2018A&A...616A...1G}, \cite{2009yCat.5133....0K} or \cite{2012ApJS..199...30P} consistently point to a cool star of \teff\ between 4000 and 4500\,K. In this paper, we analyzed ten \lamost\ MRS measured in 2018 while  \cite{Wang2020ApJ...891...23W} derived APs from two spectra measured in 2019 and one, in 2018. The APs derived by \cite{Wang2020ApJ...891...23W} from the spectrum acquired on May, 24, 2018, which was also analyzed in this paper, yielded \teff\ = 6135$\pm$119\,K, \logg\ = 4.16$\pm$0.17\,dex, and \feh\ = $-$0.09$\pm$0.12\,dex (thus the large standard errors of the means listed in Table~\ref{Tab:discrepant}), which agree with the APs we derived within the respective $1\sigma$ error bars. Since the reason for         the discrepancy in the APs of KIC\,8750025 derived from the \lamost\ MRS measured in different seasons is not clear, we suggest using them with care.

According to our AP determinations, KIC\,10526137, which is marked in Fig.~\ref{Fig:JMZ_comp_Teff}a, is a late-type giant of metallicity slightly lower than solar. For that star, the APs reported in the literature show high dispersion (see Table~\ref{Tab:discrepant}). In particular, the \logg\ determinations range from 0.19\,dex \citep{2019AaA...628A..94A} to 5.30\,dex \citep{2019ApJS..245...34X}. \cite{2018ApJ...866...99B} who used the parallaxes from the Gaia DR\,2 \citep{2018A&A...616A...1G, 2018A&A...616A...2L} and the DR25 Kepler Stellar Properties Catalog \citep{2017ApJS..229...30M} computed that radius of this star, $R = 146.830\,\rm{R_{\odot}}$, and classified it as a red giant. We find the analyses based on the Gaia parallaxes carried out by \cite{2019AaA...628A..94A} and \cite{2018ApJ...866...99B} to be conclusive and the APs of KIC\,10526137 derived in this paper to be fully reliable.

Figure~\ref{Fig:FeH_Zong_discrepant} shows the comparison of the \feh\ values derived from the MRS \lamost\ spectra by \cite{Zong2020ApJS..251...15Z} and the literature values for the 27 stars that fall into the yellow rectangle in Fig.~\ref{Fig:JMZ_comp_FeH}c. The \feh\ values derived in this paper are indicated with black open circles. Those derived from high-resolution APOGEE spectra by \cite{2020AJ....160..120J} or \cite{2020AJ....159..182O} are indicated with large purple dots. The \feh\ values derived from the MRS \lamost\ spectra by \cite{Wang2020ApJ...891...23W} are indicated with green squares and those derived from the LRS \lamost\ spectra by \cite{2019ApJS..245...34X} are indicated with blue triangles. The linear fit to the data is plotted with a solid line. The linear regression gives a slope $b=0.96$ and intercept $a=0.83$.

Both this paper and \cite{Zong2020ApJS..251...15Z} find those 27 stars to be late-type giants cooler than $\sim$4500\,K. The discrepancies in \feh\ shown in Fig.~\ref{Fig:FeH_Zong_discrepant} highlight the necessity of using high-resolution spectroscopic analyses for deriving precise and accurate values of APs; they also show that in the discussed range of the APs neither LASP nor ROTFIT provide results that are fully reliable.

\section{Online figures and tables}
\label{Appendix:online}

\begin{figure}[ht]
\begin{center}
\includegraphics[width=9cm]{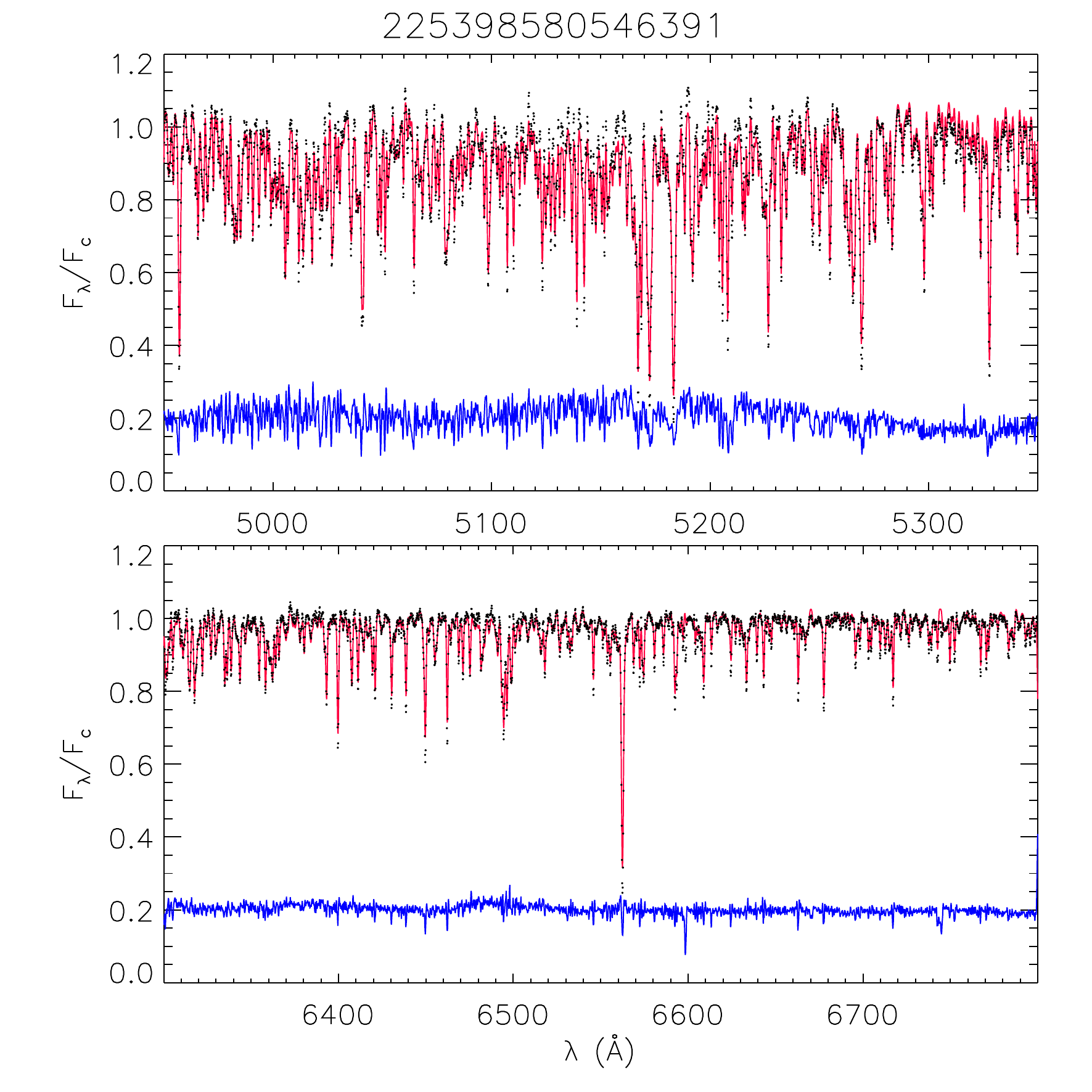}
\caption{Same as Fig.\,\ref{Fig:spectrum} but for a K0\,III star.  }
\label{Fig:spectrum3}
\end{center}
\end{figure}

\begin{figure}[ht]
\begin{center}
\includegraphics[width=9cm]{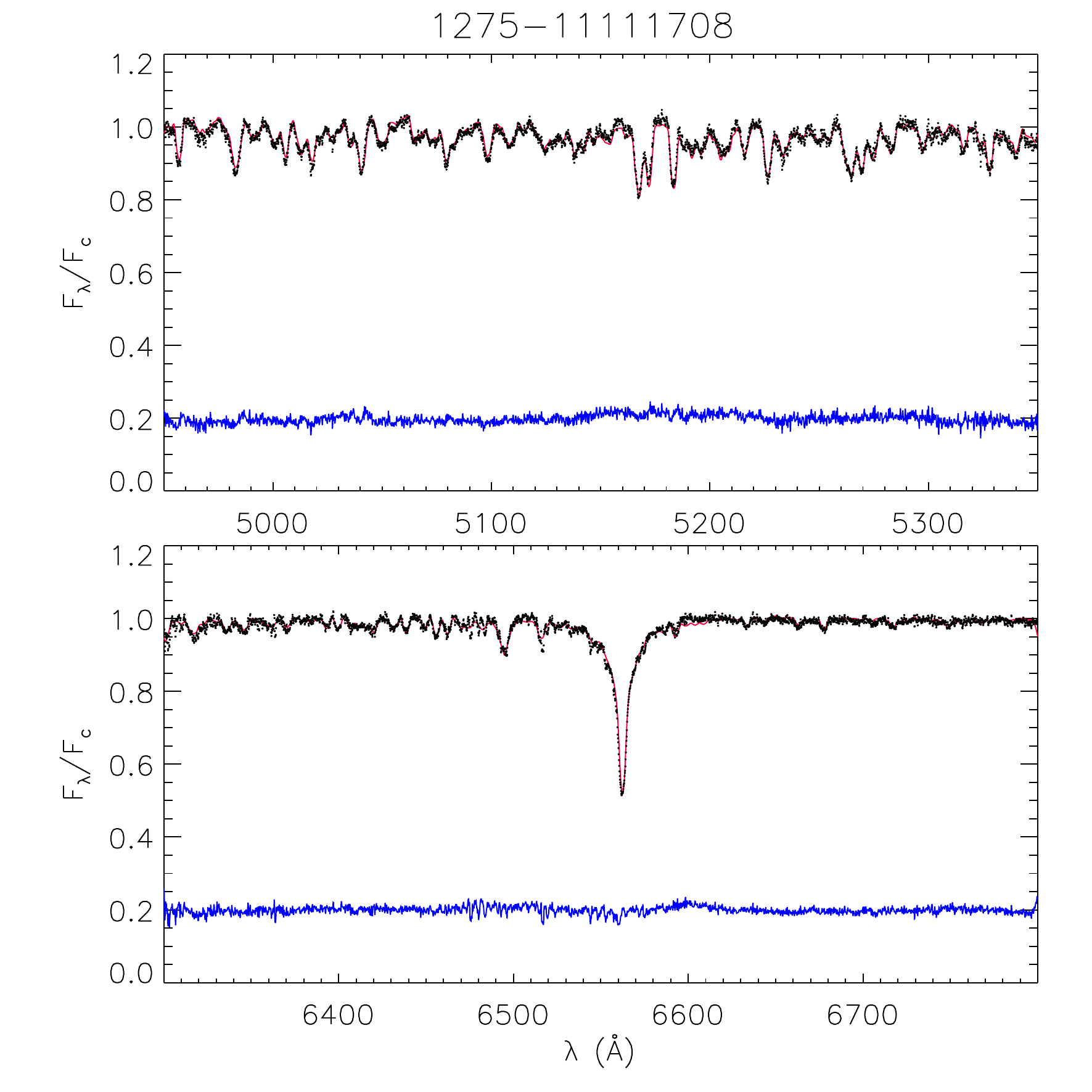}
\caption{Same as Fig.\,\ref{Fig:spectrum} but for a mid-F star rotating at about 110\,\kms.  }
\label{Fig:spectrum2}
\end{center}
\end{figure}

\begin{figure}[htb]
\includegraphics[width=8cm]{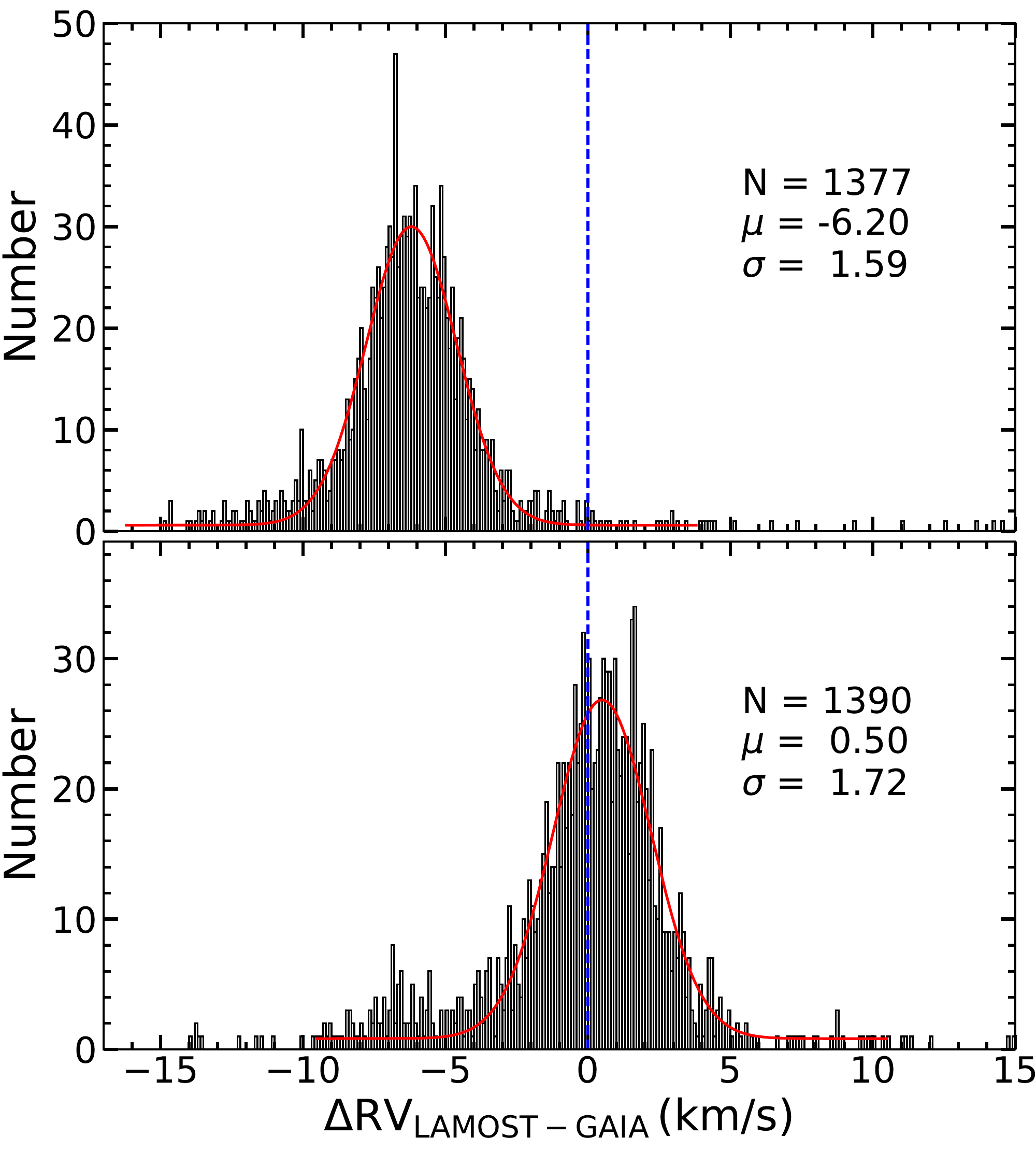}
\caption{Radial velocity difference of LK--MRS to Gaia (black histograms) for two plates calibrated with different lamps in the blue arm ({\it top:} 58267-HIPP95119KP01; {\it bottom:} 58263-HIP9511901). Red lines represent the best Gaussian fittings, whose centers and widths, $\mu$ and $\sigma$, are quoted in the legend. The vertical dotted blue line marks  $\Delta$RV=0.}
\label{Fig:RV_corr}
\end{figure}

\begin{figure}[htb]
\includegraphics[width=8cm]{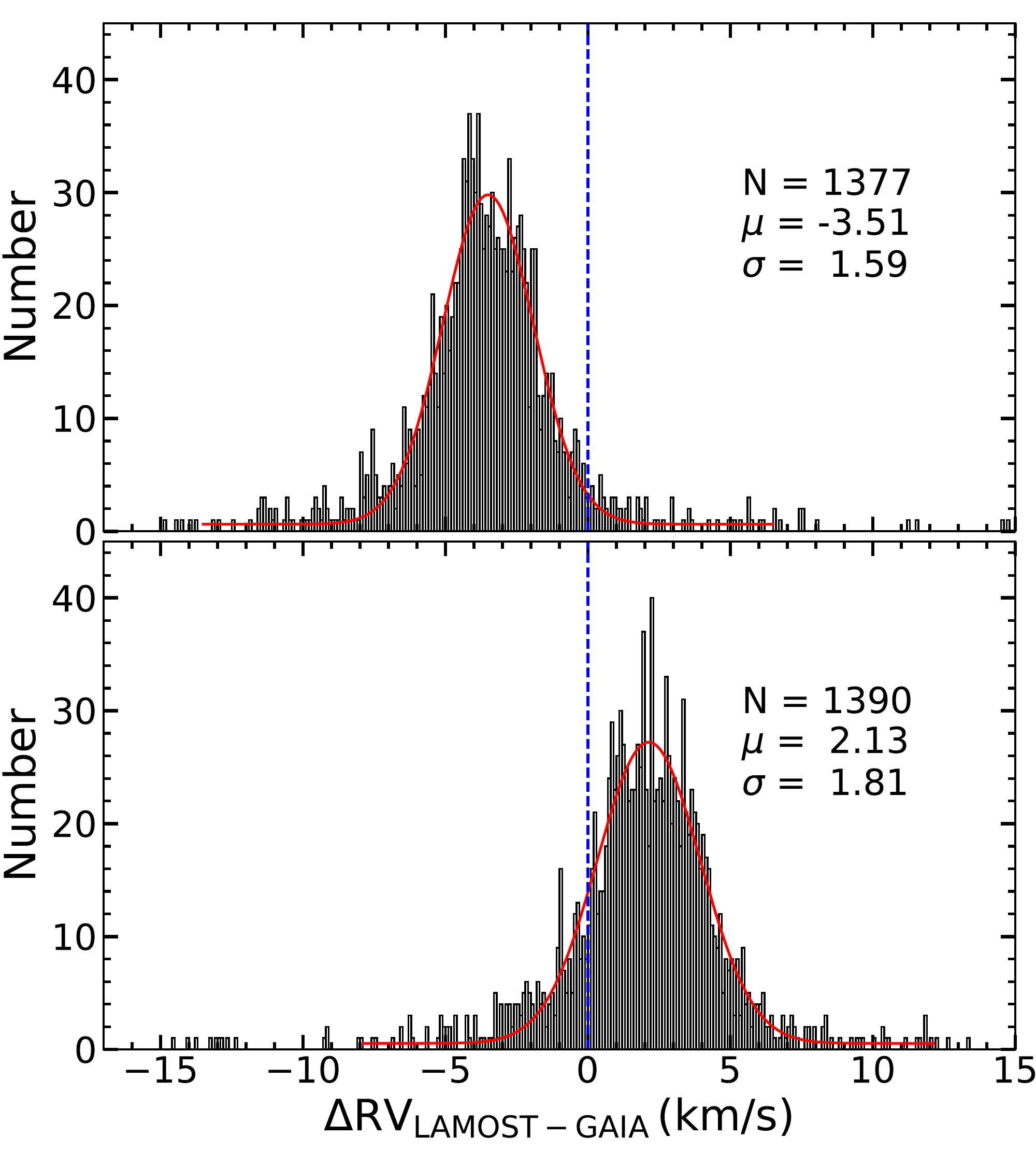}
\caption{Same as Fig.\,\ref{Fig:RV_corr} but for the red-arm RVs.}
\label{Fig:RV_corr_red}
\end{figure}

\begin{table}[tb]
\caption[ ]{Radial velocity corrections. }
\label{Tab:RV_corr}
\begin{tabular}{lrcrcr}
\hline
\hline
\noalign{\smallskip}
 RJD--Plate & $\mu_{\rm blue}$ & $\sigma_{\rm blue}$  & $\mu_{\rm red}$ & $\sigma_{\rm red}$ & $N$ \\ 
       & \multicolumn{2}{c}{(\kms)} & \multicolumn{2}{c}{(\kms)} &    \\ 
%\noalign{\smallskip} 
\hline
\noalign{\smallskip} 
58025-HIP9645901 & $-6.02$ & 1.68 & $-3.54$ & 1.69 & 827  \\
58030-HIP9286201 & $-5.69$ & 1.91 & $-3.47$ & 1.77 & 432  \\
58030-HIP9737201 & $-5.76$ & 1.88 & $-3.32$ & 1.71 & 548  \\
58031-HIP9380801 & $-6.00$ & 1.68 & $-3.27$ & 1.74 & 714  \\
58031-HIP9587901 & $-6.10$ & 1.78 & $-3.44$ & 1.69 & 756  \\
58032-HIP9448701 & $-6.19$ & 1.84 & $-3.39$ & 1.92 & 604  \\
58263-HIP9511901 & $ 0.50$ & 1.72 & $ 2.13$ & 1.81 & 1390 \\
58267-HIP95119KP01 & $-6.20$ & 1.59 & $-3.51$ & 1.59 & 1377 \\
58268-HIP95119KP01 & $-6.28$ & 1.63 & $-3.65$ & 1.58 & 1391 \\
58269-HIP95119KP01 & $-6.09$ & 1.64 & $-3.32$ & 1.55 & 1370 \\
58270-HIP95119KP01 & $-6.22$ & 1.51 & $-3.52$ & 1.53 & 1379 \\
\hline
\noalign{\smallskip} 
\end{tabular}
~\\
{\bf Notes.} RJD is the reduced Julian date. The center and width of the RV difference distributions are indicated with $\mu$ and $\sigma$, respectively. The number of stars is reported in the last column.
\end{table}

\begin{figure}[htb]
\includegraphics[width=9cm]{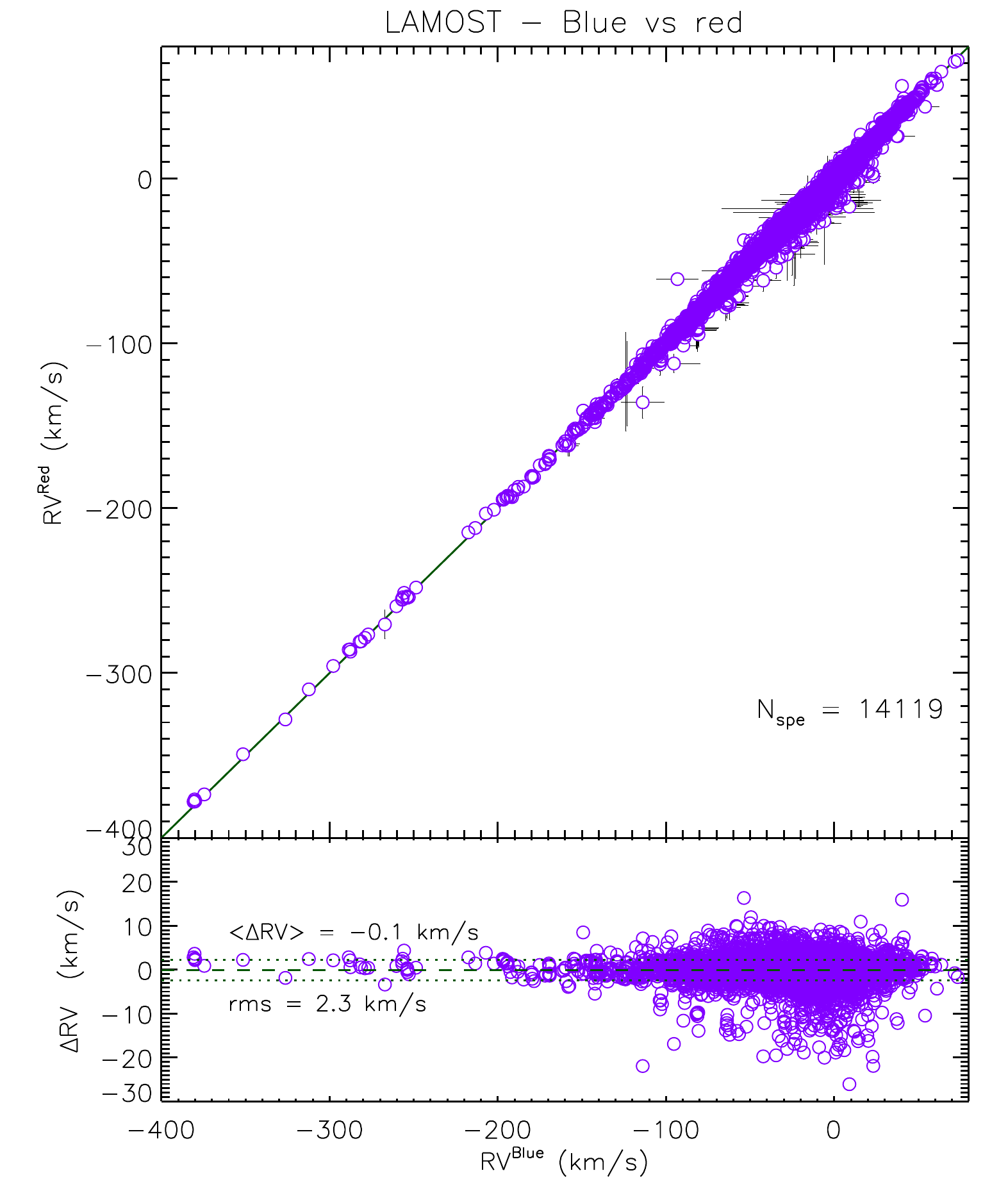}
\caption{Comparison between the RVs measured in this work on the blue-arm (Table~\ref{Tab:data_blue}) and red-arm  \lamost\ MRS (Table~\ref{Tab:data_red}).  The one-to-one relation is shown by the continuous line. The RV differences displayed in the lower box show an average value of $-0.1$\,\kms\ (dashed line) and a standard deviation of 2.3\,\kms\ (dotted lines).}
\label{Fig:RV_blue_red}
\end{figure}

\begin{figure*}  
\begin{center}
\includegraphics[width=8.8cm]{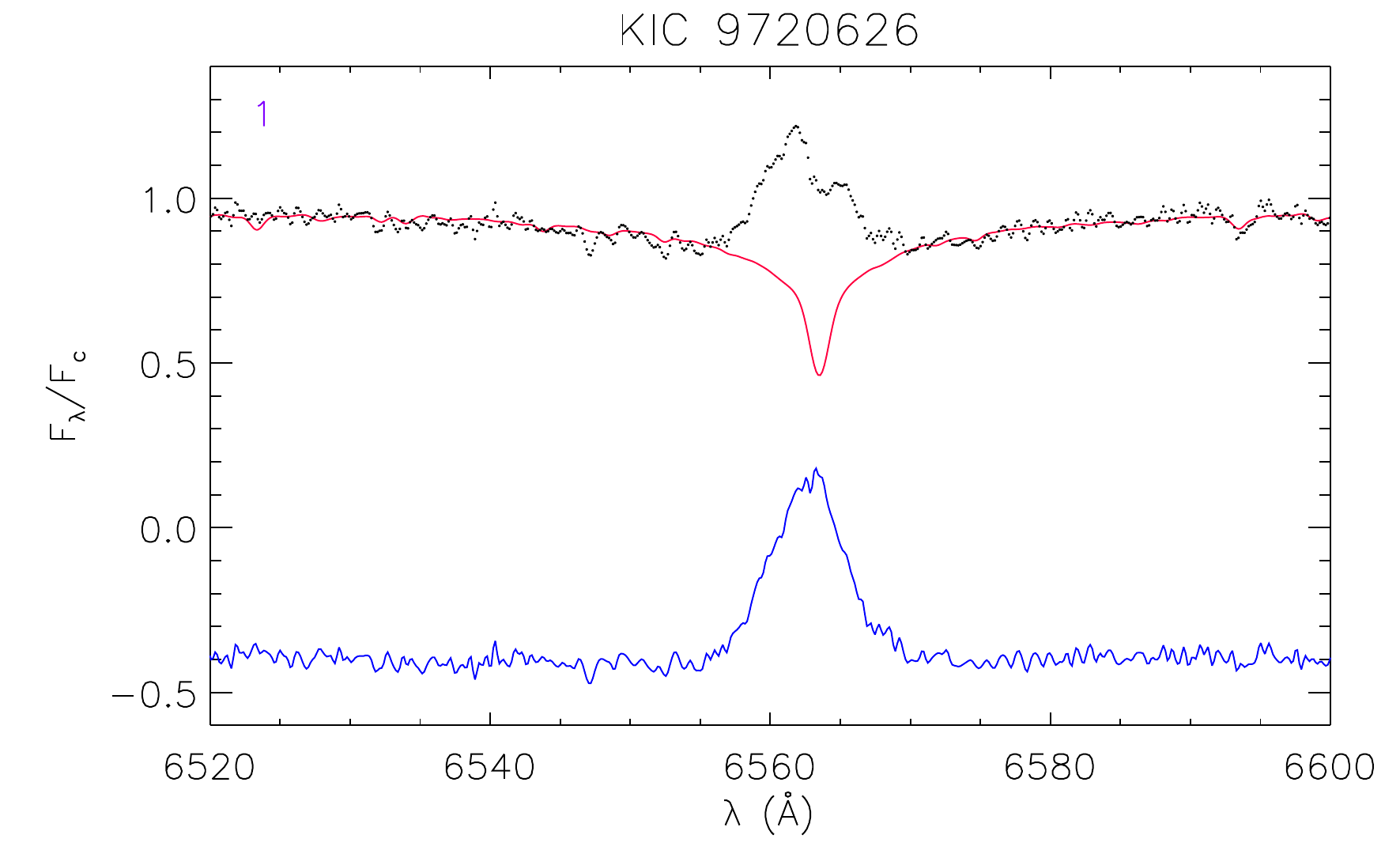}  
\includegraphics[width=8.8cm]{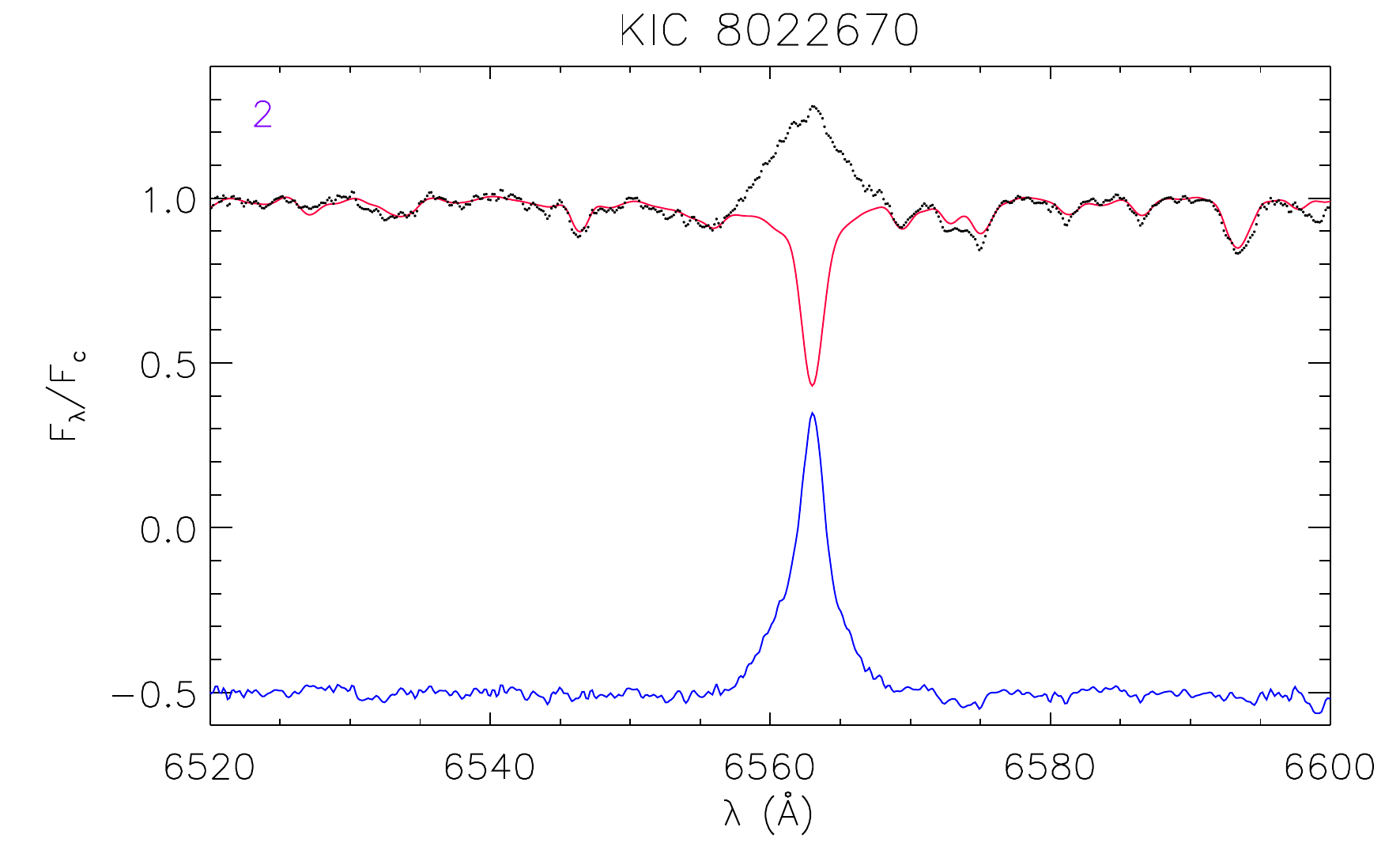}  
\includegraphics[width=8.8cm]{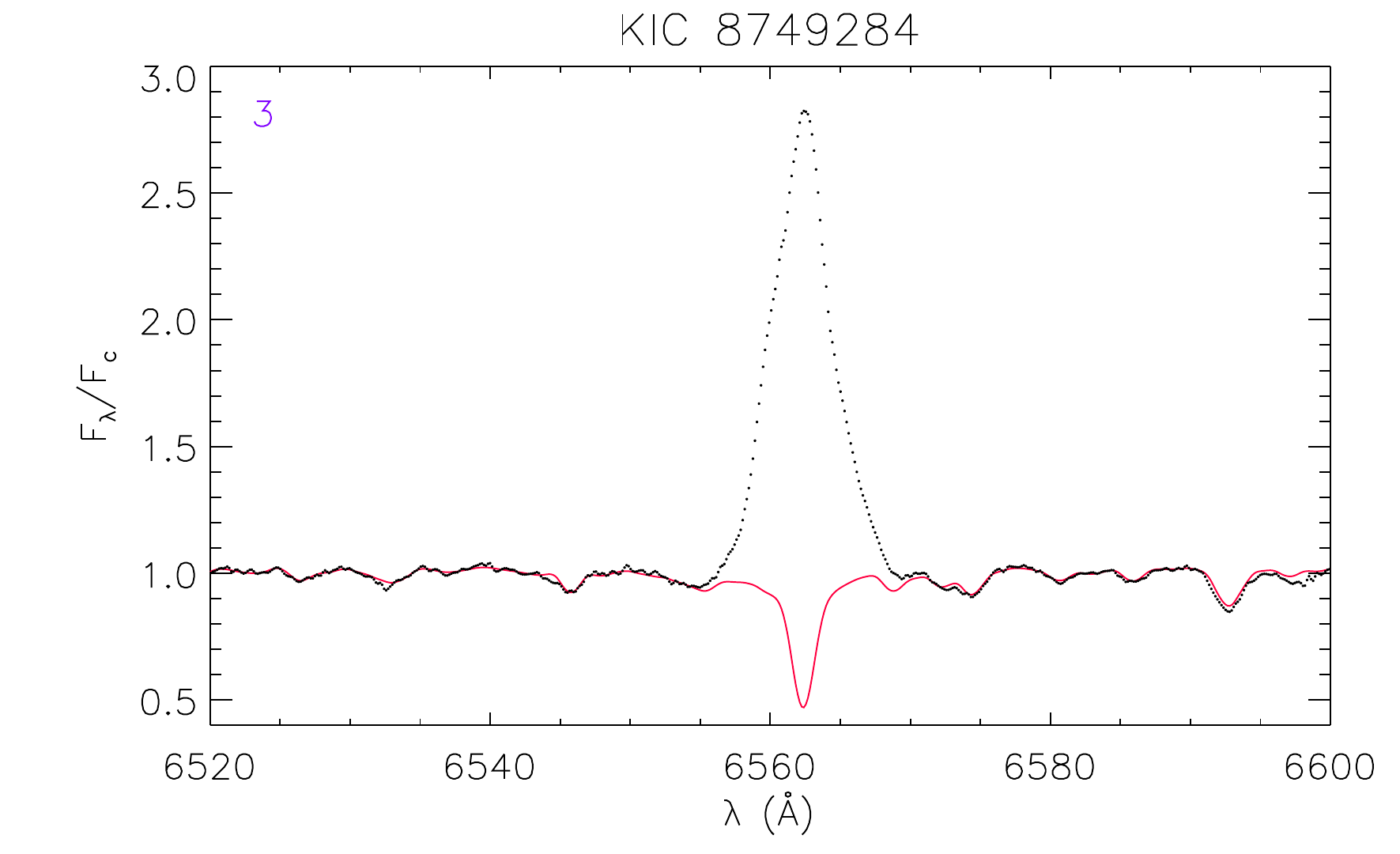}  
\includegraphics[width=8.8cm]{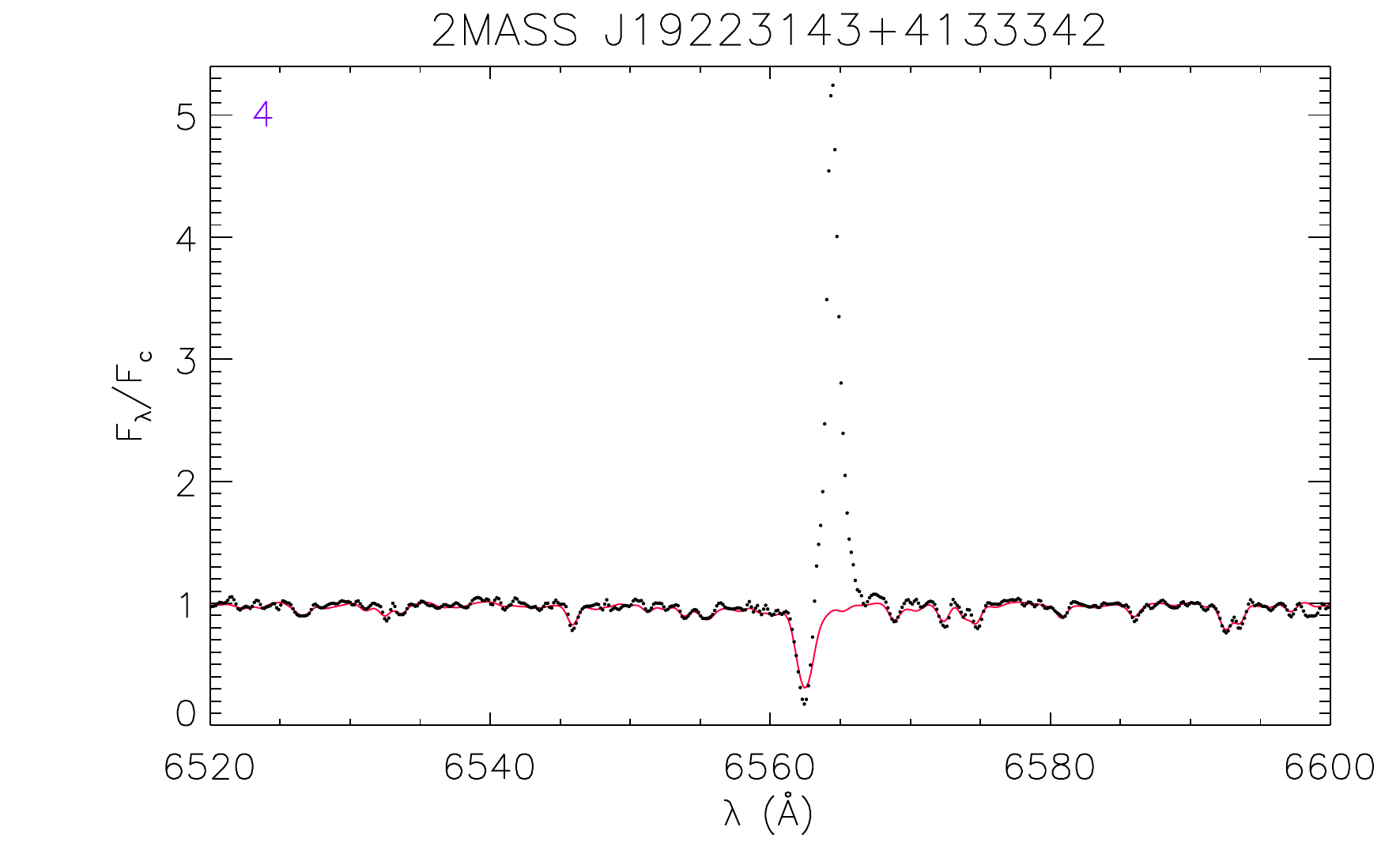}  
\includegraphics[width=8.8cm]{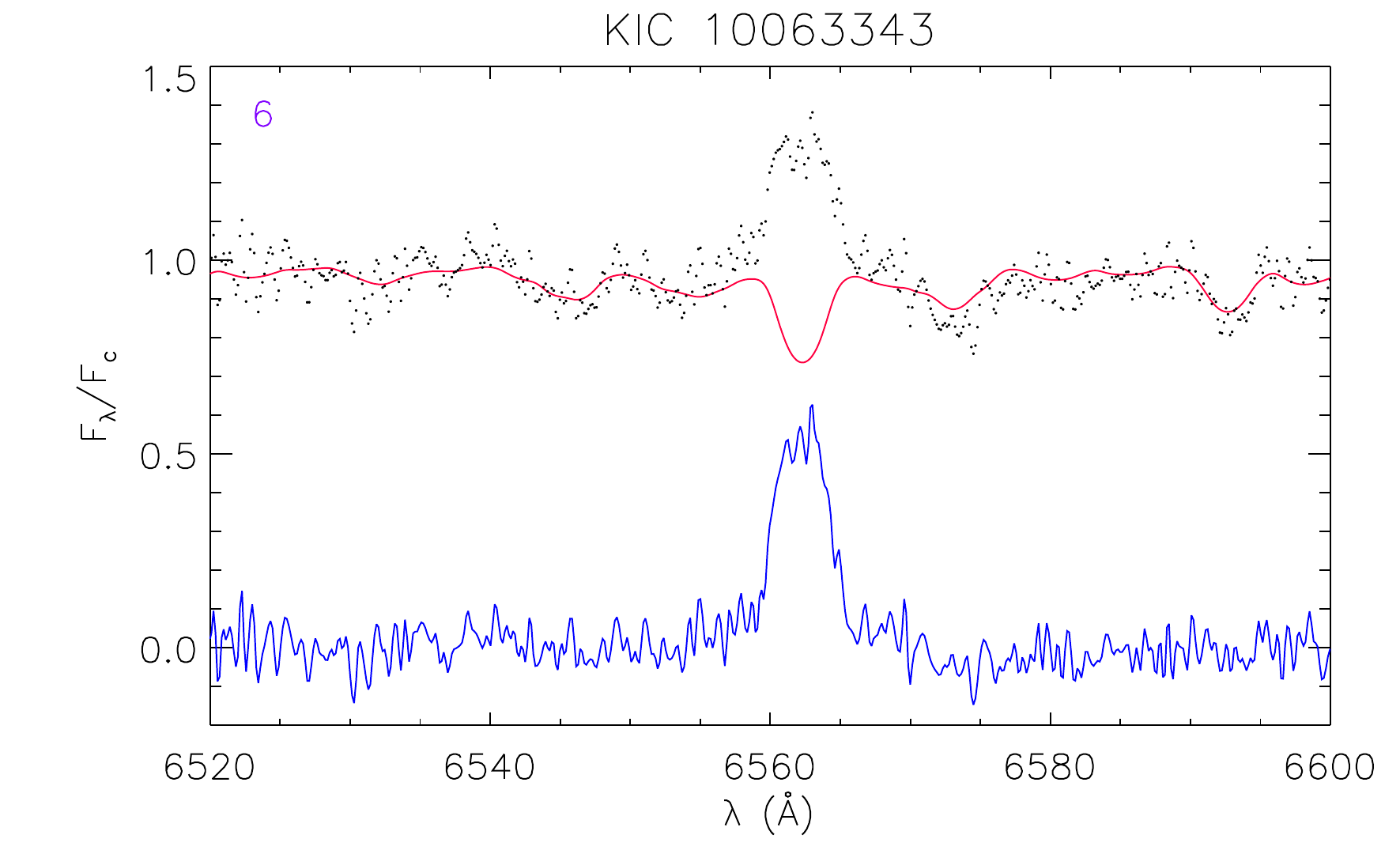}  
\includegraphics[width=8.8cm]{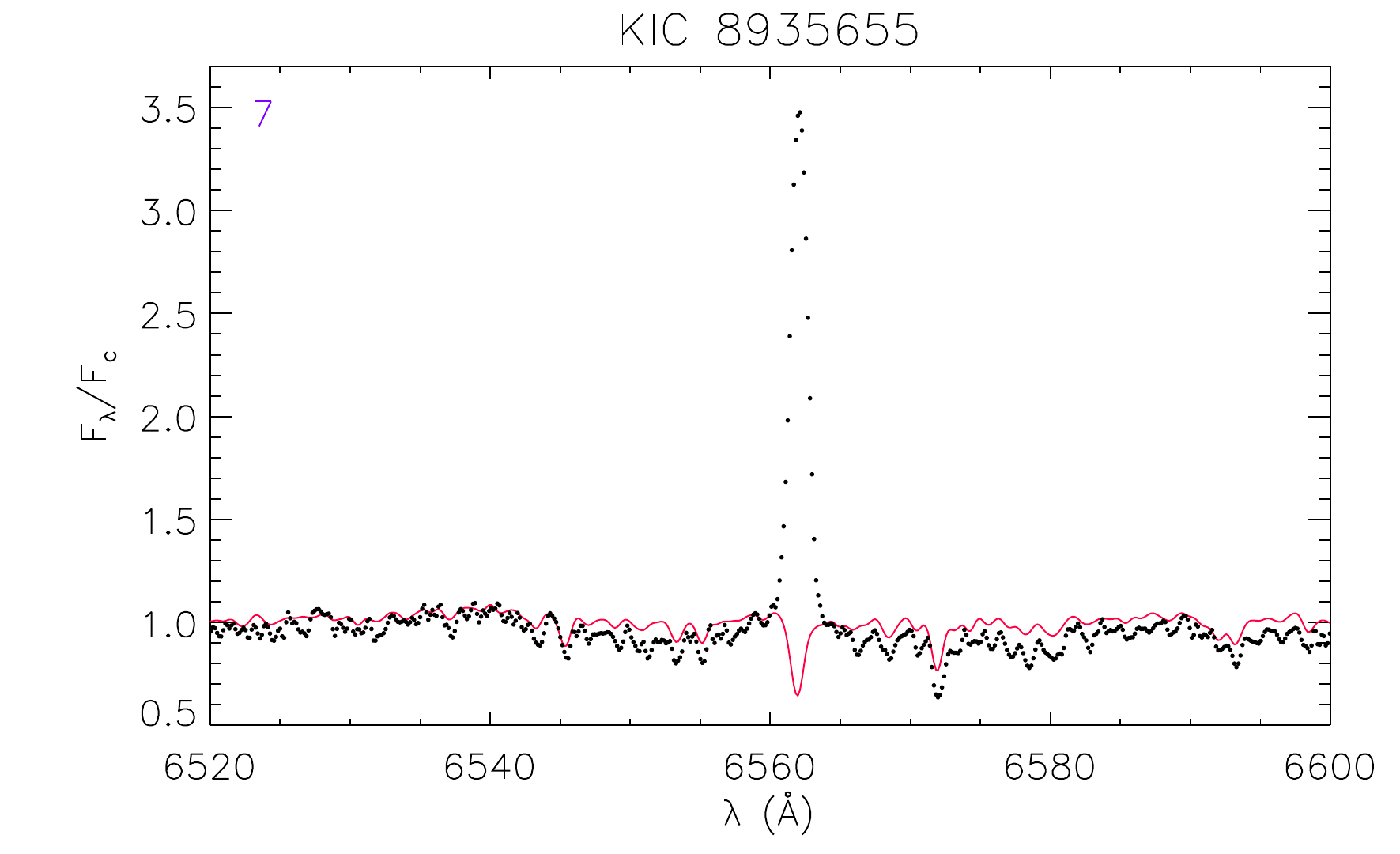}  
\caption{Observed continuum-normalized spectra of the stars with the strongest \halpha\ flux (black dots), labeled according to Fig.\,\ref{Fig:Flux_Teff}. The best fitting template is overlaid with a red line. The differences $observed-template$ are displayed in the bottom of the boxes, with blue lines for the stars with moderate emission. The \halpha\ profile of star \#5 is displayed in Fig.~\ref{Fig:Ha_Li}. }
\label{Fig:Halpha_prof}
\end{center}
\end{figure*}

\onecolumn

\begin{landscape}
\renewcommand{\tabcolsep}{0.1cm}

\vspace{0cm}
\begin{longtable}{cccccccrrrrrrrrrrrrr}
\caption{Stellar parameters and RVs derived with ROTFIT from the red-arm MRS \lamost\ spectra.}\\
\scriptsize
\hspace{-2cm}
\begin{tabular}{ccccccrrrrrrrrrrrrrr}
\hline
\hline
\noalign{\smallskip}
Designation & KIC &  RA  &   Dec.   & HJD & Spectrum & S/N  & \teff & err & \logg & err & \feh &  err & \vsini & err & RV & err & RV$_{\rm corr}$ & SpT & Rem \\ 
        &      & $\degr$ & $\degr$ &($-$2\,450\,000) &     &   & \multicolumn{2}{c}{(K)} & \multicolumn{2}{c}{dex} & \multicolumn{2}{c}{dex} & \multicolumn{2}{c}{(\kms)} & \multicolumn{3}{c}{(\kms)} & & \\ 
\hline
\noalign{\smallskip} 
 J193127.20+430204.5 & KIC07448374 & 292.863347 & 43.034586 & 58025.00354 & med-58025-HIP9645901\_sp02-003.fits  & 201.2 & 5601 & 182 & 4.02 & 0.26 & 0.20 &   0.11 & 7.0 & 3.0 & -22.11 & 1.81 & -18.57 & G6V &  ... \\    
 J193219.71+424432.2 & KIC07204500 & 293.082154 & 42.742285 & 58025.00356 & med-58025-HIP9645901\_sp02-005.fits & 105.2 & 4543 & 135 & 2.40 & 0.29 & -0.20 & 0.25 & 0.0 & 3.0 & 6.33       & 1.83 & 9.87 &  K0III & ... \\
 J193133.39+424228.5 &  KIC07203838 & 292.889145 & 42.707934 & 58025.00355 &       med-58025-HIP9645901\_sp02-010.fits & 28.7 & 4869 & 275 & 2.40 & 0.77 &  -1.51 &   0.37 & 63.4 & 28.8 & -299.30 & 2.75 & -295.76 &  G9III & ... \\        
 J193133.34+424513.5 &  KIC07203835 & 292.888950 & 42.753761 & 58025.00355      & med-58025-HIP9645901\_sp02-011.fits & 119.5 & 5766 & 86 & 4.32 & 0.13 & -0.10 &  0.21 & 3.2 & 4.7 & -23.96 & 1.73 & -20.42  & G2V &  ... \\   
 J193207.12+425118.9 & KIC07285973 & 293.029697 & 42.855268 & 58025.00356 &       med-58025-HIP9645901\_sp02-015.fits & 75.5 & 5094 & 126 & 4.53 & 0.14 & -0.07 & 0.13 & 7.0 & 3.0 & -36.47 & 1.77 & -32.93 & K2.5V &  ... \\         
  ...  &   ...  &   ...  &   ...  &   ...  &   ...  &   ...  &   ...  &   ...  &   ...  &   ...  &   ...  &   ...  &   ...  &  ...  &   ...  &   ...  &  ... & ... \\
  ...  &   ...  &   ...  &   ...  &   ...  &   ...  &   ...  &   ...  &   ...  &   ...  &   ...  &   ...  &   ...  &   ...  &  ...  & ... &  ...  &   ...  &  ... & ... \\
 J193943.93+444609.4 &  KIC08633160 &  294.933065 &  44.769293 &  58025.00365 & med-58025-HIP9645901\_sp04-238.fits & 165.6 & \dots & \dots & \dots & \dots & \dots &    \dots  & \dots & \dots & \dots & \dots & \dots & ... &  SB2 \\      
 J193912.56+444012.1 & KIC08567854 & 294.802368 & 44.670039 & 58025.00364 &       med-58025-HIP9645901\_sp04-240.fits & 115.9 & 4582 & 85 & 2.61 & 0.25 & 0.06 &   0.23 & 7.0 & 3.0 & -21.06  & 2.22 & -17.52 & K2III & ... \\      
 J194006.80+444459.3 &  KIC08633453 & 295.028356 & 44.749829 & 58025.00366 &       med-58025-HIP9645901\_sp04-245.fits  & 173.7 & 5616 & 164 & 4.30 & 0.16 & -0.45 &   0.25 & 24.2 & 4.8 & -92.67 & 1.87 & -89.13 & G2V &  ... \\      
 J193824.53+444434.9 & KIC08632169 & 294.602210 & 44.743035 & 58025.00363 &       med-58025-HIP9645901\_sp04-246.fits & 14.1 & \dots & \dots & \dots & \dots & \dots & \dots & \dots & \dots & \dots & \dots & \dots & ... & Noisy \\      
 J193953.83+443027.7 & KIC08503800 & 294.974298 & 44.507714 & 58025.00366 &       med-58025-HIP9645901\_sp04-247.fits & 109.6 & 6036 & 105 & 4.21 & 0.18 & 0.10 &   0.12 & 20.7 & 3.2 & -43.51 & 2.20 & -39.97 & G0IV & ... \\         
  ...  &   ...  &   ...  &   ...  &   ...  &   ...  &   ...  &   ...  &   ...  &   ...  &   ...  &   ...  &   ...  &   ...  &  ...  &   ...  &   ...  &  ... & ... \\
  ...  &   ...  &   ...  &   ...  &   ...  &   ...  &   ...  &   ...  &   ...  &   ...  &   ...  &   ...  &   ...  &   ...  &  ...  & ... &  ...  &   ...  &  ... & ... \\
\noalign{\smallskip}
\hline 
\noalign{\medskip}
\end{tabular}
~\\
{\bf Notes.} The full table is only available in electronic form at the CDS.
\label{Tab:data_red}
\end{longtable}

\vspace{0cm}
\begin{longtable}{cccccccrrrrrrrrrrrrr}
\caption[ ]{Stellar parameters and RVs derived with ROTFIT from the blue-arm MRS \lamost\ spectra.}\\
\scriptsize
\hspace{-2cm}
\begin{tabular}{ccccccrrrrrrrrrrrrrr}
\hline
\hline
\noalign{\smallskip}
Designation & KIC &  RA  &   Dec.   & HJD & Spectrum & S/N  & \teff & err & \logg & err & \feh &  err & \vsini & err & RV & err & RV$_{\rm corr}$ & SpT & Rem \\ 
        &      & $\degr$ & $\degr$ &($-$2\,450\,000) &     &   & \multicolumn{2}{c}{(K)} & \multicolumn{2}{c}{dex} & \multicolumn{2}{c}{dex} & \multicolumn{2}{c}{(\kms)} & \multicolumn{3}{c}{(\kms)} & & \\ 
\hline
\noalign{\smallskip} 
 J193127.20+430204.5 & KIC07448374 & 292.863347 & 43.034586 & 58025.00354 &       med-58025-HIP9645901\_sp02-003.fits & 145.9 & 5861 & 71 & 4.25 & 0.15 & 0.17 & 0.12 & 0.0 & 3.0 & -24.13 & 1.84 & -18.11 & G0 & ... \\ 
 J193219.71+424432.2 & KIC07204500 & 293.082154 & 42.742285 & 58025.00356 &       med-58025-HIP9645901\_sp02-005.fits & 52.5 & 4528 & 117 & 2.16 & 0.13 & -0.20 &    0.17 & 0.0 & 3.0 & 4.35       & 2.69 & 10.37 & K1III & ... \\ 
 J193133.39+424228.5 & KIC07203838 & 292.889145 & 42.707934 & 58025.00355 &       med-58025-HIP9645901\_sp02-010.fits & 15.5 & 5162 & 271 & 2.46 & 0.62 & -0.77 &    0.31 & 0.0 & 3.0 & -303.90 & 1.98 & -297.88 & G4III-IV & ... \\ 
 J193133.34+424513.5 & KIC07203835 & 292.888950 & 42.753761 & 58025.00355 &       med-58025-HIP9645901\_sp02-011.fits & 88.3 & 5837 & 64 & 4.43 & 0.12 & 0.06 & 0.10 & 0.0 & 3.0 & -25.61 & 1.93 & -19.59 & G1.5V &  ... \\ 
 J193207.12+425118.9 & KIC07285973 & 293.029697 & 42.855268 & 58025.00356 &       med-58025-HIP9645901\_sp02-015.fits & 50.0 & 5222 & 93 & 4.56 & 0.10 & -0.07 &    0.15 & 0.0 & 3.0 & -37.64 & 3.32 & -31.62 & K0V & ... \\ 
  ...  &   ...  &   ...  &   ...  &   ...  &   ...  &   ...  &   ...  &   ...  &   ...  &   ...  &   ...  &   ...  &   ...  &  ...  &   ...  &   ...  &  ... & ... \\
  ...  &   ...  &   ...  &   ...  &   ...  &   ...  &   ...  &   ...  &   ...  &   ...  &   ...  &   ...  &   ...  &   ...  &  ...  &   ...  &   ... & ...  &  ... & ... \\
 J193943.93+444609.4 & KIC08633160 & 294.933065 & 44.769293 & 58025.00365 &       med-58025-HIP9645901\_sp04-238.fits & 109.1 & \dots & \dots & \dots & \dots & \dots &   \dots & \dots & \dots & \dots & \dots & \dots & ... & SB2 \\
 J193912.56+444012.1 & KIC08567854 & 294.802368 & 44.670039 & 58025.00364 &       med-58025-HIP9645901\_sp04-240.fits & 49.1 & 4509 & 96 & 2.40 & 0.38 & 0.16 & 0.14 & 0.0 & 3.0 & -24.24 & 3.19 & -18.22 & K2III & ... \\ 
 J194006.80+444459.3 & KIC08633453 & 295.028356 & 44.749829 & 58025.00366 &       med-58025-HIP9645901\_sp04-245.fits & 108.9 & 5664 & 70 & 4.12 & 0.16 & -0.39 & 0.11 & 12.1 & 4.8 & -96.47 & 3.13 & -90.45 & G2V & ... \\
 J193824.53+444434.9 & KIC08632169 & 294.602210 & 44.743035 & 58025.00363 &       med-58025-HIP9645901\_sp04-246.fits & 12.2 & 6095 & 73 & 4.16 & 0.13 & 0.06 & 0.15 & 20.0 & 3.0 & -1.52      & 2.83 & 4.50 & F8IV & ... \\
 J193953.83+443027.7 & KIC08503800 & 294.974298 & 44.507714 & 58025.00366 &       med-58025-HIP9645901\_sp04-247.fits & 69.0 & 6160 & 96 & 4.00 & 0.22 & 0.06 & 0.15 & 11.4 & 4.7 & -49.25 & 2.13 & -43.23 & F8IV & ... \\
  ...  &   ...  &   ...  &   ...  &   ...  &   ...  &   ...  &   ...  &   ...  &   ...  &   ...  &   ...  &   ...  &   ...  &  ...  &   ...  & ... &  ...  &  ... & ... \\
  ...  &   ...  &   ...  &   ...  &   ...  &   ...  &   ...  &   ...  &   ...  &   ...  &   ...  &   ...  &   ...  &   ...  &  ...  &   ...  & ... &  ...  &  ... & ... \\
\noalign{\smallskip}
\hline       
\noalign{\medskip}
\end{tabular}
~\\
{\bf Notes.} The full table is only available in electronic form at the CDS.
\label{Tab:data_blue}
\end{longtable}

\end{landscape}

\normalsize

\twocolumn

\begin{table*}[tb]
\caption[ ]{Triple (SB3) and double-lined (SB2) systems. }
\label{Tab:Binaries}
\begin{tabular}{ccccc}
\hline
\hline
\noalign{\smallskip}
 Designation & KIC & RA      &  Dec.    &  Rem \\ 
             &     & $\degr$ & $\degr$ &      \\ 
%\noalign{\smallskip} 
\hline
\noalign{\smallskip} 
J190458.00+400032.0 & KIC04907577 & 286.24167 & 40.00890 & SB3   \\ 
J192828.70+413244.3 & KIC06205852 & 292.11960 & 41.54564 & SB3   \\ 
J192634.43+415753.8 & KIC06521917 & 291.64346 & 41.96495 & SB3   \\ 
J191027.56+421429.4 & KIC06764812 & 287.61487 & 42.24151 & SB3   \\ 
J193031.19+424945.8 & KIC07284688 & 292.62997 & 42.82941 & SB3   \\ 
J185057.07+425854.3 & KIC07339348 & 282.73782 & 42.98175 & SB3   \\ 
J192542.68+425524.9 & KIC07361655 & 291.42786 & 42.92361 & SB3   \\ 
\noalign{\medskip}
J190208.98+391809.7 & KIC04243697 & 285.53745 & 39.30270 & SB2   \\ 
J190254.60+395028.6 & KIC04729553 & 285.72754 & 39.84128 & SB2   \\ 
J190458.80+395454.8 & KIC04819301 & 286.24500 & 39.91523 & SB2   \\ 
J190308.39+400159.7 & KIC04906571 & 285.78499 & 40.03327 & SB2   \\ 
J192000.87+401239.3 & KIC05093345 & 290.00366 & 40.21092 & SB2   \\ 
J190542.50+401850.5 & KIC05171053 & 286.42712 & 40.31405 & SB2   \\ 
J191735.03+401907.6 & KIC05178913 & 289.39597 & 40.31879 & SB2   \\ 
J192138.02+402057.1 & KIC05182476 & 290.40842 & 40.34920 & SB2   \\ 
J190818.91+402718.2 & KIC05260850 & 287.07882 & 40.45507 & SB2   \\ 
J190950.09+402542.7 & KIC05261743 & 287.45872 & 40.42854 & SB2   \\ 
J190717.40+403047.5 & KIC05347784 & 286.82251 & 40.51322 & SB2   \\ 
J192348.33+403253.3 & KIC05359678 & 290.95142 & 40.54816 & SB2   \\ 
J192945.54+410443.7 & KIC05792900 & 292.43979 & 41.07883 & SB2   \\ 
J191851.65+411057.9 & KIC05869903 & 289.71521 & 41.18278 & SB2   \\ 
J191108.51+412036.9 & KIC06030865 & 287.78547 & 41.34360 & SB2   \\ 
J191800.13+412651.1 & KIC06116234 & 289.50058 & 41.44753 & SB2   \\ 
J192937.51+413046.9 & KIC06206751 & 292.40631 & 41.51304 & SB2   \\ 
J190932.81+414456.3 & KIC06351674 & 287.38671 & 41.74898 & SB2   \\ 
J191238.41+414608.1 & KIC06353501 & 288.16006 & 41.76892 & SB2   \\ 
J191115.71+415221.0 & KIC06431515 & 287.81546 & 41.87250 & SB2   \\ 
J191118.95+414817.2 & KIC06431545 & 287.82898 & 41.80479 & SB2   \\ 
J192928.20+415151.7 & KIC06444482 & 292.36752 & 41.86438 & SB2   \\ 
J191043.23+415524.1 & KIC06510076 & 287.68014 & 41.92339 & SB2   \\ 
J192634.43+415753.8 & KIC06521917 & 291.64346 & 41.96495 & SB2   \\ 
J193052.32+415520.8 & KIC06525196 & 292.71802 & 41.92246 & SB2/3 \\ 
J191419.64+420124.5 & KIC06595804 & 288.58185 & 42.02349 & SB2   \\ 
J193115.60+420515.3 & KIC06609506 & 292.81500 & 42.08759 & SB2   \\ 
J193249.39+421154.0 & KIC06696265 & 293.20581 & 42.19834 & SB2   \\ 
J185140.38+421310.1 & KIC06754169 & 282.91826 & 42.21950 & SB2   \\ 
J191027.56+421429.7 & KIC06764812 & 287.61487 & 42.24159 & SB2   \\ 
J191958.38+421330.1 & KIC06771591 & 289.99326 & 42.22505 & SB2   \\ 
J192127.39+421354.6 & KIC06772830 & 290.36417 & 42.23185 & SB2   \\ 
J193309.67+421940.6 & KIC06868103 & 293.29031 & 42.32796 & SB2   \\ 
J190321.09+423259.3 & KIC07017065 & 285.83789 & 42.54981 & SB2   \\ 
J192521.03+423323.3 & KIC07032180 & 291.33765 & 42.55649 & SB2   \\ 
J185823.92+424138.9 & KIC07098225 & 284.59969 & 42.69416 & SB2   \\ 
J191717.71+425329.7 & KIC07274081 & 289.32379 & 42.89161 & SB2   \\ 
J192531.78+425113.8 & KIC07280725 & 291.38245 & 42.85385 & SB2   \\ 
J192542.68+425524.9 & KIC07361655 & 291.42786 & 42.92361 & SB2   \\ 
J192741.72+425520.6 & KIC07363238 & 291.92386 & 42.92240 & SB2   \\ 
J193409.38+430042.0 & KIC07450736 & 293.53912 & 43.01168 & SB2   \\ 
J184858.09+431036.1 & KIC07502940 & 282.24204 & 43.17670 & SB2   \\ 
J185504.28+431014.1 & KIC07506263 & 283.76785 & 43.17061 & SB2   \\ 
J191207.63+431415.6 & KIC07596873 & 288.03180 & 43.23768 & SB2   \\ 
J193005.83+431237.2 & KIC07609839 & 292.52429 & 43.21034 & SB2   \\ 
J192226.99+432308.6 & KIC07679049 & 290.61249 & 43.38574 & SB2   \\ 
J190806.21+432422.6 & KIC07741048 & 287.02588 & 43.40630 & SB2   \\ 
J192622.44+432505.3 & KIC07752458 & 291.59350 & 43.41817 & SB2   \\ 
J192516.67+433543.0 & KIC07821010 & 291.31949 & 43.59529 & SB2   \\ 
J193045.56+433352.4 & KIC07824811 & 292.68985 & 43.56457 & SB2   \\ 
J185127.31+434028.2 & KIC07871438 & 282.86380 & 43.67452 & SB2   \\ 
J192606.39+434142.6 & KIC07890271 & 291.52664 & 43.69517 & SB2   \\ 
\hline
\end{tabular}
\end{table*}

\addtocounter{table}{-1}

\begin{table*}[tb]
\caption[ ]{continued}
%Triple (SB3) and double-lined (SB2) systems. {\it Continued} }
\begin{tabular}{ccccc}
\hline
\hline
\noalign{\smallskip}
 Designation & KIC & RA      &  Dec.    &  Rem \\ 
             &     & $\degr$ & $\degr$ &      \\ 
%\noalign{\smallskip} 
\hline
\noalign{\smallskip} 
J193125.77+435258.5 & KIC08031672 & 292.85738 & 43.88293 & SB2   \\ 
J193335.51+435538.0 & KIC08102109 & 293.39800 & 43.92723 & SB2   \\ 
J191108.52+440656.9 & KIC08223197 & 287.78552 & 44.11581 & SB2   \\ 
J192529.50+441146.2 & KIC08232163 & 291.37292 & 44.19617 & SB2   \\ 
J191339.56+441639.7 & KIC08290831 & 288.41486 & 44.27771 & SB2   \\ 
J192911.21+441649.9 & KIC08301013 & 292.29674 & 44.28054 & SB2   \\ 
J191405.06+441855.9 & KIC08357101 & 288.52109 & 44.31555 & SB2   \\ 
J191701.73+441856.2 & KIC08358655 & 289.25723 & 44.31563 & SB2   \\ 
J185352.74+442646.5 & KIC08413015 & 283.46977 & 44.44625 & SB2   \\ 
J193009.44+442845.5 & KIC08432609 & 292.53937 & 44.47932 & SB2   \\ 
J191247.00+443043.8 & KIC08486354 & 288.19586 & 44.51218 & SB2   \\ 
J193819.10+443441.7 & KIC08502619 & 294.57959 & 44.57826 & SB2   \\ 
J184932.35+444608.5 & KIC08605198 & 282.38480 & 44.76904 & SB2   \\ 
J192829.16+444336.4 & KIC08625668 & 292.12152 & 44.72680 & SB2   \\ 
J193943.93+444609.4 & KIC08633160 & 294.93306 & 44.76929 & SB2   \\ 
J185510.16+444908.0 & KIC08672566 & 283.79236 & 44.81890 & SB2   \\ 
J191851.01+444903.4 & KIC08683646 & 289.71255 & 44.81763 & SB2   \\ 
J193812.85+445101.0 & KIC08696442 & 294.55356 & 44.85030 & SB2   \\ 
J193846.90+444834.4 & KIC08696850 & 294.69542 & 44.80958 & SB2   \\ 
J192103.10+445906.6 & KIC08749424 & 290.26294 & 44.98518 & SB2   \\ 
J192122.52+450201.7 & KIC08814972 & 290.34384 & 45.03382 & SB2   \\ 
J193450.17+450037.8 & KIC08823666 & 293.70905 & 45.01050 & SB2   \\ 
J192243.11+450659.9 & KIC08882516 & 290.67966 & 45.11666 & SB2   \\ 
J192208.12+451212.9 & KIC08949352 & 290.53387 & 45.20359 & SB2   \\ 
J194343.50+451608.2 & KIC08965321 & 295.93128 & 45.26897 & SB2   \\ 
J185128.05+451817.1 & KIC09001663 & 282.86688 & 45.30477 & SB2   \\ 
J193754.55+452055.6 & KIC09029503 & 294.47730 & 45.34880 & SB2   \\ 
J194253.86+451913.0 & KIC09033354 & 295.72446 & 45.32029 & SB2   \\ 
J185156.63+452428.8 & KIC09071104 & 282.98599 & 45.40800 & SB2   \\ 
J193056.28+453528.0 & KIC09156839 & 292.73453 & 45.59112 & SB2   \\ 
J194228.00+461614.2 & KIC09597928 & 295.61667 & 46.27063 & SB2   \\ 
J192653.21+463646.7 & KIC09830839 & 291.72174 & 46.61299 & SB2   \\ 
J192915.94+463719.8 & KIC09832227 & 292.31644 & 46.62219 & SB2   \\ 
J190830.14+464400.5 & KIC09881258 & 287.12561 & 46.73348 & SB2/3 \\ 
J193141.21+464843.5 & KIC09953894 & 292.92172 & 46.81211 & SB2   \\ 
J193754.79+465149.7 & KIC09957668 & 294.47830 & 46.86381 & SB2   \\ 
J191128.13+465859.0 & KIC10002792 & 287.86722 & 46.98308 & SB2   \\ 
J191343.38+470425.2 & KIC10067180 & 288.43076 & 47.07367 & SB2   \\ 
J192740.73+472352.8 & KIC10274200 & 291.91974 & 47.39801 & SB2   \\ 
J191022.82+473004.9 & KIC10395363 & 287.59512 & 47.50138 & SB2   \\ 
J195534.80+473219.4 & KIC10425836 & 298.89502 & 47.53874 & SB2   \\ 
J192752.05+474154.4 & KIC10470616 & 291.96689 & 47.69846 & SB2   \\ 
J194753.94+474112.6 & KIC10484679 & 296.97476 & 47.68684 & SB2   \\ 
J191955.40+475931.6 & KIC10661074 & 289.98085 & 47.99212 & SB2   \\ 
J192451.69+480440.6 & KIC10729622 & 291.21539 & 48.07796 & SB2   \\ 
J195250.62+483332.9 & KIC11045383 & 298.21094 & 48.55916 & SB2   \\ 
J191241.95+483914.5 & KIC11077129 & 288.17481 & 48.65405 & SB2   \\ 
J194934.49+485409.7 & KIC11257457 & 297.39373 & 48.90270 & SB2   \\ 
J191401.65+490541.5 & KIC11290835 & 288.50688 & 49.09488 & SB2   \\ 
\noalign{\smallskip} 
\hline       
\noalign{\smallskip} 
\end{tabular}
\\
\end{table*}

\section{Spectral energy distribution}
\label{Appendix:SED}

\begin{figure}[ht]
\includegraphics[width=9.0cm]{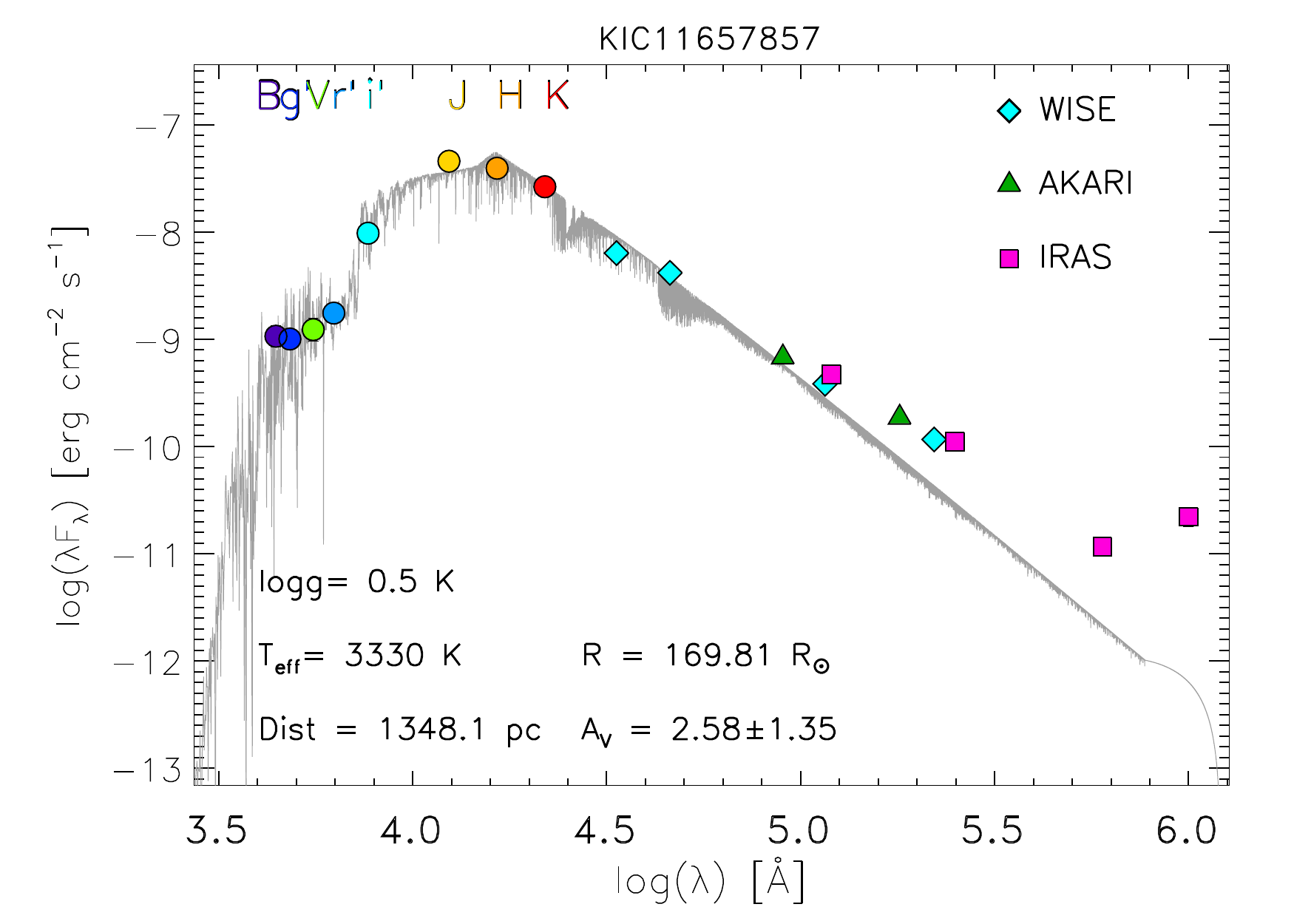}           
\caption{Spectral energy distribution of the AGB star KIC~11657857. Optical fluxes from APASS \citep{APASS} and near-infrared fluxes from the 2MASS catalog \citep{2MASS} are displayed with colored dots. Mid- and far-infrared fluxes from WISE \citep{WISE}, AKARI \citep{AKARI}, and IRAS \citep{IRAS2015} are shown with diamonds, triangles, and squares, respectively.
The BT-Settl spectrum \citep{Allard2012} that provides the best fit to the star photosphere up to the $J$ band is shown with a gray line.}
\label{Fig:SED_KIC11657857}
\end{figure}

\begin{figure}[ht]
\includegraphics[width=9.0cm]{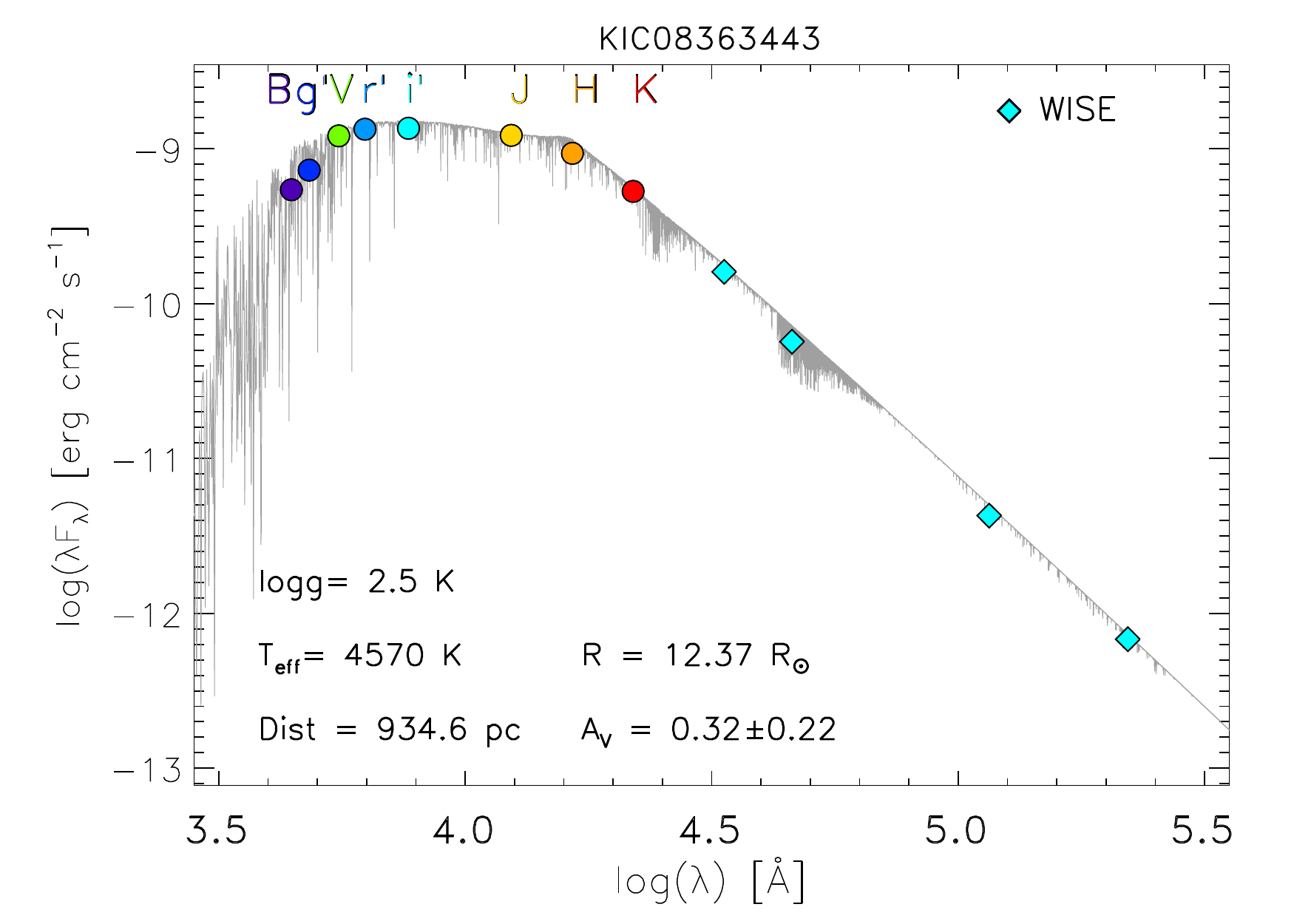}           
\caption{Spectral energy distribution of the Li-rich giant KIC~8363443 from the optical/near-infrared (colored dots) to the mid-infrared bands (diamonds). 
The BT-Settl spectrum \citep{Allard2012} that provides the best fit to the star photosphere up to the $K$ band is shown with a gray line.}
\label{Fig:SED_KIC8363443}
\end{figure}

From optical $BVg'r'i'$ photometry taken from the AAVSO Photometric All Sky Survey (APASS) catalog \citep{APASS} and near-infrared photometric data from 2MASS \citep{2MASS}, and mid-infrared WISE data \citep{WISE} we built the corresponding SED, which
was fitted with BT-Settl synthetic spectra \citep{Allard2012}. For each target, we adopted its $Gaia$ EDR3 parallax as well as the APs (\teff\ and \logg) derived in the present work, leaving the stellar radius ($R$) and the interstellar extinction $A_V$ free to vary. The values of $R$ and $A_V$ were then obtained by $\chi^2$ minimization of the flux differences between observed and synthetic SED. We fitted all the optical/near-infrared bands ($BVg'r'i'JHK$), with the exception of stars with infrared excess for which $J$ was the reddest band for the fit. Finally, the stellar luminosity was calculated as 
$L$=4\,$\pi$\,$R^2$\,$\sigma$\,$T_{\textrm{eff}}^4$. Some examples of this fitting are shown in Figs.~\ref{Fig:SED_KIC11657857}, \ref{Fig:SED_KIC8363443}, and \ref{Fig:SED_KIC8749284}.
The errors on $A_V$ and $R$ are found by the minimization procedure considering the 1$\sigma$ confidence level of the $\chi^2$ map, but we have also taken the error on \teff\ into account.

\begin{figure}[ht]
\includegraphics[width=9.0cm]{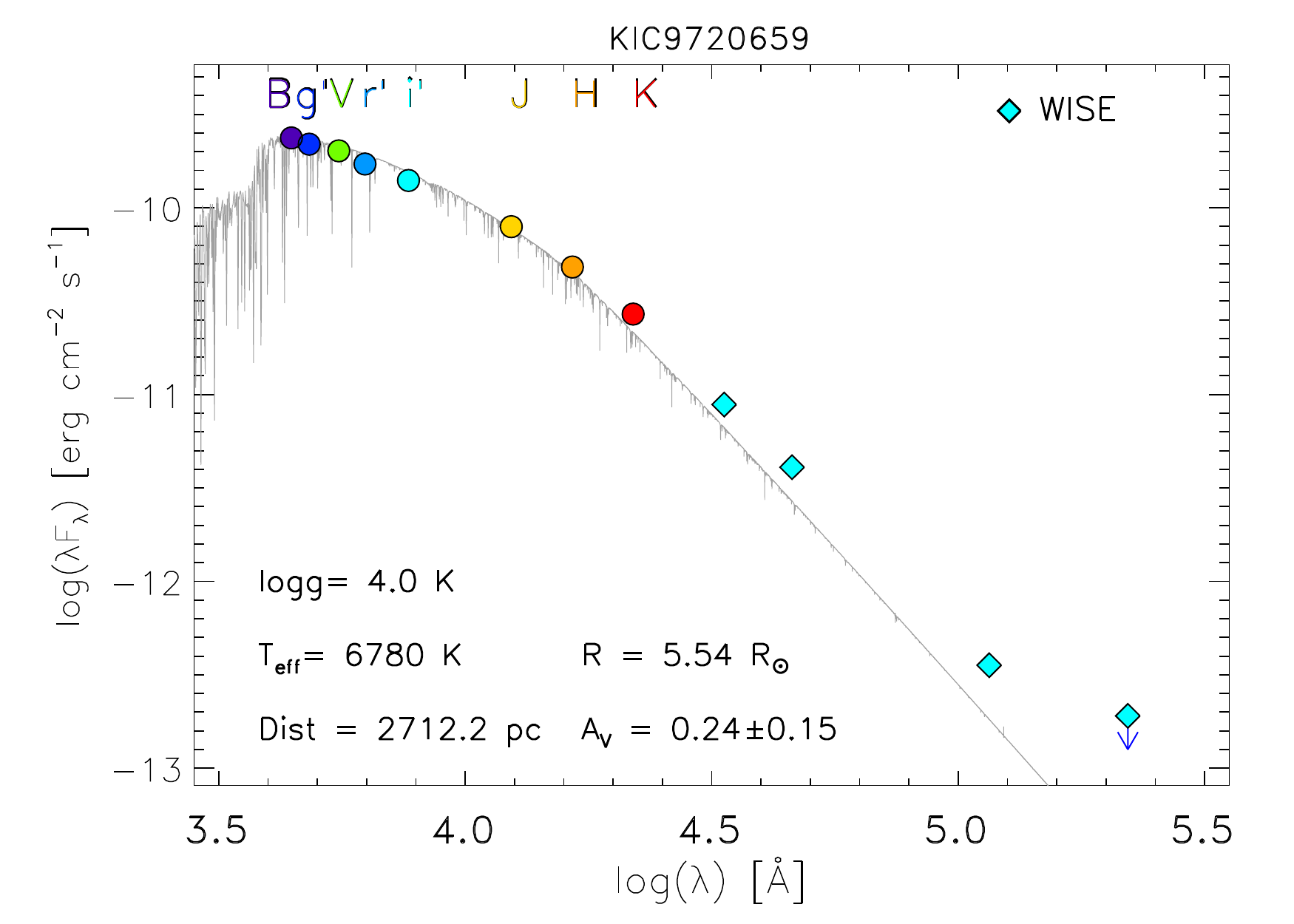}           
\caption{Spectral energy distribution of the possible Herbig star KIC~9720659 from
the optical/near-infrared (colored dots) to the mid-infrared bands (diamonds). 
The BT-Settl spectrum \citep{Allard2012} that provides the best fit to the star photosphere up to the $J$ band is shown with a gray line.}
\label{Fig:SED_KIC9720659}
\end{figure}

\begin{figure}[ht]
\includegraphics[width=9.0cm]{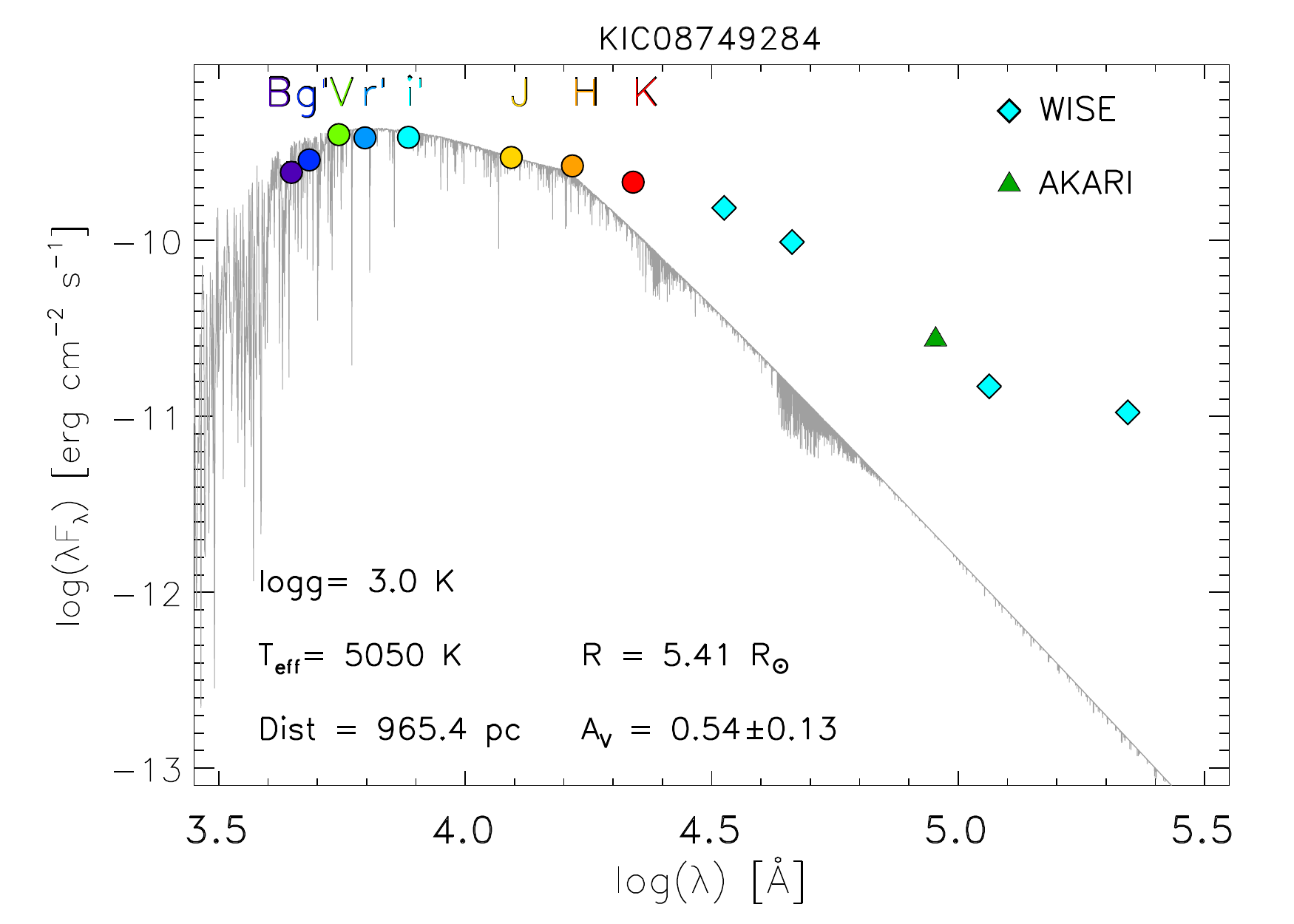}           
\caption{Spectral energy distribution of the active giant KIC~8749284. Mid-infrared fluxes are shown with different symbols (diamonds for WISE and a triangle for AKARI).
The BT-Settl spectrum \citep{Allard2012} that provides the best fit to the star photosphere up to the $J$ band is shown with a gray line.}
\label{Fig:SED_KIC8749284}
\end{figure}

\end{appendix}

\end{document}